\documentclass[prd,nofootinbib,reprint,preprintnumbers,showkeys]{revtex4-2}
\usepackage{xcolor}
\usepackage{amsmath,amsfonts,bm}
\usepackage{graphicx, epstopdf,scrextend}
\usepackage{mathrsfs,mathtools}
\usepackage{enumitem}
\usepackage{tikz}
\usepackage{tikz-feynman}

\definecolor{mygreen}{rgb}{0,0.6,0}
\definecolor{myorange}{rgb}{0.8,0.5,0}
\definecolor{darkgreen}{rgb}{0,0.4,0} 
\definecolor{darkblue}{rgb}{0,0,0.6} 
\usepackage[colorlinks=true,linkcolor=darkblue,citecolor=darkgreen]{hyperref}

\usepackage{pgfplots,pgfplotstable}
\usepgfplotslibrary{fillbetween}
\usetikzlibrary{intersections}
\pgfplotsset{compat=newest}

\newcommand{\as}{\alpha_{\mathrm{s}}}

\newcommand{\LA}{\mathrm{A}}
\newcommand{\LB}{\mathrm{B}}
\newcommand{\LC}{\mathrm{C}}
\newcommand{\scC}{\textsc{c}}

\newcommand{\scE}{\textsc{e}}
\newcommand{\LF}{\mathrm{F}}

\newcommand{\scH}{\textsc{h}}

\newcommand{\LR}{\mathrm{R}}
\newcommand{\scR}{\textsc{r}}

\newcommand{\scS}{\textsc{s}}
\newcommand{\LT}{\mathrm{T}}
\newcommand{\La}{\mathrm{a}}
\newcommand{\Lb}{\mathrm{b}}
\newcommand{\Lc}{\mathrm{c}}

\newcommand{\Lf}{\mathrm{f}}
\newcommand{\Lg}{\mathrm{g}}

\newcommand{\Lp}{\mathrm{p}}

\newcommand{\Ls}{\mathrm{s}}
\newcommand{\Lu}{\mathrm{u}}

\newcommand{\scV}{\textsc{v}}

\newcommand{\MSbar}{\overline{\mathrm{MS}}}
\newcommand{\msbar}{
  {\overline{\phantom{\raisebox{-0.6 pt}{$\scriptstyle{\textsc{ms}}$}}}
    \hskip - 8pt \scriptstyle{\textsc{ms}}}}

\newcommand{\mpone}{{m\!+\!1}}

\newcommand{\GeV}{\ \mathrm{GeV}}
\newcommand{\TeV}{\ \mathrm{TeV}}

\newcommand{\mur}{\mu_\scR}
\newcommand{\mus}{\mu_{\textsc s}}
\newcommand{\muf}{\mu_{\rm f}}
\newcommand{\muh}{\mu_{\scH}}

\newcommand{\cD}{\mathcal{D}}
\newcommand{\cE}{\mathcal{E}}
\newcommand{\cF}{\mathcal{F}}

\newcommand{\cK}{\mathcal{K}}

\newcommand{\cO}{\mathcal{O}}

\newcommand{\cS}{\mathcal{S}}

\newcommand{\cU}{\mathcal{U}}
\newcommand{\cV}{\mathcal{V}}

\newcommand{\cX}{\mathcal{X}}

\newcommand{\cZ}{\mathcal{Z}}

\definecolor{red}{rgb}{1,0,0}

\def\P#1{{\color{red}\big[}{#1}{\color{red}\big]_\mathbb{P}}}

\def\iP#1{{\color{red}[}{#1}{\color{red}]_\mathbb{P}}}

\def\mi{{\mathrm i}}

\def\ket#1{\big|{#1}\big\rangle}
\def\bra#1{\big\langle{#1}\big|}
\def\brax#1{\big\langle{#1}}   
\def\<>#1{\big\langle{#1}\big\rangle}
\def\[]#1{\big[{#1}\big]}

\def\sket#1{\big|{#1}\big)}
\def\sbra#1{\big({#1}\big|}
\def\sbrax#1{\big({#1}}        

\def\isket#1{|{#1})}
\def\isbra#1{({#1}|}
\def\isbrax#1{({#1}}

\def\iket#1{|{#1}\rangle}
\def\ibra#1{\langle{#1}|}

\usepackage{pgfplots,pgfplotstable}

\pgfplotsset{every axis/.style={
    width=8.2cm,
    height=8.2cm,
    grid=both,
    scaled ticks=false,
    yticklabel style={/pgf/number format/.cd, fixed,precision=5}
  }
}

\newif\ifusefigs
\usefigstrue

\begin{document}

\title{Multivariable evolution in parton showers with initial state partons}

\author{Zolt\'an Nagy}

\affiliation{
 Deutsches Elektronen-Synchrotron DESY, 
 Notkestr.\ 85, 22607 Hamburg, Germany
}

\email{Zoltan.Nagy@desy.de}

\author{Davison E.\ Soper}

\affiliation{
Institute for Fundamental Science,
University of Oregon,
Eugene, OR  97403-5203, USA
}

\email{soper@uoregon.edu}

\begin{abstract}
One can use more than one scale variable to define the family of surfaces in the space of parton splitting parameters that define the evolution of a parton shower. In an earlier paper, we developed this idea for electron-positron annihilation. Here, we use multiple scale variables for a parton shower with initial state partons. Then we need a more sophisticated analysis because the evolution of parton distribution functions must be coordinated with the parton shower evolution. We make the needed connections more precise than in our earlier work, even for the case of just one scale variable. Then we develop an example with three scale variables, which leads to advantages compared to the usual shower formulation with only one scale variable. We provide results for Drell-Yan muon pair production.
\end{abstract}

\date{11 April 2022}

\preprint{DESY-22-049}

\maketitle

\tableofcontents

\makeatletter
\let\toc@pre\relax
\let\toc@post\relax
\makeatother 

\section{Introduction}
\label{sec:introduction}

In a parton shower event generator, one can view the parton state as evolving according to an operator based renormalization group equation. Starting with a state with just a few partons, the shower evolves as a scale $\mus$ changes from a large value $\muh$ characteristic of the starting state to a low value $\muf$ on the order of $1 \GeV$. As the shower evolves, more and more partons are emitted. The function of the shower scale $\mus$ is to divide possible parton splittings into {\em resolvable} splittings, with scales $\mu > \mus$, and {\em unresolvable} splittings, with scales $\mu < \mus$. There is substantial freedom to choose exactly what this means. The space of possible splittings is divided into the resolvable and unresolvable regions by a surface labelled by $\mus$. Many different choices are possible for defining this surface. For instance, one can use a measure of the transverse momentum in the splitting to define the surface or one can use a measure of the virtuality in the splitting.

In this paper, we explore the possibility of using more than one variable to define a family of surfaces within the framework of our parton shower program \textsc{Deductor}. Instead of one $\mu_\scS$, we use $\vec \mu_\scS = (\mu_{\scS,1}, \mu_{\scS,2}, \dots)$. Then evolution means moving from large values of the component scales $\mu_{\scS,n}$ to small values along a path $\vec \mu_\scS(t)$. Defining this path is then part of defining the shower algorithm. 

There is an additional freedom available when multiple scales are involved. It may be possible to divide the shower splitting functions into separate terms such that one of the terms is not sensitive to one of the scales in the sense that no singularity is encountered when this scale approaches zero. When this happens, we can modify the definition of the unresolved region for this term in a way that makes this term exactly independent of this scale. This redefinition can simplify the shower evolution.

In this paper, we explore the additional freedom obtained by using three scales instead of one.

This general concept works for proton-proton, $e^\pm$-proton, and $e^+e^-$ collisions. In Ref.~\cite{angleFS}, we considered the simplest case, $e^+e^-$ collisions. In this paper, we consider collisions involving incoming hadrons, with an emphasis on hadron-hadron collisions. With incoming hadrons, the needed theoretical construction is more involved than in $e^+e^-$ annihilation. First, the representation of the cross section needs both a shower operator $\cU(t_\Lf, t_\scH)$ and an operator $\cU_\cV(t_\Lf, t_\scH)$ that sums threshold logarithms. Second, the description of initial state splittings using ``backward evolution'' requires the use of parton distribution functions (PDFs). Since the evolution of these functions must be coordinated with the evolution of the shower, they are not, in general, the $\MSbar$ PDFs used in fixed order perturbation theory. We refer to the PDFs used in the shower as shower oriented PDFs. We review this connection in some detail in this paper, expanding on our previous treatments.

Much of the analysis of this paper concerns what we call the infrared sensitive operator \cite{NSAllOrder}, $\cD(\mur, \vec \mu_\scS)$, where $\mur$ is the renormalization scale and $\vec \mu_\scS$ represents one or more shower scales that define the unresolved region. In a first order shower with just one shower scale and with $\mur = \mus$, we use the first order contribution $\cD^{[1]}(\mus, \mus)$. This can be thought of as being the more familiar shower splitting operator, $\cS^{[1]}(\mu')$ integrated over $\mu'$ from $\mu' = \mus$ down to $\mu' = 0$. This integral would be infrared divergent, but we perform the integration in $4 - 2 \epsilon$ dimensions to regulate the divergence. This makes $\cD^{[1]}$ seem like a derived quantity, but we consider $\cD^{[1]}$ to be the starting point of the definition of a shower \cite{NSAllOrder}. Then it is the shower splitting operator $\cS^{[1]}$ that is the derived quantity. By using $\cD$ as the starting point, we connect to the definitions of renormalization, factorization, and parton distribution functions used for the calculation of cross sections at any perturbative order in QCD.
 
We introduce the analysis in this paper with a review in Sec.~\ref{sec:statisticalspace} of the vector space of parton states used in the \textsc{Deductor} framework. In Sec.~\ref{sec:IRsensitiveoperator}, we sketch the role of the infrared sensitive operator $\cD$ and, with some important notation established, we provide in Sec.~\ref{sec:new} a preview of what is technically new in this paper.

We turn in Sec.~\ref{sec:cDstructure} to a detailed analysis of the structure of $\cD$. There is another operator that we need, the infrared finite operator $\cV$. We analyze its structure in Sec.~\ref{sec:IRfiniteoperator}. With the operators $\cD$ and $\cV$ available, we turn in Sec.~\ref{sec:cU} to the operator $\cU(t',t)$ that generates the probability preserving parton shower and the operator $\cU_\cV(t',t)$ that corrects for the mismatch between shower evolution and PDF evolution and thereby generates a summation of threshold logarithms. This leads us to the definition of the shower oriented PDFs in Sec.~\ref{sec:PDFs}.

In Sec.~\ref{sec:multiplescales}, we start with one scale that represents $k_\LT$ ordering, \textsc{Deductor}'s standard $\Lambda$ ordering, or angular ordering. Then we add two additional scale parameters, $\mu_\scE$ that controls soft splittings and $\mu_{\mi \pi}$ that controls the imaginary part of virtual exchange graphs. Next, we see how one can modify the definition of the unresolved region for parts of the generators of $\cU(t',t)$ and $\cU_\cV(t',t)$ that are sensitive only to some of the scale parameters but not others. In Sec.~\ref{sec:pathchoice} we suggest a special three component choice of path in the space of scale parameters. In Sec.~\ref{sec:crosssection} we show how the operators used to express an infrared safe cross section behave on the suggested path. 

We provide a numerical example in Sec.~\ref{sec:DrellYan}: the Drell-Yan cross section for muon pair production via vector boson production. We offer a summary and some comments in Sec.~\ref{sec:Conclusions}.

We also include a number of appendices, \ref{sec:ISsplittings}, \ref{sec:poles}, \ref{sec:cV}, \ref{sec:cScV}, \ref{sec:thresholdsum} and \ref{sec:twoscales}, that document derivations and formulas that are treated only briefly in the main text.

\section{Partons and the statistical space}
\label{sec:statisticalspace}

This paper generally concerns the definition of the unresolved region in the space of parton momenta in a parton shower and then how the definition of the unresolved region affects the parton shower. We will concentrate on emissions in a first order shower, but we include a discussion of the general case of a shower algorithm with splitting functions defined at an arbitrary order of perturbation theory. We use the general framework presented in Ref.~\cite{NSAllOrder}. This general framework allows for substantial freedom in choosing the functions that define a particular parton shower algorithm. We have developed a particular realization of these choices for a first order shower \cite{NSI}, a realization that is in many ways similar to other first order parton shower algorithms. It remains an open problem to realize these choices for a parton shower with splitting functions beyond order $\as$. 

The description that we use is based on linear operators on a vector space that describes the state of the partons in the shower, which we call the statistical space \cite{NSI}. We begin by recalling the nature of this space.

We consider the description of a cross section in hadron-hadron collisions. At a particular stage in the shower, there are $m$ final state partons plus two initial state partons with labels ``a'' and ``b.'' The partons have momenta and flavors $\{p,f\}_m = \{p_\La,f_\La; p_\Lb, f_\Lb; p_1,f_1; p_2,f_2,\dots,p_m,f_m\}$. We define $Q$ by
\begin{equation}
\label{eq:ISmomentumsum}
\sum_{i=1}^m p_i = Q
\;.
\end{equation}
Then also $Q = p_\La + p_\Lb$. 

The theory is expressed using linear operators that act on a vector space that we call the statistical space. Basis vectors for this space have the form $\isket{\{p,f,c, c', s, s'\}_m}$. Here $(c,c')$ and $(s,s')$ represent the quantum colors and spins of the initial and final state partons. We use the apparatus of quantum statistical mechanics, with the color and spin part of $\isket{\{p, f, c, c', s, s'\}_m}$ representing the density matrix $\iket{\{c,s\}_m} \ibra{\{c',s'\}_m}$.

In order to properly incorporate quantum mechanics into a parton shower, one must include both quantum color and spin. It is quite straightforward to include both color and spin in the evolution equations for a first order shower. For color, one then needs an approximation to obtain a description that can be implemented in a practical computer program. \textsc{Deductor} uses the ``LC+'' approximation \cite{NScolor}. Thus our description in this paper uses full color, with the LC+ approximation playing a special role in Secs.~\ref{sec:softsplittings}, \ref{sec:pathchoice}, and \ref{sec:crosssection}. A proper implementation of spin \cite{JCCspin, NSspin} is replaced by simple spin averaging in \textsc{Deductor}. Because of this, we omit spin quantum numbers in everything that follows in this paper. This results in a simpler notation without seriously obscuring conceptual issues. Using quantum color but not spin, the basis vectors for the statistical space have the form $\isket{\{p, f, c, c'\}_m}$, where the color quantum numbers represent the density matrix $\iket{\{c\}_m} \ibra{\{c'\}_m}$. We use the trace basis for the vectors $\iket{\{c\}_m}$ \cite{NSI}.

In order to use the statistical space to express a cross section, we need the probability associated with a vector $\isket{\rho}$ in the statistical space. We define a vector $\isbra{1}$ so that the probability associated with  $\isket{\rho}$ is $\isbrax{1}\isket{\rho}$ \cite{NSI}. We define $\isbra{1}$ by using the probability associated with a basis state $\isket{\{p, f, c, c'\}_m}$:
\begin{equation}
\begin{split}
\label{eq:probabilitydef}
\sbrax{1}\sket{\{p, f,  c, c'\}_m} 
= \sbrax{1_\mathrm{pf}}\sket{\{p, f\}_m}
\sbrax{1_\mathrm{color}}\sket{\{c, c'\}_m} 
\end{split}
\end{equation}
with
\begin{equation}
\begin{split}
\label{eq:probabilitypartsdef}
\sbrax{1_\mathrm{pf}}\sket{\{p, f\}_m} 
={}& 1
\;,\\
\sbrax{1_\mathrm{color}}\sket{\{c, c'\}_m} 
={}& \brax{\{c'\}_m}\ket{\{c\}_m}
\;.
\end{split}
\end{equation}
%

\section{The infrared sensitive operator}
\label{sec:IRsensitiveoperator}

The shower description in Ref.~\cite{NSAllOrder} begins with an operator on the statistical space, $\cD(\mu_\scR^2,\mu_\scS^2)$, that we call the infrared sensitive operator. This operator includes a specification of what one means by an unresolved splitting in the shower. In this section, we introduce this operator and the operators that are derived from it.

The operator $\cD(\mu_\scR^2,\mu_\scS^2)$ describes the soft and collinear singularities of QCD. Here $\mu_\scR$ is the standard renormalization scale and $\mu_\scS$ is called the shower scale. In this paper, we contemplate the possibility of having more than one independent shower scale, $\vec \mu_\scS = (\mu_{\scS,1}, \mu_{\scS,2}, \dots)$. The infrared sensitive operator is expanded in operators $\cD^{[n]}(\mu_\scR, \vec\mu_\scS)$ that are proportional to $\as^n$:
\begin{equation}
\label{eq:cDfullexpansion1}
\cD(\mu_\scR, \vec\mu_\scS) ={} 1 +
\sum_{n = 1}^\infty \cD^{[n]}(\mu_\scR, \vec\mu_\scS)
\;.
\end{equation}
The operator $\cD^{[n]}$ is further expanded as
\begin{equation}
\label{eq:cDfullexpansion2}
\cD^{[n]}(\mu_\scR, \vec\mu_\scS) ={} 
\mathop{\sum_{n_\scR = 0}^n \sum_{n_\scV = 0}^n}_{n_\scR + n_\scV = n}\,
\cD^{[n_\scR,n_\scV]}(\mu_\scR, \vec\mu_\scS)
\;.
\end{equation}
Here $\cD^{[n_\scR,n_\scV]}$ creates $n_\scR$ real emissions and $n_\scV$ virtual exchanges. After one of these operators acts on a state $\isket{\{p, f, c, c'\}_m}$, we have partons with momenta and flavors $\{\hat p,\hat f\}_{\hat m}$ with $\hat m \ge m$. The new total momentum is
\begin{equation}
\label{eq:newISmomentumsum}
\sum_{i=1}^{\hat m} \hat p_i = \hat Q
\;.
\end{equation}

The first order contribution to $\cD$ forms the basis for a first order shower. It consists of
\begin{equation}
\begin{split}
\label{eq:cD1decomposition}
\cD^{[1]}(\mu_\scR, \vec\mu_\scS)
={}& \sum_l
\cD^{[1]}_l(\mu_\scR, \vec\mu_\scS)
\\
={}& \sum_l \left[\cD^{[1,0]}_l(\mu_\scR, \vec\mu_\scS)
+ \cD^{[0,1]}_l(\mu_\scR, \vec\mu_\scS)
\right]
\;.
\end{split}
\end{equation}
In $\cD^{[1,0]}_l$, the parton labelled $l$ splits into two partons, one labelled $l$ and one labelled $\mpone$. Here $l \in \{\La, \Lb, 1, \dots, m\}$. The operator $\cD^{[0,1]}_l$ corresponds to virtual graphs, either self-energy graphs for parton $l$ or graphs in which a gluon is exchanged between parton $l$ and another parton.  The virtual exchange  part has both real and imaginary contributions, $\cD^{[0,1]}_l = \mathrm{Re}\,\cD^{[0,1]}_l + \mi\,\mathrm{Im}\,\cD^{[0,1]}_l$. We examine the imaginary contributions in Sec.~\ref{sec:ImcD01}. Until then, we mostly concentrate on the real contributions.

The operators $\cD^{[1,0]}_l$ and the real parts of $\cD^{[0,1]}_l$ for $l \in \{1, \dots, m\}$, as well as their connection to the parton shower, were described in some detail in Ref.~\cite{angleFS}. This leaves $l \in \{\La,\Lb\}$. The case $l = \Lb$ is the same as $l = \La$ with the replacement $\La \leftrightarrow \Lb$, so it suffices to describe the case $l = \La$.

The infrared sensitive operator $\cD$ provides the basis for the definition of the probability preserving parton shower evolution operator (\cite{NSLogSum}, Eqs.~(27), (51), and (54), or \cite{NSAllOrder})
\begin{equation}
\label{eq:cUfromcS}
\cU(t_2,t_1) = \mathbb{T}\exp\!\left(\int_{t_1}^{t_2}\!
dt\, \cS(t)\right)
\;.
\end{equation}
Here the scale parameters evolve along a path $\mur(t), \vec \mu_s(t)$ as a function of the {\em shower time} $t$ and $\mathbb{T}$ indicates ordering in $t$. The generator operator $\cS$ at lowest perturbative order is related to parton distribution functions supplied by an operator $\cF(\mur^2) = \cF_{\!\La}(\mur^2) \cF_{\!\Lb}(\mur^2)$ defined in Eq.~(\ref{eq:Fdef}) and to the infrared sensitive operator at lowest perturbative order, $\cD^{[1]}$. 

The first order generator of the probability preserving shower consists of three terms,
\begin{equation}
\label{eq:cS1decomposition1}
\cS^{[1]}(t) = \cS^{[1,0]}(t) - \P{\cS^{[1,0]}(t)}
+ \mi\pi\cS^{[0,1]}_{\mi\pi}(t)
\;.
\end{equation}
The operator in the first term adds one emitted parton to the statistical state. It is obtained by differentiating $\cD^{[1,0]}$:
\begin{equation}
\begin{split}
\label{eq:cS10}
\cS^{[1,0]}(t) ={}& - 
\cF(\mur^2(t))
\sum_i
\\&\times \frac{d\mu_{\scS,i}^2(t)}{dt}
\frac{\partial \cD^{[1,0]}(\mur(t),\vec\mu_\scS(t))}{\partial \mu_{\scS,i}^2}
\\&\times
\cF^{-1}(\mur^2(t))
\;.
\end{split}
\end{equation}
The operator in the second term, which we denote by $\P{\cS^{[1,0]}(t)}$, leaves the number of partons and their momenta and flavors unchanged. As described in Appendix \ref{sec:ISsplittings}, it is determined from the real emission operator $\cS^{[1,0]}(t)$ by integrating over the splitting variables for a splitting that might have happened, but did not. With this definition, $\cU(t_2,t_1)$ conserves the total probability of the statistical state: $\sbra{1}\cU(t_2,t_1)= \sbra{1}$. The operator in the third term also leaves the number of partons and their momenta and flavors unchanged. It is obtained by differentiating the imaginary part of the virtual exchange operator $\cD^{[0,1]}$:
\begin{equation}
\begin{split}
\label{eq:SiPi}
\mi\pi &\cS^{[0,1]}_{\mi\pi}(t) 
\\& = - 
\sum_i\frac{d\mu_{\scS,i}^2(t)}{dt}
\frac{\partial\, \mi\,\mathrm{Im}\,\cD^{[0,1]}(\mur(t),\vec\mu_\scS(t))}
{\partial \mu_{\scS,i}^2}
\;.
\end{split}
\end{equation}
This operator does not change probabilities: $\sbra{1}\cS^{[0,1]}_{\mi\pi}(t) = 0$.
We return to these relations in more detail in Sec.~\ref{sec:cU}.

The infrared sensitive operator $\cD$ also provides the basis for the definition of the inclusive infrared finite operator $\cV(\mur,\vec \mu_\scS)$ \cite{NSAllOrder, NSThresholdII}, which we examine starting in Sec.~\ref{sec:IRfiniteoperator}. The structure of this operator is crucial to the definition of shower oriented parton distribution functions, examined in Sec.~\ref{sec:PDFs}. It is also needed for the definition of the operator $\cU_\cV(t,t')$ that produces the summation of threshold logarithms and is examined in Sec.~\ref{sec:cU}.

\section{New features in this paper}
\label{sec:new}

The structure of the shower cross section with initial state splittings is more complicated than with just final state splittings because this structure has to include parton distribution functions and the interplay between PDF evolution and shower evolution. For this reason, we provide a rather detailed analysis in what follows. We extend the analysis of Refs.~\cite{ISevolve, NSAllOrder, NSThreshold, NSThresholdII} in several ways. First, we allow for the presence of more than one independent shower scale: $\mu_\scS \to \vec \mu_\scS$. Second, we present a more precise derivation of the needed relation between standard $\MSbar$ PDFs and the PDFs needed internally in the shower. Third, we provide an improved determination of the shower oriented PDFs from the $\MSbar$ PDFs. Fourth, we write the real emission operator $\cD^{[1,0]}_\La(\mu_\scR, \vec\mu_\scS)$ in what we think is a clearer notation. Fifth, we provide a simpler definition of the real part of the virtual exchange operator $\mathrm{Re}\,\cD^{[0,1]}_\La(\mu_\scR, \vec\mu_\scS)$ than appears in Refs.~\cite{NSAllOrder, NSThreshold, NSThresholdII}.  Sixth, we provide a simpler definition of the imaginary part of the virtual exchange operator $\mathrm{Im}\,\cD^{[0,1]}(\mu_\scR, \vec\mu_\scS)$ that appears in Refs.~\cite{NScolor, NSColoriPi}.

\section{Structure of the infrared sensitive operator}
\label{sec:cDstructure}

The infrared sensitive operator includes a specification of what one means by an unresolved splitting in the shower. We cover what this means in several steps in this section.

\subsection{Kinematics of parton splitting at first order}
\label{sec:kinematics}

In the real emission operator $\cD^{[1,0]}_\La(\mu_\scR, \vec\mu_\scS)$, initial state parton ``a'' splits, in the sense of backward evolution \cite{backwardevolution}. Before the splitting, the momentum of this parton is
\begin{equation}
\label{eq:etadef}
p_\La = \eta_\La p_\LA
\;,
\end{equation}
where $p_\LA$ is the momentum of the incoming hadron (with the approximation $p_\LA^2 = 0$) and $\eta_\La$ is the momentum fraction carried by the incoming parton. The new initial state parton carries momentum 
\begin{equation}
\label{eq:hatetadef}
\hat p_\La = \hat \eta_\La p_\LA
\;.
\end{equation}
A new final state parton carrying momentum $\hat p_\mpone$ with $\hat p_\mpone^2 = 0$, is produced. We define the momentum fraction in the splitting by
\begin{equation}
\label{eq:zdef}
z = \frac{\eta}{\hat \eta}
\;.
\end{equation}
We define a dimensionless virtuality variable for the splitting by
\begin{equation}
\label{eq:ydef}
y = \frac{2 \hat p_\La \cdot \hat p_\mpone}{2 p_\La \cdot Q}
=\frac{(\hat p_\La + \hat p_\mpone)^2}{Q^2}
\;.
\end{equation}
We also denote by $\phi$ the azimuthal angle of $\hat p_\mpone$ around the direction of $p_\La$ in the rest frame of $Q$. The three splitting variables $y,z,\phi$ suffice to define the splitting. The momenta $\{\hat p\}_\mpone$ of the partons after the splitting are determined by $y,z,\phi$ and the momenta $\{p\}_m$, as described in Appendix \ref{sec:ISsplittings}.

It is sometimes useful to define a transverse momentum variable for an initial state splitting,
\begin{equation}
\label{eq:kTdef}
k_\LT^2 = (1-z) y Q^2
\;,
\end{equation}
as in Eq.~(A.8) of Ref.~\cite{NSThreshold}. This is approximately the absolute value of the square of the part of $\hat p_\mpone$ transverse to $p_\La$ and $Q$.\footnote{The exact value is $-(\hat p_\mpone^\perp)^2 = y (1-z - zy)Q^2$.}

We will specify $\cD^{[1,0]}_\La(\mu_\scR, \vec\mu_\scS)$ in some detail in Appendix \ref{sec:ISsplittings}, but for now these details do not matter. What is important is that $\cD^{[1,0]}_\La$ exhibits collinear and soft singularities. To describe these, it is useful to define an angular variable
\begin{equation}
\label{eq:varthetafromy}
\vartheta = \frac{y z}{1-z}
\;.
\end{equation}
This is $(1-\cos\theta)/2$ where $\theta$ is the angle between $\hat p_\La$ and $\hat p_{\mpone}$ as measured in the rest frame of $\hat Q$. The operator $\cD^{[1,0]}_\La$ is singular in the collinear limit $\vartheta \to 0$ with fixed $z$, in the soft limit $(1-z) \to 0$ with fixed $\vartheta$, and in the soft$\times$collinear limit $(1-z) \to 0$ and $\vartheta \to 0$. In the integrations in $\cD^{[1,0]}_\La$, these singularities are regulated with dimensional regularization.

We can also write $\vartheta$ as a function of $z$ and $k_\LT^2$,
\begin{equation}
\label{eq:varthetakTdef}
\vartheta = a_\perp(z,k_\LT^2)
\;,
\end{equation}
where 
\begin{equation}
\label{eq:aPerpdef}
a_\perp(z,k_\LT^2) = \frac{z k_\LT^2 }{(1-z)^2 Q^2}
\;.
\end{equation}
We will use the variables $z$ and $\vartheta$ to describe a splitting. Then $y$ and $k_\LT^2$ can be determined from $(z,\vartheta)$ using
\begin{equation}
\begin{split}
\label{eq:thetatoyandktsq}
y ={}& \frac{(1-z)\,\vartheta}{z}
\;,
\\
k_\LT^2 ={}& \frac{(1-z)^2\,\vartheta Q^2}{z}
\;.
\end{split}
\end{equation}
%

\subsection{The unresolved region}
\label{sec:unresolvedregion}

The idea of the singular operator $\cD(\mu_\scR, \vec\mu_\scS)$ is that, when applied to a parton basis state $\isket{\{p, f, c, c'\}_m}$, it produces approximated cut Feynman diagrams that have the same infrared singularities as actual cut Feynman diagrams. Of course, it does not match the behavior of Feynman diagrams when the momenta are not close to one of these singularities. For this reason, we define an {\em unresolved region}, a bounded region in the parton momentum space that surrounds the singularities of interest. For a parton splitting included in $\cD^{[1,0]}_\La(\mu_\scR, \vec\mu_\scS)$, the unresolved region is a region $U(\vec\mu_\scS)$ in the space of the splitting variables $(z,\vartheta)$. Since the angle variable $\vartheta$ defined in Eq.~(\ref{eq:varthetafromy}) has the range $0 < \vartheta < 1$, we always take the unresolved region $U(\vec\mu_\scS)$ to be a subset of the region 
\begin{equation}
\begin{split}
\label{eq:physicalregion}
0 <{}& z < 1 \;,
\\
0 <{}& \vartheta < 1
\;.
\end{split}
\end{equation}
The shape of the region depends on parameters $\vec \mu_\Ls$. We define a family of unresolved regions by letting $\vec \mu_\scS$ depend on a parameter $t$ with $0 < t < t_\Lf$ such that the resolved region shrinks as $t$ increases. Often $t_\Lf = \infty$ is a convenient choice. The singular lines $0<z<1$ at $\vartheta = 0$ and $0<\vartheta<1$ at $z = 1$ must be included in $U(\vec\mu_\scS(t))$ for all $t$. This concept is discussed in some detail, with examples, in Ref.~\cite{angleFS}.

In this paper, we will impose a fixed cutoff on $k_\LT^2$ using a parameter $m_\perp^2$ that is on the order of $1 \GeV^2$. A splitting in which 
\begin{equation}
\vartheta < a_\perp(z,m_\perp^2)
\end{equation}
will always be counted as unresolved. This cutoff will serve to keep resolved parton splittings in a range for which perturbation theory is applicable.

For an initial state splitting involving a heavy quark or antiquark, we impose an additional restriction on the unresolved region. Here we count c and b quarks as heavy with $m_\perp^2 < m_\Lc^2 < m_\Lb^2$, while we consider other quarks to be massless. When a massive parton with flavor $a$ turns into a gluon in the sense of backward evolution (so, going forward in time, $\Lg \to a$ with an emitted $\bar a$) we count the splitting as unresolved if $k_\LT^2 < m_a^2$. This is consistent with switching from a 5 flavor scheme for $m_\Lb^2 < \mu^2$ to a 4 flavor scheme for $m_\Lc^2 < \mu^2 < m_\Lb^2$ and then a 3 flavor scheme $\mu^2 < m_\Lc^2$ in $\MSbar$ parton distribution functions. Although the definition of the unresolved region involves the masses of heavy quarks, the shower splitting functions approximate the quark masses by zero.\footnote{It would be better to use on-shell charm and bottom quarks with masses in the shower evolution and then use corresponding quark mass dependence in the PDF evolution equations \cite{NSpartons}. However, the current version of \textsc{Deductor} treats charm and bottom quarks as massless except in the functions that define the unresolved region for initial state splittings. With this approach, the parton shower misses important effects when the scales $\vec\mu$ are not much greater than the quark masses.} We can summarize this by saying that an initial state splitting with $a \to \hat a$ in backward evolution is always counted as unresolved when
\begin{equation}
\label{eq:unresolvedcutoff}
\vartheta < a_\perp(z,m_\perp^2(a,\hat a))
\;,
\end{equation}
where
\begin{equation}
\label{eq:mperpahata}
m_\perp^2(a,\hat a) = 
\max(m_\perp^2, m_a^2 - m_{\hat a}^2)
\;.
\end{equation}

In order to further specify the family of unresolved regions, define a function $a_\mathrm{cut}(z,\vec\mu_\scS(t))$ such that $(z,\vartheta) \in U(\vec\mu_\scS(t))$ if $0<z<1$, $0 < \vartheta < 1$, and
\begin{equation}
\label{eq:acutamperp}
\vartheta < \max\left\{a_\mathrm{cut}(z,\vec\mu_\scS(t)), 
a_\perp(z,m_\perp^2(a,\hat a)\right\}
\end{equation}
As $t$ increases, $a_\mathrm{cut}(z,\vec\mu_\scS(t))$ decreases. In this paper, we impose some conditions on this function. We require that $a_\mathrm{cut}(z,\vec\mu_\scS(t)) \to 0$ for $t \to t_\Lf$.  We further require that for $t \to t_\Lf$, $a_\mathrm{cut}$ takes a simpler limiting form,
\begin{equation}
\label{eq:ylimdef0}
a_\mathrm{cut}(z,\vec\mu_\scS(t)) \sim 
a_\mathrm{lim}(z,\mu_\mathrm{lim}^2(\vec\mu_\scS(t))
\;,
\end{equation}
where $\mu_\mathrm{lim}$ is a function $\mu_\mathrm{lim}(\vec\mu_\scS)$ of the scales $\vec \mu_\scS$, with
\begin{equation}
\label{eq:muCgeneraldef}
\mu_\mathrm{lim}^2(t) = \mu_\mathrm{lim}^2(\vec\mu_\scS(t))
\to 0
\end{equation}
for $t \to t_\Lf$. We require that the limiting function $a_\mathrm{lim}(z,\mu_\mathrm{lim}^2(t))$ takes the factored form
\begin{equation}
\label{eq:alimdef}
a_\mathrm{lim}(z,\mu_\mathrm{lim}^2) = 
f(z)\, \frac{\mu_\mathrm{lim}^2}{Q^2}
\;.
\end{equation}

Thus the scale parameter $\mu_\mathrm{lim}$ controls the collinear limit of  $a_\mathrm{cut}$ when $\vartheta \to 0$ with fixed $z$. It is useful to choose
\begin{equation}
\label{eq:murfrommusdef}
\mur^2(t) \sim \mu_\mathrm{lim}^2(t)
\end{equation}
for large $t$. 

The function $a_\mathrm{lim}(z,\mu_\mathrm{lim}^2(t))$, together with the choice of $\mur^2(t)$, specifies the unresolved region and its relation to the renormalization scale in the collinear limit, in which $a_\mathrm{cut} \ll 1$ for fixed $z$. For the definitions of the unresolved region considered in this paper, $a_\mathrm{lim}(z,\mu_\mathrm{lim}^2(t))$ has the very simple form shown in Eq.~(\ref{eq:alimdef}). This simple form enables us to define shower oriented PDFs as a function of a single scale $\mur^2$. For not-so-large values of $t$, or not-so-small values of $a_\mathrm{cut}$, we can allow a more general structure. First, we can maintain the relation (\ref{eq:murfrommusdef}) in the large $t$ limit while modifying $\mur^2(t)$ away from this limit. Additionally, for small $(1-z)$ with $t$ not large, we can have  $a_\mathrm{lim}(z,\mu_\scC^2(t)) > 1$. We therefore allow a more general form for $a_\mathrm{cut}(z,\vec\mu_\scS(t))$:
\begin{equation}
\label{eq:ycutylimyDelta}
a_\mathrm{cut}(z,\vec\mu_\scS(t)) = 
a_\mathrm{lim}(z,\mu_\mathrm{lim}^2(t))\,
\exp(h(z,\vec\mu_\scS(t))
\;.
\end{equation}
All that we need is that $h(z,\vec\mu_\scS(t)) \to 0$ at least as fast as $\mur^2(t)/Q^2$ in the large $t$, small $\mur^2(t)/Q^2$ limit.

\subsection{Simple cases for the unresolved region}
\label{sec:simpleordering}

Before we examine the relation between $\cD^{[1,0]}_\La(\mu_\scR, \vec\mu_\scS)$ and the shower, it may be helpful to provide some examples of the unresolved region in cases with only one shower scale, $\mu_\scC^2$. With only one scale, we let $\mu_\mathrm{lim}(\mu_\scC) = \mu_\scC$ and
\begin{equation}
\begin{split}
a_\mathrm{cut}(z,\mu_\scC^2) ={}& a_\mathrm{lim}(z,\mu_\scC^2) = a_\scC(z,\mu_\scC^2)
\;,
\end{split}
\end{equation}
where we let $a_\scC(z,\mu_\scC^2)$ be given by one of three choices, $a_\perp(z,\mu_\perp^2)$, $a_\Lambda(z,\mu_\Lambda^2)$, or $a_\angle(z,\mu_\angle^2)$. That is, in $a_\scC(z,\mu_\scC^2)$, $\LC = \perp$, $\Lambda$, or $\angle$. In each case, we take the corresponding renormalization scale to be $\mur^2 = \mu_\scC^2$.

$k_\LT$ {\em definition}:
We can base the unresolved region on the transverse momentum variable defined in Eq.~(\ref{eq:kTdef}). A splitting is unresolved if $k_\LT^2 < \mu_\perp^2$, where $\mu_\perp^2$ is the shower scale. We choose $\mur^2 = \mu_\perp^2$. Accounting also for the fixed $m_\perp^2(a,\hat a)$ cutoff, we adopt the definition that $(z,\vartheta) \in U(\mu_\perp^2)$ if $0 < z < 1$, $0 < \vartheta < 1$ and 
\begin{equation}
\begin{split}
\label{eq:UforkT}
\vartheta <{}& 
\max\left\{
a_\perp(z, \mu_\perp^2),
a_\perp(z,m_\perp^2(a,\hat a))\right\}
\;,
\end{split}
\end{equation}
where $a_\perp(z, \mu^2)$ is defined in Eq.~(\ref{eq:aPerpdef}).

$\Lambda$ {\em definition}: The default ordering variable in \textsc{Deductor} \cite{ShowerTime} is
\begin{equation}
\label{eq:Lambdadef}
\Lambda^2 = \frac{2\hat p_\La \cdot \hat p_\mpone}{2\,p_\La\cdot Q_0} \,Q_0^2
\;,
\end{equation}
where $Q_0$ is the total momentum of the final state partons at the start of the shower. Thus, letting $\eta_\La^{(0)}$ and $\eta_\Lb^{(0)}$ be the momentum fractions of the initial state partons at the start of the shower,
\begin{equation}
\label{eq:Lambda2}
\Lambda^2 = yQ^2\,\frac{\eta_\La^{(0)}  \eta_\Lb^{(0)} s}
{\eta_\La  \eta_\Lb^{(0)} s}
= \frac{y\,Q^2}{r_\La}
\;,
\end{equation}
where
\begin{equation}
\label{eq:rLadef}
r_\La = \frac{\eta_\La}{\eta_\La^{(0)}}
\;.
\end{equation}

A splitting is considered unresolved if $\Lambda^2 < \mu_\Lambda^2$, where $\mu_\Lambda^2$ is the shower scale. We choose $\mur^2 = \mu_\Lambda^2$. Accounting also for the fixed $m_\perp^2(a,\hat a)$ cutoff, we say that $(z,\vartheta) \in U(\mu_\Lambda^2)$ if $0 < z < 1$, $0 < \vartheta < 1$ and 
\begin{equation}
\begin{split}
\label{eq:UforLambda}
\vartheta <{}& 
\max\left\{
a_\Lambda(z, \mu_\Lambda^2),
a_\perp(z,m_\perp^2(a,\hat a))\right\}
\;,
\end{split}
\end{equation}
where $a_\perp(z, \mu^2)$ is defined in Eq.~(\ref{eq:aPerpdef}) and
\begin{equation}
\label{eq:aLambdadef}
a_\Lambda(z,\mu^2) = \frac{z r_\La \mu^2}{(1-z) Q^2}
\;.
\end{equation}
%

{\em $\vartheta$ definition}: We can use the angle variable $\vartheta$ as the ordering variable.\footnote{We could use the angle measured in the fixed reference frame defined by $Q_0$, but then the formulas are more complicated.} Then we consider a splitting to be unresolved if $\vartheta\,  Q^2 < \mu_\angle^2$, where $\mu_\angle^2$ is the shower scale. We choose $\mur^2 = \mu_\angle^2$. Accounting also for the fixed $m_\perp^2(a,\hat a)$ cutoff, we say that $(z,\vartheta) \in U(\mu_\angle^2)$ if  $0 < z < 1$, $0 < \vartheta < 1$ and
\begin{equation}
\begin{split}
\label{eq:UforAngle}
\vartheta <{}& 
\max\left\{
a_\angle(z, \mu_\angle^2),
a_\perp(z,m_\perp^2(a,\hat a))\right\}
\;,
\end{split}
\end{equation}
where $a_\perp(z, \mu^2)$ is defined in Eq.~(\ref{eq:aPerpdef}) and
\begin{equation}
\label{eq:aAngledef}
a_\angle(z,\mu^2) = \frac{\mu^2}{Q^2}
\;.
\end{equation}
%

\subsection{Splitting operator for real emissions}
\label{sec:realemissions}

Real parton splittings at first order in $\as$ are created by the operator $\cF_{\!\La} {\cal D}^{[1,0]}_\mathrm{a}\cF_{\!\La}^{-1}$ applied to a state $\isket{\{p,f,c,c'\}_{m}}$. We can now state very briefly what $\cF_{\!\La} {\cal D}^{[1,0]}_\La \cF_{\!\La}^{-1}$ contains. The details are in Appendix \ref{sec:ISsplittings}.

In the context of a shower cross section, the state is always accompanied by  PDFs that are part of the probability associated with that state. The PDFs are supplied by an operator $\cF_{\!\La}(\mur^2)\cF_{\!\Lb}(\mur^2)$, Eq.~(\ref{eq:Fdef}). For initial state splittings of parton ``a'' realized by backward evolution, we need to remove the prior PDF factor by applying the operator $\cF_{\!\La}^{-1}(\mur^2)$. Then after the splitting we supply the new PDF factor by applying the operator $\cF_{\!\La}(\mur^2)$. Thus it is useful to consider the operator $\cF_{\!\La} {\cal D}^{[1,0]}_\mathrm{a}\cF_{\!\La}^{-1}$.

We apply $\cD^{[1,0]}_\La(\mur,\vec \mu_\scS^2)$ with its accompanying PDF operators to an $m$-parton state and write the result in the form
\begin{equation}
\begin{split}
\label{eq:cDandhatD}
\cF_{\!\La}(\mur^2)&
{\cal D}^{[1,0]}_\mathrm{a}(\mur,\vec \mu_\scS)
\cF_{\!\La}^{-1}(\mur^2)\sket{\{p,f,c,c'\}_{m}}
\\
={}& 
\int\!d\{\hat p,\hat f\}_\mpone\
\sket{\{\hat p,\hat f\}_{\mpone}} 
\\&\times
\frac{\as(\mur^2)}{2\pi}
\sum_{\hat a}\int_0^1\!\frac{dz}{z}\
\frac{
f_{\hat a/A}(\eta_{\La}/z,\mur^2)}
{f_{a/A}(\eta_{\La},\mur^2)}
\\&\times
\hat {\bm D}^\La_{a \hat a}(z;\{\hat p,\hat f\}_\mpone,\{p,f\}_{m})
\sket{\{c,c'\}_{m}}
\;.
\end{split}
\end{equation}
Here ${\cal D}^{[1,0]}_\mathrm{a}$ adds one new parton and we integrate over the momenta and flavors $\{\hat p,\hat f\}_\mpone$ of the partons after the splitting. There are dimensionally regulated singularities, so this integration is in $4 - 2 \epsilon$ dimensions for each momentum. Then $\hat {\bm D}^\La_{a \hat a}$ is a function of the momenta and flavors before and after the splitting and is an operator that maps the $m$-parton color space to the $\mpone$ parton color space. The index $a$ in $\hat {\bm D}^\La_{a \hat a}$ is the flavor of the incoming parton ``a.'' The operator $\hat {\bm D}^\La_{a \hat a}$ also depends on the scales $\mur^2$ and $\vec \mu_\scS$ and it depends on $\epsilon$, but here we do not make that dependence explicit in the notation.

The operator $\hat {\bm D}^\La_{\hat a a}$ takes the form
\begin{align}
\label{eq:hatDstructure}
\notag
\hat {\bm D}^\La_{a \hat a}&(z;\{\hat p,\hat f\}_\mpone,\{p,f\}_{m})
\\\notag ={}&
\frac{z^\epsilon}{(1-z)^{2\epsilon}}
\left(\frac{\mur^2}{Q^2}\right)^{\!\epsilon}
\frac{(4\pi)^\epsilon}{\Gamma(1-\epsilon)}\,
\int_0^1\!\frac{d\vartheta}{\vartheta}\
[\vartheta(1-\vartheta)]^{-\epsilon}
\\ &\times
\int\!\frac{d^{1-2\epsilon}\phi}{S(2-2\epsilon)}\,
\Theta\big((z,\vartheta) \in U(\vec \mu_\scS)\big)\,
\\\notag &\times
\delta\big(\{\hat p, \hat f\}_\mpone 
- R_\La(z,\vartheta,\phi,\hat a;\{p,f\}_{m})\big)
\\\notag&\times
\sum_{k}\frac{1}{2}\bigg[
\theta(k = \La)\,
\frac{1}{N(a ,\hat a)}\, \hat P_{a \hat a}(z, \vartheta,\epsilon)
\\\notag&\qquad -
\theta(k\ne \La)\,
\frac{2 \delta_{a \hat a}}{1-z}\, 
W_0\!\left(\xi_{\La k}, z, \vartheta, \phi - \phi_k\right)
\bigg]
\\\notag&\times
\{
t^\dagger_\La(f_\La \to \hat f_\La\!+\!\hat f_\mpone)\otimes
t_k(f_k \to \hat f_k \!+\! \hat f_\mpone)
\\\notag&\quad
+
t^\dagger_k(f_k \to \hat f_k \!+\! \hat f_\mpone)\otimes
t_\La(f_\La \to \hat f_\La \!+\! \hat f_\mpone)
\}
\;.
\end{align}
The pieces in this formula are described in Appendix \ref{sec:ISsplittings}. Here we mention only the most important features. The operator $\hat {\bm D}^\La_{a \hat a}$ depends on the splitting variable $z$ and on $\hat a$, which is the flavor of the incoming parton after the splitting (in the sense of backward evolution). The right-hand side of Eq.~(\ref{eq:hatDstructure}) begins with integrations over the other two splitting variables $\vartheta$ and $\phi$, with appropriate dependence on the dimensional regulation parameter $\epsilon$. There is a theta function that requires $(z,\vartheta)$ to be in the unresolved region $U(\vec \mu_\scS)$ according to the shower scales $\vec \mu_\scS$. After the integrations, there is a delta function that sets $\{\hat p, \hat f\}_\mpone$ to the momenta and flavors obtained from a splitting with variables $(z,\vartheta,\phi,\hat a)$ applied to partons with momenta and flavors $\{p,f\}_{m}$ according to \textsc{Deductor} conventions. 

\textsc{Deductor} is a dipole shower. There is a sum over the index $k \in \{\La,\Lb,1,\dots,m\}$ of a dipole partner parton for the splitting of parton ``a.'' For the case that $k = \La$, there is a splitting probability $\hat P_{a \hat a}(z, \vartheta,\epsilon)$, Eqs.~(\ref{eq:Paanotation}), (\ref{eq:hatP}), (\ref{eq:Pepsilon}), and (\ref{eq:hatPzeroy}). For $\epsilon = \vartheta = 0$, this is the standard Dokshitzer-Gribov-Lipatov-Alterelli-Parisi (DGLAP) splitting function. The function $N(a ,\hat a)$ provides a color factor, Eq.~(\ref{eq:naahat}). For $k \ne \La$, we have a gluon emitted from parton ``a'' in the ket state interfering with a gluon emitted from parton $k$ in the bra state, or the same configuration with bra and ket interchanged. The function $W_0$ depends on $\xi_{\La k} = p_\La \cdot p_k/p_k \cdot Q$. This function is described in detail in Appendix \ref{sec:ISsplittings}. It has the important property that $W_0/[(1-z)\vartheta]$ is singular when the gluon is soft, but not when it is collinear to $p_\La$.

The final factor in Eq.~(\ref{eq:hatDstructure}) contains color operators that act on the ket color state $\isket{\{c,c'\}_m}$ to give the linear combination of new color states $\isket{\{\hat c, \hat c'\}_\mpone}$ that one gets after emitting the new parton $\mpone$. In $t_\La^\dagger \otimes t_k$, $t_\La^\dagger$ acts on the ket color state $\iket{\{c\}_m}$ to give a new ket color state $t_\La^\dagger \iket{\{c\}_m}$ and $t_k$ acts on the bra color state $\ibra{\{c'\}_m}$ to give a new bra color state $\ibra{\{c'\}_m}t_k$. When parton $\mpone$ is a gluon, the color operators obey the identity
\begin{equation}
\label{eq:coloridentity}
\sum_{k} t_k(f_k \to f_k\!+\!\Lg)
= 0
\;.
\end{equation}
We have used this identity to rewrite the operator $\cD_\La^{[1,0]}$ used in \textsc{Deductor} \cite{NSThreshold} in what we think is a more transparent form. 

\begin{widetext}

\subsection{Inclusive probability for real emission}
\label{sec:inclusiverealemissions}

We have presented the matrix element to obtain a particular state $\isket{\{\hat p,\hat f, \hat c, \hat c'\}_\mpone}$ produced by $\cD^{[1,0]}_\La$. We also need the inclusive probability for a splitting starting from the state $\sket{\{p, f, c, c'\}_m}$. Using Eqs.~(\ref{eq:probabilitydef}) and (\ref{eq:probabilitypartsdef}), the probability corresponding to $\cF_{\!\La} {\cal D}^{[1,0]}_\mathrm{a}\cF_{\!\La}^{-1}$ applied to the state $\sket{\{p,f,c,c'\}_{m}}$ is
\begin{align}
\label{eq:1D10begin}
\sbra{1}\cF_{\!\La}(\mur^2)&
{\cal D}^{[1,0]}_\mathrm{a}(\mur^2,\vec \mu_\scS)
\cF_{\!\La}^{-1}(\mur^2)\sket{\{p,f,c,c'\}_{m}}
\\ \notag
={}& 
\int\!d\{\hat p,\hat f\}_\mpone\
\frac{\as(\mur^2)}{2\pi}
\sum_{\hat a}\int_0^1\!\frac{dz}{z}\
\frac{
f_{\hat a/A}(\eta_{\La}/z,\mur^{2})}
{f_{a/A}(\eta_{\La},\mur^{2})}
\sbra{1_\textrm{color}}
\hat {\bm D}^\La_{a \hat a}(z;\{\hat p,\hat f\}_\mpone,\{p,f\}_{m})
\sket{\{c,c'\}_{m}}
\;.
\end{align}

We write this using another operator $\hat {\bm P}^\La_{a \hat a}$ as
\begin{equation}
\begin{split}
\label{eq:1D10}
\sbra{1}\cF_{\!\La}(\mur^2)
{\cal D}^{[1,0]}_\mathrm{a}(\mur^2,\vec \mu_\scS)&
\cF_{\!\La}^{-1}(\mur^2)\sket{\{p,f,c,c'\}_{m}}
\\
={}& 
\frac{\as(\mur^2)}{2\pi}
\sum_{\hat a}\int_0^1\!\frac{dz}{z}\
\frac{
f_{\hat a/A}(\eta_{\La}/z,\mur^{2})}
{f_{a/A}(\eta_{\La},\mur^{2})}
\sbra{1_\textrm{color}}
\hat {\bm P}^\La_{a \hat a}(z;\{p\}_{m})
\sket{\{c,c'\}_{m}}
\;,
\end{split}
\end{equation}
where $\isbra{1_\textrm{color}}$ times the operator $\hat {\bm P}^\La_{a \hat a}$ is
\begin{equation}
\begin{split}
\label{eq:1PA}
\sbra{1_\mathrm{color}}&
\hat {\bm P}^\La_{a \hat a}(z;\{p\}_{m})
\sket{\{c,c'\}_{m}}
\\={}&
\frac{z^\epsilon}{(1-z)^{2\epsilon}}
\left(\frac{\mur^2}{Q^2}\right)^{\!\epsilon}
\frac{(4\pi)^\epsilon}{\Gamma(1-\epsilon)}\,
\int_0^1\!\frac{d\vartheta}{\vartheta}\
[\vartheta(1-\vartheta)]^{-\epsilon}
\int\!\frac{d^{1-2\epsilon}\phi}{S(2-2\epsilon)}\
\Theta\big((z,\vartheta) \in U(\vec \mu_\scS)\big)\,
\\&\times
\sum_{k}\frac{1}{2}\bigg[
\theta(k = \La)\,
\frac{1}{N(a ,\hat a)}\, \hat P_{a \hat a}(z,\vartheta,\epsilon)
-
\theta(k\ne \La)\,
\frac{2\delta_{a \hat a}}{1-z}\, 
W_0\!\left(\xi_{\La k}, z, \vartheta, \phi - \phi_k\right)
\bigg]
\\&\times
\bra{\{c'\}_{m}}
t_k(f_k \to \hat f_k \!+\! \hat f_\mpone)
t^\dagger_\La(f_\La \to \hat f_\La\!+\!\hat f_\mpone)
+
t_\La(f_k \to \hat f_k \!+\! \hat f_\mpone)
t^\dagger_k(f_\La \to \hat f_\La\!+\!\hat f_\mpone)
\ket{\{c\}_{m}}
\;.
\end{split}
\end{equation}
\end{widetext}
Here we have used the momentum conserving delta function in $\hat {\bm D}^\La_{a \hat a}$ to eliminate the integration over $\{\hat p,\hat f\}_\mpone$. In the color factor, we have used the instruction in Eq.~(\ref{eq:probabilitypartsdef}) to take the trace of the color density matrix after the splitting. The flavors in the color factor are determined by the flavor indices $a$ and $\hat a$. The argument $\{p\}_{m}$ of $\hat {\bm P}^\La_{a \hat a}$ refers to the dependence of $W_0$ on the variables $\xi_{\La k}$.

We can simplify the color here. In the case that $k = \La$,
\begin{equation}
\begin{split}
t_\La(f_\La \to \hat f_\La \!+\! \hat f_\mpone)
t^\dagger_\La(f_\La \to \hat f_\La\!+\!\hat f_\mpone)
= N(a,\hat a)
\;,
\end{split}
\end{equation}
where $N(a,\hat a)$ is the Casimir eigenvalue (\ref{eq:naahat}) appropriate to the flavor content of the splitting. When $k \ne a$, the emitted parton is always a gluon. The gluon line attaches to line ``a'' with a color generator matrix $T_\La^c$ in the  $\bm 8$, $\bm 3$ or $\bar {\bm 3}$ representation according to the flavor of parton $a$. The gluon line attaches to line $k$ with the appropriate generator matrix $T_k^c$. Then we sum over the gluon color index $c$. The result can be denoted by $\bm T_k \cdot \bm T_\La$. Thus for $k \ne \La$,
\begin{equation}
\begin{split}
t_k(f_k \to \hat f_k &\!+\! \hat f_\mpone)
t^\dagger_\La(f_\La \to \hat f_\La\!+\!\Lg)
\\
={}& t_\La(f_\La \to \hat f_\La \!+\! \Lg)
t^\dagger_k(f_k \to \hat f_k\!+\!\Lg)
\\
={}& \bm T_k \cdot \bm T_\La
\;.
\end{split}
\end{equation}
These simplifications give us
\begin{widetext}
\begin{equation}
\begin{split}
\label{eq:1PB}
\sbra{1_\mathrm{color}}
\hat {\bm P}^\La_{a \hat a}(z;&\{p\}_{m})
\sket{\{c,c'\}_{m}}
\\={}&
\frac{z^\epsilon}{(1-z)^{2\epsilon}}
\left(\frac{\mur^2}{Q^2}\right)^{\!\epsilon}
\frac{(4\pi)^\epsilon}{\Gamma(1-\epsilon)}\,
\int_0^1\!\frac{d\vartheta}{\vartheta}\
[\vartheta(1-\vartheta)]^{-\epsilon}
\int\!\frac{d^{1-2\epsilon}\phi}{S(2-2\epsilon)}\
\Theta\big((z,\vartheta) \in U(\vec\mu_\scS)\big)\,
\\&\times
\bigg[
\hat P_{a \hat a}(z,\vartheta,\epsilon)\,
\brax{\{c'\}_{m}}\ket{\{c\}_{m}}
-
\sum_{k\ne \La}\,
\delta_{a \hat a}\,
\frac{2}{1-z}\, 
W_0\!\left(\xi_{\La k}, z, \vartheta, \phi - \phi_k\right)
\bra{\{c'\}_{m}}\bm T_k \cdot \bm T_\La\ket{\{c\}_{m}}
\bigg]
\;.
\end{split}
\end{equation}
This specifies $\sbra{1_\mathrm{color}} \hat {\bm P}^\La_{a \hat a}(z;\{p\}_{m})$ but not the operator $\hat {\bm P}^\La_{a \hat a}(z;\{p\}_{m})$. We note that
\begin{equation}
\begin{split}
\bra{\{c'\}_{m}}\bm T_k \cdot \bm T_\La\ket{\{c\}_{m}}
 ={}& \mathrm{Tr}\left[
 \bm T_k \cdot \bm T_\La\ket{\{c\}_{m}}\bra{\{c'\}_{m}}
 \right]
=
\mathrm{Tr}\left[
 \ket{\{c\}_{m}}\bra{\{c'\}_{m}}\bm T_k \cdot \bm T_\La
 \right]
 \;.
\end{split}
\end{equation}
Thus we can define the color content of $\hat {\bm P}^\La_{a \hat a}(z;\{p\}_{m})$ by
\begin{equation}
\begin{split}
\label{eq:1PC}
\hat {\bm P}^\La_{a \hat a}(z;\{p\}_{m})
={}&
\frac{z^\epsilon}{(1-z)^{2\epsilon}}
\left(\frac{\mur^2}{Q^2}\right)^{\!\epsilon}
\frac{(4\pi)^\epsilon}{\Gamma(1-\epsilon)}\,
\int_0^1\!\frac{d\vartheta}{\vartheta}\
[\vartheta(1-\vartheta)]^{-\epsilon}
\int\!\frac{d^{1-2\epsilon}\phi}{S(2-2\epsilon)}\
\Theta\big((z,\vartheta) \in U(\vec \mu_\scS)\big)\,
\\&\times
\bigg[
\hat P_{a \hat a}(z,\vartheta,\epsilon)\, 
[1 \otimes 1]
-
\sum_{k\ne \La}\,
\delta_{a \hat a}\,
\frac{2}{1-z}\, 
W_0\!\left(\xi_{\La k}, z, \vartheta, \phi - \phi_k\right)
\frac{1}{2}\left\{
[\bm T_k \cdot \bm T_\La \otimes 1]
+[1 \otimes \bm T_k \cdot \bm T_\La]
\right\}
\bigg]
\;.
\end{split}
\end{equation}
\end{widetext}
This particular choice of the color operator in the second term in $\hat {\bm P}^\La_{a \hat a}$ gives this operator the color structure of a virtual gluon exchange between partons ``a'' and $k$.

\subsection{Real part of the virtual exchange operator}
\label{sec:RecD01}

According to Eq.~(\ref{eq:cD1decomposition}), the singular operator $\cD^{[1]}_\La$  consists of two terms, $\cD^{[1,0]}_\La$ that specifies real splittings of parton ``a'' and $\cD^{[0,1]}_\La$, in which a virtual parton is exchanged. We have described $\cD^{[1,0]}_\La$. We now would like to define the real part of ${\cal D}^{[0,1]}_\La(\mur,\vec \mu_\scS)$. We consider the imaginary part of ${\cal D}^{[0,1]}(\mur,\vec \mu_\scS)$, for exchanges involving both initial state and final state partons, in the following subsection.

The operator $\cD^{[0,1]}_\La$ comes from virtual graphs, in which we integrate over a momentum $q$ that flows around a loop. This operator describes the infrared singularity structure when $q \to 0$ or $q$ becomes collinear with $p_\La$ \cite{NSAllOrder}. Since ${\cal D}^{[0,1]}_\La$ simply captures the singularities, it is defined to leave parton momenta and flavors unchanged. The operator ${\cal D}^{[0,1]}_\La$ does, however, change colors. First, it contains terms from self-energy insertions on one of the parton legs. These terms are proportional to the unit operator on the color space times $C_\LF$ or $C_\LA$. Second, there are terms coming from gluon exchanges between two parton legs. These terms are proportional to either $[\bm T_k \cdot \bm T_\La \otimes 1]$ for a virtual graph on the ket amplitude or $[1 \otimes \bm T_k \cdot \bm T_\La]$ for a virtual graph on the bra amplitude. The virtual graphs have $1/\epsilon^2$ and $1/\epsilon$ poles. By using the identity (\ref{eq:coloridentity}), we can arrange that the terms proportional to the unit operator on the color space have $1/\epsilon^2$ and $1/\epsilon$ poles, while the terms with $[\bm T_k \cdot \bm T_\La \otimes 1]$ and $[1 \otimes \bm T_k \cdot \bm T_\La]$ color operators have only $1/\epsilon$ poles that arise from the exchange of a soft gluon.

Since $\mathrm{Re}\, {\cal D}^{[0,1]}_\La$ leaves parton momenta and flavors unchanged but can change the $m$-parton color state, it has the form
\begin{equation}
\begin{split}
\label{eq:Gammaadef}
\mathrm{Re}\,&{\cal D}^{[0,1]}_\La(\mur,\vec \mu_\scS)
\sket{\{p,f,c,c'\}_{m}}
\\& =  \sket{\{p,f\}_{m}}
\frac{\as(\mur^2)}{2\pi}\,
\bm \Gamma_a(\{p\}_m) \sket{\{c,c'\}_{m}}
\;,
\end{split}
\end{equation}
where $a$ is the flavor of the incoming parton ``a.'' The operator $\bm \Gamma_a(\{p\}_m)$ has a soft divergence that should be regulated. As we will see, this divergence is cancelled.

Now, we need to define $\bm \Gamma_a$. The probability associated with $\mathrm{Re}\,{\cal D}^{[0,1]}_\La$ is
\begin{equation}
\begin{split}
\label{eq:1D01Re}
\sbra{1}&\mathrm{Re}\,{\cal D}^{[0,1]}_\La(\mur,\vec \mu_\scS)
\sket{\{p,f,c,c'\}_{m}}
\\& = 
\frac{\as(\mur^2)}{2\pi}\,
\sbra{1_\mathrm{color}}
\bm \Gamma_a(\{p\}_m) \sket{\{c,c'\}_{m}}
.
\end{split}
\end{equation}
The probability associated with $\mathrm{Im}\,{\cal D}^{[0,1]}_\La$ vanishes \cite{NSThreshold}. Thus
\begin{equation}
\begin{split}
\label{eq:1D01}
\sbra{1}& {\cal D}^{[0,1]}_\La(\mur,\vec \mu_\scS)
\sket{\{p,f,c,c'\}_{m}}
\\& = 
\frac{\as(\mur^2)}{2\pi}\,
\sbra{1_\mathrm{color}}
\bm \Gamma_a(\{p\}_m) \sket{\{c,c'\}_{m}}
\;.
\end{split}
\end{equation}

We can combine the probabilities (\ref{eq:1D10}) associated with ${\cal D}^{[1,0]}_\mathrm{a}$ and (\ref{eq:1D01}) associated with ${\cal D}^{[0,1]}_\mathrm{a}$  to obtain the probability associated with ${\cal D}^{[1]}_\mathrm{a}$ (noting that $\cF_{\!\La}$ commutes with ${\cal D}^{[0,1]}_\mathrm{a}$):
\begin{equation}
\begin{split}
\label{eq:1D10and01}
\sbra{1}\cF_{\!\La}(\mur^2)&
{\cal D}^{[1]}_\mathrm{a}(\mur^2,\vec\mu_\scS)
\cF_{\!\La}^{-1}(\mur^2)
\sket{\{p,f,c,c'\}_{m}}
\\
={}& 
\frac{\as(\mur^2)}{2\pi}
\sum_{\hat a}\int_0^1\!\frac{dz}{z}\
\frac{
f_{\hat a/A}(\eta_{\La}/z,\mur^2)}
{f_{a/A}(\eta_{\La},\mur^2)}
\\&\times
\sbra{1_\textrm{color}}
{\bm P}^\La_{a \hat a}(z;\{p\}_{m})
\sket{\{c,c'\}_{m}}
\;,
\end{split}
\end{equation}
where
\begin{equation}
\begin{split}
\label{eq:bmGammadef}
{\bm P}^\La_{a a'}(z; \{p\}_m)
={}&
\hat{\bm P}^\La_{a a'}(z; \{p\}_m)
\\&
+ \delta_{a a'}\,\delta(1-z)\,
\bm \Gamma_{a'}(\{p\}_m)
\;.
\end{split}
\end{equation}

We now propose a definition for the operator $\bm \Gamma_a$. Both the real and virtual terms in ${\bm P}^\La_{a a'}(z; \{p\}_m)$ contain $1/\epsilon^2$ and $1/\epsilon$ poles. However, as we will see in Appendix \ref{sec:poles}, collinear factorization and real-virtual cancellations lead to 
\begin{equation}
\label{eq:polestructure}
\left[\sum_a \int_0^1\! dz\ z\, {\bm P}^\La_{a a'}(z; \{p\}_m)
\right]_\mathrm{poles} = 0
\;.
\end{equation}
This fixes the pole part of $\bm \Gamma_a(\{p\}_m,\epsilon)$ but leaves its finite part undefined. It is not straightforward to impose an ultraviolet cutoff on the unresolved region for virtual graphs that matches the cutoff that we used for real emission graphs. In Ref.~\cite{NSThreshold} we proposed a method for this. Here, we propose a simpler method that gives essentially the same result. We impose a {\em momentum sum rule} on ${\bm P}^\La_{a a'}(z; \{p\}_m)$:
\begin{equation}
\label{eq:MSR}
\sum_a \int_0^1\! dz\ z\, {\bm P}^\La_{a a'}(z; \{p\}_m) = 0
\;.
\end{equation}
This defines $\bm \Gamma_{a}(\{p\}_m)$:
\begin{equation}
\bm \Gamma_{a'}(\{p\}_m)
= - \sum_{c} \int_0^1\! dz\ z\, \hat{\bm P}^\La_{c a'}(z; \{p\}_m)
\;.
\end{equation}

We can write the value of $\bm P$ with the virtual contributions determined by the momentum sum rule (MSR) as
\begin{equation}
\label{eq:usingMSR}
{\bm P}^\La_{a a'}(z; \{p\}_m)
=
\left[\hat{\bm P}^\La_{a a'}(z; \{p\}_m)\right]_\textsc{msr}
\;,
\end{equation}
where $[\hat{\bm P}^\La]_\textsc{msr}$ is defined using a subtraction at $z = 1$ on $\hat{\bm P}^\La$ considered as a matrix in $a a'$. For a generic operator ${\bm A}_{a a'}(z; \{p\}_m)$, the definition is
\begin{equation}
\label{eq:MSR1}
\big[{\bm A}_{a a'}(z; \{p\}_m)\big]_\textsc{msr} =
{\bm A}_{a a'}(z; \{p\}_m)
\;,\ a \ne a' 
\end{equation}
and
\begin{equation}
\begin{split}
\label{eq:MSR2}
\big[{\bm A}_{a a}(z;& \{p\}_m)\big]_\textsc{msr} 
\\={}& 
\frac{1}{z} \left[z\,{\bm A}_{a a}(z; \{p\}_m)\right]_+
\\&
- \delta(1-z) \sum_{c \ne a}\int_0^1\!d\bar z\ 
\bar z\,{\bm A}_{c a}(\bar z; \{p\}_m)
\;.
\end{split}
\end{equation}
Here $[\cdots]_+$ denotes the usual $+$ prescription,
\begin{equation}
\int_0^1\!dz\ [F(z)]_+ h(z) = \int_0^1\!dz\ F(z) \{h(z) - h(1)\}
\;.
\end{equation}
%

\subsection{Imaginary part of the virtual exchange operator}
\label{sec:ImcD01}

The operator $\cD^{[0,1]}$, in which a virtual parton is exchanged between two partons, has an imaginary part when the two partons are either two final state partons or the two initial state partons. Our calculation here builds on that in Ref.~\cite{NScolor}. We write the imaginary part of ${\cal D}^{[0,1]}$ as
\begin{equation}
\begin{split}
\mi\,\mathrm{Im}\,&{\cal D}^{[0,1]}(\mur,\vec \mu_\scS)
\sket{\{p,f,c,c'\}_{m}}
\\& =  \sket{\{p,f\}_{m}}
\frac{\as(\mur^2)}{2\pi}
\\&\times
\Bigg[
\bm G_{\La \Lb}(\{p\}_m)
+
\sum_{l=1}^{m-1} \sum_{k=l+1}^m
\bm G_{lk}(\{p\}_m)
\Bigg]
\\&\times\sket{\{c,c'\}_{m}}
\;.
\end{split}
\end{equation}
Here we sum over pairs of distinct final state partons and over the single possible pair of initial state partons, $\{l,k\} = {\La,\Lb}$. The operator ${\cal D}^{[0,1]}(\mur,\vec \mu_\scS)$ corresponds to the graph in which a gluon is exchanged between partons. The gluon exchange is approximated as not changing the parton momenta, so that when ${\cal D}^{[0,1]}$ is applied to a state $\sket{\{p,f,c,c'\}_{m}}$, the momenta and flavors of the resulting state remain $\{p,f\}_{m}$. However, ${\cal D}^{[0,1]}$ can change the parton color state.

\begin{figure}
\begin{center}
\includegraphics[width = 5.5 cm]{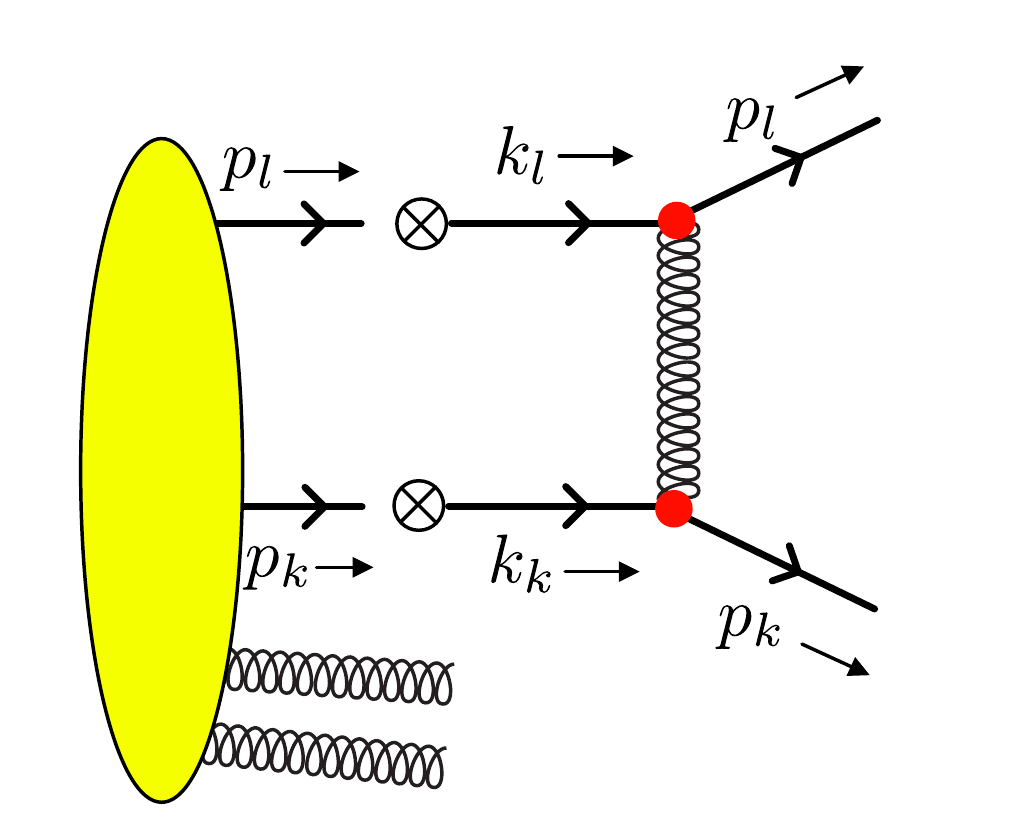} 
\end{center}
\caption{
Virtual gluon exchange between partons $l$ and $k$, leading to an imaginary part of the virtual exchange operator.
}
\label{fig:ImD}
\end{figure}

For a final state interaction, illustrated in Fig.~\ref{fig:ImD}, let us denote the parton momenta before the exchange by $k_l$ and $k_k$. For an initial state scattering, $k_l$ and $k_k$ denote the parton momenta before the exchange in the sense of backward evolution. The partons with momenta $k_l$ and $k_k$ connect to a graph representing much harder interactions in which $k_l$ and $k_k$ are approximated as $p_l$ and $p_k$. 

To obtain the imaginary part of the graph, we replace the propagators for the partons with momenta $k_l$ and $k_k$ by delta functions, $i/(k_l^2 + \mi \epsilon) \to 2\pi \delta(k_l^2)$ and $i/(k_k^2 + \mi \epsilon) \to 2\pi \delta(k_k^2)$. Thus the imaginary part of the diagram corresponds to elastic parton-parton scattering. We integrate over the scattering angle $\theta$ in the $p_l + p_k$ rest frame. The integral is logarithmically divergent at $\theta \to 0$. We regulate the divergence using dimensional regularization with a scale $\mu_\mathrm{IR}^2$ and impose a cut
\begin{equation}
1 - \cos \theta < x_{lk}
\;,
\end{equation}
so that we integrate over an ``unresolved'' interval with boundary $x_{lk}$ that surrounds the singularity. Then, neglecting contributions suppressed by a power of $x_{lk}$, the integration gives
\begin{equation}
\begin{split}
\bm G_{lk}&(\{p\}_m) 
\\={}& 2\pi\mi\,
\Big\{
\frac{(4\pi)^{\epsilon}}{\Gamma(1-\epsilon)} 
\left[\frac{2\mu_\mathrm{IR}^2}{p_l\cdot p_k}\right]^\epsilon
\frac{1}{\epsilon}
- \log(x_{lk})
\Big\}
\\&\times
[\bm{T}_l\cdot \bm{T}_k \otimes 1 - 1 \otimes \bm{T}_l\cdot \bm{T}_k]
\;.
\end{split}
\end{equation}
There are two terms, proportional to $\bm{T}_l\cdot \bm{T}_k \otimes 1$ and $1 \otimes \bm{T}_l\cdot \bm{T}_k$, which correspond to a gluon exchange in the ket state and in the bra state, respectively. If we had color factors $1 \otimes 1$, these terms would cancel. With the color factors that we have, $\mathrm{Im}\,{\cal D}^{[0,1]}$ can potentially have an effect.

Now we can define a scale parameter $\mu_{\mi\pi}^2(\vec \mu_\scS)$ that depends on the shower scale or scales $\vec \mu_\scS(t)$, so that then $\mu_{\mi\pi}^2$ depends on $t$. If there is only one scale, $\mu_\scC$, as described in Sec.~\ref{sec:simpleordering} for $k_\LT$ ordering, $\Lambda$ ordering, or $\vartheta$ ordering, then we can make the simplest choice, $\mu_{\mi\pi}^2 = \mu_\scC^2$. We let the scattering angle cutoff $x_{lk}$ be related to $\mu_{\mi\pi}^2$ by
\begin{equation}
\label{eq:muiPi}
x_{lk} = 
\mu_{\mi\pi}^2\,f_{\mi\pi}^{lk}(\{p\}_m)
\;,
\end{equation}
where $f_{\mi\pi}^{lk}$ is any function of the external parton momenta that we choose. Then
\begin{align}
\notag
\bm G_{lk}&(\{p\}_m) 
\\ \notag
={}& 2\pi\mi\,
\Big\{
\frac{(4\pi)^{\epsilon}}{\Gamma(1-\epsilon)} 
\left[\frac{2\mu_\mathrm{IR}^2}{p_l\cdot p_k}\right]^\epsilon
\frac{1}{\epsilon}
- \log(\mu_{\mi\pi}^2\,f_{\mi\pi}^{lk}(\{p\}_m))
\Big\}
\\&\times
[\bm{T}_l\cdot \bm{T}_k \otimes 1 - 1 \otimes \bm{T}_l\cdot \bm{T}_k]
\;.
\end{align}

We can use the relation (\ref{eq:SiPi}) between $\mathrm{Im}\,\cD^{[0,1]}$ and the corresponding contribution $\cS^{[0,1]}_{\mi\pi}$ to the generator of shower evolution to find
\begin{equation}
\begin{split}
\label{eq:SiPiResult0}
\mi\pi \cS^{[0,1]}_{\mi\pi}(t) \hskip - 0.5 cm&
\\={}& \mi\pi\,\frac{\as(\mur^2)}{2\pi}
\frac{1}{\mu_{\mi\pi}^2(t)}\,
\frac{d\mu_{\mi\pi}^2(t)}{dt}
\\& \times 2\bigg\{
[\bm{T}_\La\cdot \bm{T}_\Lb \otimes 1 - 1 \otimes \bm{T}_\La\cdot \bm{T}_\Lb]
\\&\quad +
\sum_{l=1}^{m-1} \sum_{k=l+1}^m
[\bm{T}_l\cdot \bm{T}_k \otimes 1 - 1 \otimes \bm{T}_l\cdot \bm{T}_k]
\bigg\}
\;.
\end{split}
\end{equation}
Notice that $\cS^{[0,1]}_{\mi\pi}$ does not depend on the functions $f_{\mi\pi}^{lk}(\{p\}_m)$ in Eq.~(\ref{eq:muiPi}). The color operator simplifies \cite{NScolor}, leaving
\begin{equation}
\begin{split}
\label{eq:SiPiResult}
\mi\pi \cS^{[0,1]}_{\mi\pi}(t)
={}& \mi\pi\,\frac{\as(\mur^2)}{2\pi}
\frac{1}{\mu_{\mi\pi}^2(t)}\,
\frac{d\mu_{\mi\pi}^2(t)}{dt}
\\& \times 
4\big[\bm{T}_\La\cdot \bm{T}_\Lb \otimes 1 - 1 \otimes \bm{T}_\La\cdot \bm{T}_\Lb
\big]
\;.
\end{split}
\end{equation}
%

\section{The inclusive infrared finite operator}
\label{sec:IRfiniteoperator}

In order to define a parton shower operator $\cU(t_\Lf,t_\scH)$ starting with the operator $\cD(\mur, \vec \mu_\scS)$ that describes the infrared singularities of QCD, the construction of Ref.~\cite{NSAllOrder} defines, in its Eqs.~(79) and (92), an inclusive infrared finite operator $\cV(\mur,\vec \mu_\scS)$.\footnote{In papers prior to Ref.~\cite{NSAllOrder}, we used the notation $\cV$ to denote a different operator.} The operator $\cV$ leaves the number of partons and their momenta and flavors unchanged and is related to $\cD$ by
\begin{equation}
\begin{split}
\label{eq:cVdef}
\sbra{1}\cV(\mur,\vec \mu_\scS) ={}& 
\sbra{1}
\big[\cF(\mur^2)\circ\cK(\mur^2)\circ\cZ_F(\mur^2)\big]
\\&\times
\cD(\mur,\vec \mu_\scS)\cF^{-1}(\mur^2)
\;.
\end{split}
\end{equation}
The operator $\cV$ at first order contains contributions from emissions from each of the two initial state partons: $\cV^{[1]} = \cV^{[1]}_\La + \cV^{[1]}_\Lb$. There is no contribution from emissions from final state partons because of cancellations between $\cD^{[1,0]}$ and $\cD^{[0,1]}$ \cite{angleFS}. The factor $[\cF(\mur^2)\circ\cK(\mur^2)\circ\cZ_F(\mur^2)]$ is an operator that multiplies by the bare PDFs. This factor, omitting $\cK$, would be present in any perturbative calculation of a cross section in which PDFs are factored from the hard scattering cross section \cite{NSAllOrder}. The circles, $a \circ b$, represent convolutions in the PDF momentum fraction variables.  The operator that performs $\MSbar$ subtractions of $1/\epsilon^n$ poles is $\cZ_F(\mur^2)$. The factor $\cK(\mur^2)$ transforms from the $\MSbar$ scheme for the PDFs to the shower scheme used in the PDFs represented by $\cF(\mur^2)$, Eq.~(\ref{eq:Fdef}). We will see in what follows why we need a shower scheme instead of the $\MSbar$ scheme for the PDFs.

The right-hand-side of Eq.~(\ref{eq:cVdef}) is infrared finite, even though $\cD(\mur,\vec \mu_\scS)$ is singular, for two reasons. First, we form an inclusive probability by multiplying by $\sbra{1}$, which allows real-virtual cancellations. Second, the initial state singularities that do not cancel in this fashion are removed by $\cZ_F(\mur^2)$ in the PDF factor. 

As required in Ref.~\cite{NSAllOrder}, we need to ensure that $\cV(\mur,\vec \mu_\scS)$ is not only finite after dimensional regularization is removed, but also that it approaches the unit operator in the limit of small scale parameters. Specifically, we need $\cV(\mur,\vec \mu_\scS) \to 1$  when $\mur \to 0$, $\vec \mu_\scS \to 0$  and the infrared cutoff parameter is removed, $m_\perp^2(a,\hat a) \to 0$, while $\mur$ and $\vec \mu_\scS$ are related as in Eq.~(\ref{eq:murfrommusdef}) and $\as(\mur^2)$ is held constant. Without these conditions, the operator $\cU_{\cV}$ that is constructed from $\cV$ and sums threshold logarithms \cite{NSAllOrder} would be contaminated by effects with a 1 GeV scale. We will see that these conditions can be achieved by making a suitable choice of $\cK(\mur)$.

The operator $\cV$ can operate non-trivially on the parton color space \cite{NSAllOrder}. We will define the color content of the first order contribution to $\cV$ in what follows.

We will need the detailed formula for the first order contribution to $\cV$ from emissions from parton ``a'' in \textsc{Deductor}. We will write this formula in the form
\begin{equation}
\begin{split}
\label{eq:cV1integral}
\cV^{[1]}_\mathrm{a}(&\mur, \vec\mu_\scS)
\sket{\{p,f,c,c'\}_{m}}
\\={}&
\sket{\{p,f\}_{m}}
\frac{\as(\mur^2)}{2\pi}\,
\sum_{a'}
\int_0^1\!\frac{dz}{z}\,
\frac{
f_{a'/A}(\eta_{\La}/z,\mur^{2})}
{f_{a/A}(\eta_{\La},\mur^{2})}
\\&\times
{\bm V}^\La_{a a'}(z; \{p\}_m)
\sket{\{c,c'\}_m}
\;.
\end{split}
\end{equation}
Here ${\bm V}^\La_{a a'}(z; \{p\}_m)$ is a function of the splitting variable $z$ and the momentum variables $\{p\}_{m}$ of the statistical state $\sket{\{p,f,c,c'\}_{m}}$ and is an operator on the color part $\sket{\{c,c'\}_m}$ of the statistical state. 

We can find what ${\bm V}^\La_{a a'}(z; \{p\}_m)$ is by writing out Eq.~(\ref{eq:cVdef}) at order $\as^1$. We provide details in Appendix~\ref{sec:cV} and give just a brief summary here. Eq.~(\ref{eq:bmVstart}) gives
\begin{equation}
\begin{split}
\label{eq:bmVstartbis}
{\bm V}^\La_{a a'}(z; \{p\}_m) ={}& 
\lim_{\epsilon \to 0}\big\{
K_{a a'}(z,\mur^2)
\\&
+ 
\frac{1}{\epsilon}\frac{(4\pi)^\epsilon}{\Gamma(1-\epsilon)}\,
P_{a a'}(z)
\\ & +
{\bm P}^\La_{a a'}(z;\{p\}_{m};\epsilon)
\big\}
\;.
\end{split}
\end{equation}
Here the term proportional to $P_{a a'}(z)/\epsilon$ is the first order contribution from $\cZ_F(\mur)$. The term $K_{a a'}(z,\mur^2)$ allows the shower to use a shower-oriented PDF instead of an $\MSbar$ PDF:
\begin{align}
\label{eq:Kuse}
\notag
g_{a/A}^{\msbar}(\eta,\mur^2) ={}&
f_{a/A}(\eta,\mur^2)
\\ \notag &
+
\frac{\as(\mur^2)}{2\pi}\sum_{a'}
\int_0^1\!\frac{dz}{z}\
f_{a'/A}(\eta/z,\mur^2)
\\&\quad\times
K_{a a'}(z,\mur^2)
+ \cO(\as^2)
\;.
\end{align}
Here $g_{a/A}^{\msbar}(\eta,\mur^2)$ obeys the $\MSbar$ evolution equation. The remaining term in Eq.~(\ref{eq:bmVstartbis}), ${\bm P}^\La_{a a'}$, gives $\isbra{1}\cF \cD^{[1]}_\La \cF^{-1}$, Eq.~(\ref{eq:1D10and01}). 

\begin{widetext}
After some manipulation of ${\bm P}^\La_{a a'}$ in Eq.~(\ref{eq:bmVstartbis}), we reach a result given in Eq.~(\ref{eq:bfPresult}). Using this result, 
Eq.~(\ref{eq:bmVstartbis}) becomes
\begin{equation}
\begin{split}
\label{eq:bfVresult}
{\bm V}^\La_{a a'}(z;\{p\}_{m}) ={}&
K_{a a'}(z,\mur^2)
+ \left[\hat P_{a a'}^{(\epsilon)}(z)\right]_\textsc{msr}
\\ &
+ \left[\hat P_{a a'}(z)\,
\log\!\left(
\max\left\{
\frac{(1-z)^2 Q^2}{z\mur^2}\,
a_\mathrm{lim}(z,\mu_\mathrm{lim}^2(\vec \mu_\scS)),
\frac{m_\perp^2(a,a')}{\mur^2}
\right\}\right)\right]_\textsc{msr}
+ {\bm P}^\mathrm{a,NS}_{a a'}(z;\{p\}_{m})
\;.
\end{split}
\end{equation}
The kernel $\hat P_{a a'}^{(\epsilon)}(z)$ is given in Eq.~(\ref{eq:Pepsilon}). The contribution proportional to $P_{a a'}(z)/\epsilon$ in Eq.~(\ref{eq:bmVstartbis}) has cancelled an identical singular term in ${\bm P}^\La_{a a'}(z;\{p\}_{m};\epsilon)$ that appears in Eq.~(\ref{eq:bfPresult}). After this cancellation, the second and third terms in Eq.~(\ref{eq:bfVresult}) remain and an infrared ``non-sensitive'' operator ${\bm P}^\mathrm{a,NS}_{a a'}$ remains. The operator ${\bm P}^\mathrm{a,NS}_{a a'}$ is an operator on the color space, while the first three terms in Eq.~(\ref{eq:bfPresult}) are proportional to the unit operator $[1\otimes  1]$ in color. The operator ${\bm P}^\mathrm{a,NS}_{a a'}$ is given in Eqs.~(\ref{eq:hatbfPNS}) and (\ref{eq:bmPNSmsr}). It has the crucial property that it has no $1/\epsilon$ poles and vanishes in the limit of small scales $\mur^2$, $\vec \mu_\scS$ and small infrared cutoff $m_\perp^2(a,a')$ when integrated against a fixed test function $h(z)$. This insensitivity to effects at small momentum scales is exactly what we want in ${\bm V}^\La_{a a'}(z;\{p\}_{m})$.

We are left with the second and third terms in Eq.~(\ref{eq:bfVresult}). These terms are not singular, but they are not small when the scales $\mur^2$, $\vec \mu_\scS$ and the infrared cutoff $m_\perp^2(a,a')$ become small.  In order to cancel these infrared sensitive terms, we set
\begin{equation}
\begin{split}
\label{eq:Kdef}
K_{a a'}(z,\mur^2)  
={}&
-\left[\hat P_{a a'}^{(\epsilon)}(z)\right]_\textsc{msr}
-\left[\hat P_{a a'}(z)\,
\log\left(
\max\left\{
\frac{(1-z)^2 Q^2}{z\mur^2}\,a_\mathrm{lim}(z,\mu_\mathrm{lim}^2(\vec \mu_\scS)),
\frac{m_\perp^2(a,a')}{\mur^2}
\right\}\right)\right]_\textsc{msr}
.
\end{split}
\end{equation}
Then
\begin{equation}
\begin{split}
\label{eq:bfVresult2}
{\bm V}^\La_{a a'}(z;\{p\}_{m}) 
={}& 
{\bm P}^\mathrm{a,NS}_{a a'}(z;\{p\}_{m})
\end{split}
\end{equation}
is insensitive to effects from small momentum scales, as desired. We will see in Sec.~\ref{sec:PDFs} how Eqs.~(\ref{eq:Kuse}) and (\ref{eq:Kdef}) can be used to produce PDFs that can be used in the shower evolution.

\end{widetext}
\section{The shower and threshold operators}
\label{sec:cU}

We now outline, very briefly, how the operators introduced above are used to define a parton shower.

First define (\cite{NSLogSum}, Eq.~(18), or \cite{NSAllOrder},)
\begin{equation}
\begin{split}
\label{eq:cX1def}
\cX_1(\mur,\vec\mu_\scS) ={}&
\big[\cF(\mur^2)\circ\cK(\mur^2)\circ\cZ_F(\mur^2)\big]
\\&\times
\cD(\mur,\vec\mu_\scS)
\cF^{-1}(\mur^2)\
\cV^{-1}(\mur,\vec\mu_\scS)
\;.
\end{split}
\end{equation}
Then the shower evolution operator is (\cite{NSLogSum}, Eqs.~(27), (51), and (54), or \cite{NSAllOrder})
\begin{equation}
\begin{split}
\label{eq:cU}
\cU(t_2,& t_1) 
\\
={}& \lim_{\epsilon \to 0}
\cX_1^{-1}(\mur(t_2),\vec\mu_\scS(t_2))
\cX_1(\mur(t_1),\vec\mu_\scS(t_1))
\\
={}& \mathbb{T}\exp\!\left(\int_{t_1}^{t_2}\!
dt\, \cS(t)\right)
\;.
\end{split}
\end{equation}
Here $(\mur(t),\vec\mu_\scS(t))$ defines a path in the space of the scale variables as in Sec.~\ref{sec:unresolvedregion} and $\mathbb{T}$ indicates ordering in the path parameter $t$, the shower time. The operator $\cS(t)$ is the generator of shower evolution, defined by
\begin{equation}
\label{eq:cS}
\cS(t)
= 
-\cX_1^{-1}(\mur(t),\vec\mu_\scS(t)) 
\frac{d}{d t}
\cX_1(\mur(t),\vec\mu_\scS(t))
\;.
\end{equation}
To use this in a first order shower, one evaluates $\cS(t)$ at order $\as$ and then exponentiates it.

We also define the threshold operator (\cite{NSLogSum}, Eqs.~(56), (58), and (60), or \cite{NSAllOrder}),
\begin{equation}
\begin{split}
\label{eq:cUcV}
\cU_\cV(t_\Lf,& t_{\scH}) 
\\
={}& 
\cV^{-1}(\mur(t_\Lf),\vec\mu_\scS(t_\Lf))\,
\cV(\mur(t_\scH),\vec\mu_\scS(t_\scH))
\\
={}& \mathbb{T}\exp\!\left(\int_{t_\scH}^{t_\Lf}\!
dt\, \cS_\cV(t)\right)
\;,
\end{split}
\end{equation}
where
\begin{equation}
\label{eq:cScV}
\cS_\cV(t)
= 
-\cV^{-1}(\mur(t),\vec\mu_\scS(t)) 
\frac{d}{d t}
\cV(\mur(t),\vec\mu_\scS(t))
\;.
\end{equation}
Here $t_\scH$ is the shower time at the start of the shower, corresponding to scale parameters comparable to the scale of the hard interaction considered. Then $t_\Lf$ is the shower time at the end of the shower, corresponding to scales on the order of $1 \GeV$. It is important that $\cV(\mur(t_\Lf),\vec\mu_\scS(t_\Lf)) \approx 1$. Then the integration in the exponent of Eq.~(\ref{eq:cUcV}) is dominated by scales near the hard scale, while contributions from scales near $1 \GeV$ are power suppressed. To use $\cU_\cV(t_\Lf, t_{\scH})$ in a first order shower, one evaluates $\cS_\cV(t)$ at order $\as$ and then exponentiates it. The operator $\cS_\cV(t)$ is analyzed in some detail in Appendix \ref{sec:cScV}.

With the use of these operators, the cross section for an infrared safe observable is, as described in some detail in Ref.~\cite{NSAllOrder},
\begin{equation}
\begin{split}
\label{eq:sigmaJshower}
\sigma ={}& \sbra{1}
\cO_J\,
\cU(t_\Lf, t_{\scH})\,
\cU_\cV(t_\Lf, t_{\scH})
\cF(\muh^2)
\sket{\rho_\scH}
\;.
\end{split}
\end{equation}
Here $\sket{\rho_\scH}$ is the parton statistical state at the hard interaction and $\cF(\muh^2)$ supplies the parton distribution function factor, using PDFs that match the organization of the parton shower. Then $\cU_\cV(t_\Lf, t_{\scH})$  provides a factor that sums threshold logarithms. It does not change the number of partons or their momenta or flavors. Next, $\cU(t_\Lf, t_{\scH})$ provides a parton shower, creating many partons, while preserving the total probability obtained by summing inclusively over the parton states. Finally, $\cO_J$ makes the desired measurement on the resulting parton state, and $\sbra{1}$ specifies an inclusive sum over the parton state variables.

\section{Shower oriented PDFs}
\label{sec:PDFs}

We now examine the shower oriented parton distribution functions $f_{a/A}(\eta,\mur^2)$ defined in Eq.~(\ref{eq:Kuse}) in the simple cases of $k_\LT$ ordering, $\Lambda$ ordering, and $\vartheta$ ordering as specified in Sec.~\ref{sec:simpleordering}. The shower oriented PDFs are different for different choices of shower ordering. This difference plays a role in how the cross section sums threshold logarithms, as outlined in Appendices \ref{sec:cScV} and \ref{sec:thresholdsum}. Ref.~\cite{Jadach:2015mza} presents a different view of the connection between PDFs and parton showers.

\subsection{$k_\LT$ ordering}
\label{sec:PDFskT}

We define shower oriented parton distribution functions $f_{a/A}(\eta,\mur^2;\lambda)$, where $\lambda = 0$ refers to $k_\LT$ ordering. Rearranging Eq.~(\ref{eq:Kuse}), we have, neglecting terms of order $\as^2$,
\begin{align}
\label{eq:KusekT1}
\notag
f_{a/A}(\eta,\mur^2;0) ={}&
g_{a/A}^{\msbar}(\eta,\mur^2)
\\& \notag
-
\frac{\as(\mur^2)}{2\pi}\sum_{a'}
\int_0^1\!\frac{dz}{z}\
g_{a'/A}^{\msbar}(\eta/z,\mur^2)
\\&\quad \times
K_{a a'}(z, \eta, \mur^2;0)
+ \cO(\as^2)
\;.
\end{align}
Using Eq.~(\ref{eq:aPerpdef}) in Eq.~(\ref{eq:Kdef}) with $a_\mathrm{lim} = a_\perp$ and $\mu^2_\mathrm{lim} = \mu_\perp^2 = \mur^2$, we have
\begin{equation}
\begin{split}
\label{eq:Kdef0}
K_{a a'}(z, \eta, \mur^2;0)  
={}&
-\left[\hat P_{a a'}^{(\epsilon)}(z)\right]_\textsc{msr}
+ \widetilde K_{a a'}(z,\mur^2;0)
\;,
\end{split}
\end{equation}
where
\begin{align}
\label{eq:tildeK0}
\widetilde K_{a a'}&(z,\mur^2,0) 
\\ \notag
={}& -\left[ \hat P_{a a'}(z)\,
\max\!\left[0,\log\!\left(\frac{m_\perp^2(a,a')}{\mur^2}\right)\right]
\right]_\textsc{msr}
.
\end{align}

Consider first the case that $\mu_\scR^2 > m_\Lb^2$, so that $\mur^2 > m_\perp^2(a,a')$ for all choices of $(a,a')$. Then $\widetilde K_{a a'}(z,\mur^2,0) = 0$ so
\begin{equation}
\begin{split}
\label{eq:Kdef0A}
K_{a a'}(z, \eta, \mur^2;0)  
={}&
-\left[\hat P_{a a'}^{(\epsilon)}(z)\right]_\textsc{msr}
\;.
\end{split}
\end{equation}
Thus $f_{a/A}(\eta,\mur^2;0)$ is close to $g_{a'/A}^{\msbar}(\eta,\mur^2)$, but there is a small change because we do not renormalize the $k_T$-ordered parton distribution functions by simply integrating over all $k_\LT$ and subtracting poles.

Now consider the case that $m_\perp^2 < m_\Lc^2 < \mur^2 < m_\Lb^2$. Then there is a small complication. We take the operator $\cZ_F(\mur^2)$ in Eq.~(\ref{eq:cVdef}) to be the operator that renormalizes parton distribution functions according to the $\MSbar$ prescription in five flavor QCD. Then $g_{a/A}^{\msbar}(\eta,\mur^2)$ is the five flavor $\MSbar$ parton distribution function for flavor $a$. With the conventional definition in which there are no intrinsic b-quarks, $g_{\Lb/A}^{\msbar}(\eta,m_\Lb^2) = 0$, but (applying five flavor evolution) $g_{\Lb/A}^{\msbar}(\eta,\mur^2) < 0$ for $\mur^2 < m_\Lb^2$. We then have
\begin{align}
\label{eq:tildeKb}
\widetilde K_{\Lb \Lg}(z,\mur^2,0) 
={}& - \left[ \hat P_{\Lb \Lg}(z)\,
\log\!\left( \frac{m_\Lb^2}{\mur^2}\right)\right]_\textsc{msr}
.
\end{align}
Since $P_{a a'}(z)$ is the kernel for PDF evolution in $\mu^2$, this contribution to  $K_{a a'}(z, \eta, \mur^2;0)$ cancels the evolution of the five flavor $\MSbar$ PDF in the region $\mur^2 < m_\Lb^2$ and gives us 
\begin{equation}
f_{\Lb/A}(\eta,\mur^2;0) = 0 + \cO(\as^2)
\end{equation}
for $\mur^2 < m_\Lb^2$. That is, $f_{a/A}(\eta,\mur^2;0)$ is the four flavor $\MSbar$ PDF as conventionally defined for $m_\perp^2 < m_\Lc^2 < \mur^2 < m_\Lb^2$. 

Similarly, for  $m_\perp^2 < \mur^2 < m_\Lc^2$, $f_{\Lc/A}(\eta,\mur^2;0) = 0 + + \cO(\as^2)$.

\subsection{$\Lambda$ ordering}
\label{sec:PDFsLambda}

We now generalize the definitions in the previous subsection by defining  shower oriented parton distribution functions $f_{a/A}(\eta,\mur^2;\lambda)$, where $\lambda = 0$ refers to $k_\LT$ ordering and $\lambda = 1$ refers to $\Lambda$ ordering. Rearranging Eq.~(\ref{eq:Kuse}) gives
\begin{align}
\label{eq:Kuse2}
\notag
f_{a/A}(\eta,\mur^2;\lambda) ={}&
g_{a/A}^{\msbar}(\eta,\mur^2)
\\& \notag
-
\frac{\as(\mur^2)}{2\pi}\sum_{a'}
\int_0^1\!\frac{dz}{z}\
g_{a'/A}^{\msbar}(\eta/z,\mur^2)
\\&\quad \times
K_{a a'}(z, \eta, \mur^2;\lambda)
+ \cO(\as^2)
\;.
\end{align}
Using Eqs.~(\ref{eq:aPerpdef}) and (\ref{eq:aLambdadef}) in Eq.~(\ref{eq:Kdef}), we now have,
\begin{equation}
\begin{split}
\label{eq:Kdeflambda}
K_{a a'}&(z, \eta, \mur^2;\lambda)  
\\ ={}&
-\left[\hat P_{a a'}^{(\epsilon)}(z)\right]_\textsc{msr}
- \lambda \log(r_\La)\,  P_{a a'}(z)
\\  &
+ \widetilde K_{a a'}(z, r_\La^\lambda \mur^2;\lambda)
\;,
\end{split}
\end{equation}
where
\begin{align}
\label{eq:Winterpolation}
\widetilde K_{a a'}&(z,r_\La^\lambda \mur^2,\lambda) 
\\ \notag
={}& -\left[ \hat P_{a a'}(z)\,
\log\!\left(\max\!\left[(1-z)^\lambda,\frac{m_\perp^2(a,a')}{r_\La^\lambda \mur^2}\right]\right)\right]_\textsc{msr}
.
\end{align}
Eq.~(\ref{eq:Kdeflambda}) follows from Eqs.~(\ref{eq:aPerpdef}) and (\ref{eq:aLambdadef}) for $\lambda = 0$ and $\lambda = 1$. For $0 < \lambda < 1$, it is a simple interpolation.

We recall from Eq.~(\ref{eq:rLadef}) that $r_\La = \eta/\eta_\La^{(0)}$, where $\eta_\La^{(0)}$ is the momentum fraction of parton ``a'' at the start of the shower. The appearance of $r_\La$ has two effects. First, it makes the kernel $K$ depend on both $\eta$ and $z$ instead of just $z$. Second, it introduces an external parameter  $\eta_\La^{(0)}$ into $K$.

For $\Lambda$ ordering, we need $f_{a/A}(\eta,\mur^2;\lambda)$ at $\lambda = 1$. To find this starting with $f_{a/A}(\eta,\mur^2;\lambda)$ at $\lambda = 0$, we consider $\lambda$ in the range $0 \le \lambda \le 1$. To find $f_{a/A}(\eta,\mur^2;\lambda)$ while working around the complications produced by the presence of $r_\La$ in $K$, we first define
an auxiliary scale variable,
\begin{equation}
\mu^2 = r_\La^\lambda \mur^2
\;.
\end{equation}
Then we define an auxiliary parton distribution function that is, in the end, simpler than $f_{a/A}(\eta,\mur^2;\lambda)$,
\begin{equation}
\label{eq:tildef}
\tilde f_{a/A}(\eta,\mu^2;\lambda) = f_{a/A}(\eta,r_\La^{-\lambda} \mu^2;\lambda)
\;.
\end{equation}
From its definition, we have
\begin{equation}
\label{eq:tildef0}
\tilde f_{a/A}(\eta,\mu^2;0) = f_{a/A}(\eta, \mu^2;0)
\;.
\end{equation}
For $\lambda > 0$, we use Eqs.~(\ref{eq:Kuse2}), (\ref{eq:Kdeflambda}), and (\ref{eq:tildef}) to give
\begin{align}
\label{eq:tildef1}
\notag
\tilde f_{a/A}&(\eta,\mu^2;\lambda)
\\\notag  ={}&
g_{a/A}^{\msbar}(\eta,r_\La^{-\lambda} \mu^2)
\\& \notag
+ \log(r_\La^\lambda)\,
\frac{\as}{2\pi}
\sum_{a'}\int_0^1\!\frac{dz}{z}\
g_{a'/A}^{\msbar}(\eta/z,r_\La^{-\lambda} \mu^2)
P_{a a'}(z)
\\ \notag &
+\frac{\as}{2\pi}\sum_{a'}
\int_0^1\!\frac{dz}{z}\
g_{a'/A}^{\msbar}(\eta/z,r_\La^{-\lambda} \mu^2)
\left[\hat P_{a a'}^{(\epsilon)}(z)\right]_\textsc{msr}
\\ \notag &
-\frac{\as}{2\pi}\sum_{a'}
\int_0^1\!\frac{dz}{z}\
g_{a'/A}^{\msbar}(\eta/z,r_\La^{-\lambda} \mu^2)\,
\widetilde K_{a a'}(z,\mu^2,\lambda)
\\&
+ \cO(\as^2)
\;.
\end{align}
Since $P_{a a'}(z)$ is the generator of scale changes for $g^{\msbar}$, the sum of the first two terms is $g_{a/A}^{\msbar}(\eta,\mu^2)$ up to order $\as^2$ corrections. In the remaining terms, we can replace the scale $r_\La^{-\lambda} \mu^2$ by just $\mu^2$ at leading order in $\as$. This gives
\begin{align}
\label{eq:tildef2}
\notag
\tilde f_{a/A}&(\eta,\mu^2;\lambda)
\\\notag  ={}&
g_{a/A}^{\msbar}(\eta,\mu^2)
\\ \notag &
+\frac{\as}{2\pi}\sum_{a'}
\int_0^1\!\frac{dz}{z}\
g_{a'/A}^{\msbar}(\eta/z,\mu^2)
\left[\hat P_{a a'}^{(\epsilon)}(z)\right]_\textsc{msr}
\\ \notag &
-\frac{\as}{2\pi}\sum_{a'}
\int_0^1\!\frac{dz}{z}\
g_{a'/A}^{\msbar}(\eta/z,\mu^2)\,
\widetilde K_{a a'}(z,\mu^2,\lambda)
\\&
+ \cO(\as^2)
\;.
\end{align}
The sum of the first two terms and the $\lambda = 0$ contribution from $\widetilde K_{a a'}$ gives $\tilde f_{a/A}(\eta,\mu^2;\lambda)$ at $\lambda = 0$, while in the last term, we can replace  $g^{\msbar}$ by $\tilde f_{a/A}(\eta,\mu^2;0)$ at leading order in $\as$. This gives us
\begin{align}
\label{eq:tildef3}
\notag
\tilde f_{a/A}&(\eta,\mu^2;\lambda)
\\\notag  ={}&
\tilde f_{a/A}(\eta,\mu^2;0)
\\ \notag &
-\frac{\as}{2\pi}\sum_{a'}
\int_0^1\!\frac{dz}{z}\
\tilde f_{a'/A}(\eta/z,\mu^2;0)
\\ \notag &\quad\times
[\widetilde K_{a a'}(z,\mu^2,\lambda)
- \widetilde K_{a a'}(z,\mu^2,0)]
\\&
+ \cO(\as^2)
\;.
\end{align}

Now, differentiating with respect to $\lambda$ gives the differential equation
\begin{align}
\label{eq:dtildefdlambda}
\notag
\frac{d\tilde f_{a/A}(\eta,\mu^2;\lambda)}{d\lambda}\hskip - 1.0 cm& 
\\\notag ={}& 
-\sum_{a'}
\int_0^1\!\frac{dz}{z}\
\frac{\as(\mu^2)}{2\pi}
\tilde f_{a'/A}(\eta/z,\mu^2;\lambda)
\\ \notag &\times
\frac{\partial}{\partial \lambda}
\widetilde K_{a a'}(z,\mu^2,\lambda)
\\ &
+ \cO(\as^2)
\;,
\end{align}
where
\begin{align}
\label{eq:lambdageneratorA}
-\frac{\partial}{\partial \lambda}&
\widetilde K_{a a'}(z,\mu^2,\lambda) 
\\ \notag ={}& 
\Bigg[
\log(1-z)
\hat P_{a a'}(z)\,
\theta\!\left((1-z)^\lambda > \frac{m_\perp^2(a,a')}{\mu^2}\right)
\Bigg]_\textsc{msr}
.
\end{align}
We can solve Eq.~(\ref{eq:dtildefdlambda}) for $\tilde f_{a/A}(\eta,\mu^2;1)$ with $\tilde f_{a/A}(\eta,\mu^2;0)$ as the boundary value. Then the shower oriented PDF at $\lambda = 1$ is
\begin{equation}
\label{eq:tildef4}
f_{a/A}(\eta,\mur^2;1)
=
\tilde f_{a/A}(\eta,\mur^2 r_\La;1)
\;.
\end{equation}

Eq.~(\ref{eq:dtildefdlambda}) results from the interpolation that we chose in Eq.~(\ref{eq:Kdeflambda}) between $\cK^{[1]}(\mu^2,0)$ and $\cK^{[1]}(\mu^2,1)$. It is of interest to know how much the solution of Eq.~(\ref{eq:dtildefdlambda})  for $\tilde f_{a/A}(\eta,\mu^2;1)$ depends on this choice of interpolation. To this end, we note that $\tilde f_{a/A}(\eta,\mu^2;1)$ can be expressed, using a more compressed operator notation, as
\begin{equation}
\begin{split}
\label{eq:f1exponential}
\tilde f(\mu^2,1) ={}& 
\tilde f(\mu^2,0) 
\\&\circ
\mathbb{T}
\exp\!\left[
-\frac{\as}{2\pi}\int_0^1\!d\lambda\,  
\frac{d \widetilde K(\mu^2,\lambda)}{d\lambda}  
\right]
,
\end{split}
\end{equation}
where we exponentiate using the convolution product $\circ$ and $\mathbb{T}$ represents ordering in $\lambda$. We have noted that the evolution kernel is the derivative of the first order term of the operator $\widetilde K$. We have included the ordering instruction $\mathbb{T}$ because $d\widetilde K(\mu^2,\lambda)/d\lambda$ is a matrix in the parton flavor space and the matrices for different values of $\lambda$ do not commute. However, at $\lambda = 0$ and $\lambda = 1$, the most important terms in $\widetilde K$ are the terms with $1/(1-z)$ singularities. These terms are diagonal in flavor and {\em do} commute. Assuming the terms with $1/(1-z)$ singularities in $\widetilde K$ for $0 < \lambda < 1$ are diagonal in flavor, we can conclude that to a reasonable approximation, the ordering instruction $\mathbb{T}$ is not needed and, for $\mu^2 > m_\perp^2(a,a')$,
\begin{equation}
\begin{split}
\label{eq:f1exponential3}
\tilde f(\mu^2,1) \approx{}& 
\tilde f(\mu^2,0) 
\\&\circ
\exp\!\left[
\frac{\as}{2\pi}
\left\{\widetilde K(\mu^2,0) 
- \widetilde K(\mu^2,1)\right\}  
\right]
\\
= {}&
\tilde f(\mu^2,0) 
\circ
\exp\!\left[-
\frac{\as}{2\pi}
\widetilde K(\mu^2,1) 
\right]
.
\end{split}
\end{equation}
Thus, to a reasonable approximation, only the endpoints matter, not the interpolation.

How does $\tilde f(\mu^2,1)$ evolve under changes of $\mu^2$? At leading order in $\as$, we can omit the ordering instruction $\mathbb{T}$ in Eq.~(\ref{eq:f1exponential}), so that we can use Eq.~(\ref{eq:f1exponential3}) for $\tilde f(\mu^2,1)$. For $\tilde f(\mu^2,0)$, we have the ordinary first order DGLAP equation, 
\begin{equation}
\mu^2\frac{d}{d\mu^2}\tilde f(\mu^2,0) = \tilde f(\mu^2,0)\circ \frac{\as}{2\pi} P
+ \cO(\as^2)
\;.
\end{equation}
Differentiating Eq.~(\ref{eq:f1exponential3}) then gives 
\begin{equation}
\label{eq:tildefmuevolution0}
\mu^2\frac{d}{d\mu^2}\tilde f(\mu^2,1) = \tilde f(\mu^2,1)\circ \frac{\as}{2\pi} \widetilde P(\mu^2)
+ \cO(\as^2)
\;,
\end{equation}
where
\begin{align}
\widetilde P_{a a'}(z,&\mu^2)
\\\notag
={}& P_{a a'}(z) - 
\mu^2\frac{\partial \widetilde K_{a a'}(z,\mu^2,1)}{d\mu^2} 
\\\notag
={}&
\left[ \hat P_{a a'}(z) \left(1 - \theta((1-z)\mu^2 < m_\perp^2(a,a'))\right)
\right]_\textsc{msr}
.
\end{align}
Thus
\begin{equation}
\begin{split}
\label{eq:evolutionkernel1}
\widetilde P_{a a'}(z,\mu^2)
={}& 
\left[ \hat P_{a a'}(z)\,
\theta((1-z)\mu^2 > m_\perp^2(a,a'))\right]_\textsc{msr} 
.
\end{split}
\end{equation}
That is, the first order kernel for evolution in $\mu^2$ of $\tilde f(\mu^2,1)$ is the familiar DGLAP kernel but with a cut $k_\LT^2 > m_\perp^2(a,a')$, where $k_\LT^2 = (1-z) \mu^2$. The subtraction term in $\tilde P_{a a'}(z,\mu^2)$ proportional to $\delta(1-z)$ is determined by the momentum sum rule. 

We write Eq.~(\ref{eq:tildefmuevolution0}) in detail as
\begin{align}
\label{eq:dtildefdlogmusq}
\notag
\mu^2 \frac{d\tilde f_{a/A}(\eta,\mu^2;1)}{d\mu^2}\hskip - 1.0 cm& 
\\\notag ={}& 
\sum_{a'}
\int_0^1\!\frac{dz}{z}\
\frac{\as(\mu^2)}{2\pi}
\tilde f_{a'/A}(\eta/z,\mu^2;1)
\\ \notag &\times
\widetilde P_{a a'}(z,\mu^2)
\\ &
+ \cO(\as^2)
\;.
\end{align}
The kernel $\widetilde P_{a a'}(z,\mu^2)$ in Eq.~(\ref{eq:evolutionkernel1}) was introduced in Ref.~\cite{NSThreshold}, but without enforcing the momentum sum rule. 

This gives us two ways to determine $\tilde f_{a/A}(\eta,\mu^2;\lambda)$ at $\lambda = 1$. We can start with $\tilde f_{a/A}(\eta,\mu^2;0)$, which is $g_{a/A}^{\msbar}(\eta,\mu^2)$ with a small correction given by Eq.~(\ref{eq:Kdef0A}). For $g_{a/A}^{\msbar}(\eta,\mu^2)$ one can use a PDF set that is fit to data and uses at least NLO evolution in $\mu^2$. Then we can solve the $\lambda$-evolution equation (\ref{eq:dtildefdlambda}) to find $\tilde f_{a/A}(\eta,\mu^2;1)$. We know the kernel for this differential equation only at first order, Eq.~(\ref{eq:lambdageneratorA}), but the change in $\lambda$ is not large. Alternatively, we can note that $\tilde f_{a/A}(\eta,\mu^2;1) = \tilde f_{a/A}(\eta,\mu^2;0)$ at $\mu^2 = m_\perp^2$. Thus $\tilde f_{a/A}(\eta, m_\perp^2;1)$ is known and we can use the $\mu^2$-evolution equation (\ref{eq:dtildefdlogmusq}) to determine $\tilde f_{a/A}(\eta, \mu^2;1)$ at larger values of $\mu^2$. We know the kernel for this differential equation only at first order, Eq.~(\ref{eq:evolutionkernel1}), but the lack of higher order corrections should not be a problem if we are interested in values of $\log(\mu^2/m_\perp^2)$ that are not too large.

\subsection{$\vartheta$ ordering}
\label{sec:PDFsAngular}

We define shower oriented parton distribution functions $f_{a/A}(\eta,\mur^2;\lambda)$, where $\lambda = 0$ refers to $k_\LT$ ordering and $\lambda = 1$ refers now to $\vartheta$ ordering. Rearranging Eq.~(\ref{eq:Kuse}), we have, up to order $\as^2$,
\begin{align}
\label{eq:KuseAngle}
\notag
f_{a/A}(\eta,\mur^2;\lambda) ={}&
g_{a/A}^{\msbar}(\eta,\mur^2)
\\& \notag
-
\frac{\as(\mur^2)}{2\pi}\sum_{a'}
\int_0^1\!\frac{dz}{z}\
g_{a'/A}^{\msbar}(\eta/z,\mur^2)
\\&\quad \times
K_{a a'}(z, \mur^2;\lambda)
+ \cO(\as^2)
\;.
\end{align}
Here $K$ is a different kernel than the one used for $\Lambda$ ordering. Using Eqs.~(\ref{eq:aPerpdef}) and (\ref{eq:aAngledef}) in Eq.~(\ref{eq:Kdef}), we have,
\begin{align}
\label{eq:KdefAngular}
K_{a a'}&(z, \mur^2;\lambda)  
\\\notag ={}&
-\left[\hat P_{a a'}^{(\epsilon)}(z)\right]_\textsc{msr}
\\\notag &
-\Bigg[
\max\!\left\{\!
\lambda\log\!\left(\frac{(1-z)^2}{z}\!\right),
\log\!\left(\frac{m_\perp^2(a,a')}{\mur^2}\!\right)
\!\right\}
\\\notag&\quad\times
\hat P_{a a'}(z)
\Bigg]_\textsc{msr}
\!.
\end{align}
Now, differentiating $f$ with respect to $\lambda$ gives the differential equation
\begin{align}
\label{eq:fdlambdaAngular}
\notag
\frac{d f_{a/A}(\eta,\mur^2;\lambda)}{d\lambda} 
={}& 
-\frac{\as(\mur^2)}{2\pi}\sum_{a'}
\int_0^1\!\frac{dz}{z}\,
f_{a'/A}(\eta/z,\mur^2;\lambda)
\\ \notag &\quad\times
\frac{d}{d\lambda}\,
K_{a a'}(z,\mur^2;\lambda)
\\ &
+ \cO(\as^2)
\;,
\end{align}
where
\begin{align}
\label{eq:A1angular}
-\frac{d}{d\lambda}\,
K_{a a'}(z,&\mur^2;\lambda) 
\\ \notag ={}& 
\Bigg[
\log\!\left(\!\frac{(1-z)^2}{z}\!\right)\!
\hat P_{a a'}(z)
\\\notag &\quad\times
\theta\!\left(\left[\frac{(1-z)^2}{z}\right]^\lambda 
> \frac{m_\perp^2(a,a')}{\mur^2}\!\right)\!
\Bigg]_\textsc{msr}
.
\end{align}
We can solve this for $f_{a/A}(\eta,\mur^2;1)$ with $f_{a/A}(\eta,\mur^2;0)$. 

As with $\Lambda$ ordering, we can also write a first order equation for the evolution of $f_{a/A}(\eta,\mur^2;1)$ as $\mur^2$ varies with $\vartheta$ ordering:
\begin{align}
\label{eq:dtildefdlogmusqangle}
\notag
\mur^2\frac{df_{a/A}(\eta,\mur^2;1)}{d\mur^2}\hskip - 1.0 cm& 
\\\notag ={}& 
\sum_{a'}
\int_0^1\!\frac{dz}{z}\
\frac{\as(\mur^2)}{2\pi}
f_{a'/A}(\eta/z,\mur^2;1)
\\ \notag &\times
\widetilde P_{a a'}(z,\mu^2)
\\ &
+ \cO(\as^2)
\;.
\end{align}
Here for $\vartheta$ ordering $\widetilde P$ is
\begin{align}
\label{eq:evolutionkernel1angle}
\widetilde P_{a a'}(z,&\mur^2)
\\\notag 
={}& 
\left[ \hat P_{a a'}(z)\,
\theta\!\left(\frac{(1-z)^2\mur^2}{z} > m_\perp^2(a,a')\right)
\right]_\textsc{msr} 
\!.
\end{align}
%

\subsection{Modified argument of $\as$}
\label{sec:asmodification}

In the differential equation (\ref{eq:dtildefdlambda}) for the $\lambda$ dependence $\tilde f_{a/A}(\eta,\mu^2;\lambda)$, we have taken the argument of $\as$ to be the scale $\mu^2$. We made the same choice in Eq.~(\ref{eq:dtildefdlogmusq}) for the $\mu^2$ evolution of $\tilde f_{a/A}(\eta,\mu^2;\lambda)$ at $\lambda = 1$. Similarly, in Eqs.~(\ref{eq:fdlambdaAngular}) and (\ref{eq:dtildefdlogmusqangle}), we evaluated $\as$ at the scale $\mur^2$ of the shower oriented parton distribution functions. These are the natural choices for a formalism that is designed to make sense at any perturbative order. However, in this paper, we have only a first order shower. We can modify the argument of $\as$ so as to include some desired higher order corrections in the evolution kernels. 

In Eq.~(\ref{eq:dtildefdlambda}), we include a factor
\begin{equation}
R_{\alpha_s} = \frac{\as^{(K_\Lg)}(k_\LT^2)}{\as(\mu^2)}
\end{equation}
in the kernel $\partial \widetilde K_{a a'}(z,\mu^2,\lambda) /\partial \lambda$. Here  $k_\LT^2 = (1-z)^\lambda \mu^2$ and 
\begin{equation}
\begin{split}
\label{eq:asKg}
\as^{(K_\Lg)}(k_\LT^2) = {}&
\as(k_\LT^2)\left[
1 + K_\Lg \frac{\as(k_\LT^2)}{2\pi}
\right]
\;.
\end{split}
\end{equation}
Here $K_\Lg$ is the standard factor \cite{CMW},
\begin{equation}
\label{eq:Kg}
K_\Lg = C_\LA \frac{67 - 3 \pi^2}{18} - T_\LR \frac{10 n_\Lf}{9}
\;,
\end{equation}
with $n_\Lf$ set to the number of active quark flavors at scale $k_\LT^2$. We note that $\partial \widetilde K_{a a'}(z,\mu^2,\lambda) /\partial \lambda$ contains a factor $\theta(k_\LT^2 > m_\perp^2(a,a'))$, which protects us from a singularity of $\as(k_\LT^2)$. This substitution, based on Ref.~\cite{CMW}, can help to sum large logarithms \cite{NSLogSum}. An example, for threshold logarithms, is analyze in Appendix \ref{sec:thresholdsum}.

We make the analogous adjustments in Eqs.~(\ref{eq:dtildefdlogmusq}), (\ref{eq:fdlambdaAngular}) and (\ref{eq:dtildefdlogmusqangle}).

\subsection{Numerical values of the PDFs}
\label{sec:PDFsresults}

In this subsection, we look at the shower oriented PDFs in a proton. We obtain these PDFs from the corresponding $\MSbar$ PDF by first applying the $P^{(\epsilon)}$ transformation (\ref{eq:Kdef0A}) to obtain the $k_\LT$ ordered PDF. Then we solve either Eq.~(\ref{eq:dtildefdlambda}) to obtain the $\Lambda$ ordered PDF or Eq.~(\ref{eq:fdlambdaAngular}) to obtain the $\vartheta$ ordered PDF. In the case of $\Lambda$ ordering, we choose $r_\La = 1$ or, alternatively, examine the PDF $\tilde f_{a/\Lp}(\eta,\mu^2)$ with a modified scale parameter, Eq.~(\ref{eq:tildef4}).

\begin{figure}
\ifusefigs 

\begin{tikzpicture}
  \begin{semilogxaxis}[title = {up-quark distribution, 
    $\mu \approx 1\ \mathrm{TeV}$},
    xlabel={$\eta$}, ylabel={$\eta\,f_{\Lu/\Lp}(\eta,\mu^2)$},
    xmin=0.01, xmax=1,
    legend cell align=left,
    every axis legend/.append style = {
    at={(0.6,0.96)},
    anchor=north west}
    ]
  \pgfplotstableread{PDFPlots/up.dat}\up
  
  \addplot [black, dashed,restrict x to domain = -4.5:-0.356675] 
  table [x index=1, y expr=\thisrow{8}]{\up};
  \addlegendentry{$\MSbar$}

  \addplot [black, thick,restrict x to domain = -4.5:-0.356675] 
  table [x index=1, y expr=\thisrow{2}]{\up};
  \addlegendentry{$k_T$ ordered}
  
  \addplot [red,thick,restrict x to domain = -4.5:-0.356675] 
  table [x index=1, y expr=\thisrow{3}]{\up};
  \addlegendentry{$\Lambda$ ordered}

  \addplot [blue, thick,restrict x to domain = -4.5:-0.356675] 
  table [x index=1, y expr=\thisrow{4}]{\up};
  \addlegendentry{$\vartheta$ ordered}

  \end{semilogxaxis}
\end{tikzpicture}

\else 
\begin{center}
\includegraphics[width = 5.5 cm]{sample.pdf}
\end{center}
\fi

\caption{Up-quark distributions $\eta\,f_{\Lu/\Lp}(\eta,\mu^2)$ at $\mu = 1007\ \mathrm{GeV}$: the $\MSbar$ distribution and the shower oriented distributions for $k_\LT$ ordering, $\Lambda$ ordering, and $\vartheta$ ordering.
}
\label{fig:up1TeV}
\end{figure}

\begin{figure}
\ifusefigs 

\begin{tikzpicture}
  \begin{semilogxaxis}[title = {$f_{\Lu/\Lp}/f^\msbar_{\Lu/\Lp}$, 
  $\mu \approx 1\ \mathrm{TeV}$},
    xlabel={$\eta$}, ylabel={$f_{\Lu}(\eta,\mu^2)/f_{\Lu}^{\msbar}(\eta,\mu^2)$},
    xmin=0.01, xmax=1,
    ymin=0.85, ymax= 1.4,
    legend cell align=left,
    every axis legend/.append style = {
    at={(0.1,0.96)},
    anchor=north west}
    ]
  \pgfplotstableread{PDFPlots/up.dat}\up

  \addplot [black, thick,restrict x to domain = -4.5:-0.356675] 
  table [x index=1, y expr=\thisrow{2}/\thisrow{8}]{\up};
  \addlegendentry{$k_T$ ordered}
  
  \addplot [red,thick,restrict x to domain = -4.5:-0.356675] 
  table [x index=1, y expr=\thisrow{3}/\thisrow{8}]{\up};
  \addlegendentry{$\Lambda$ ordered}

  \addplot [blue, thick,restrict x to domain = -4.5:-0.356675] 
  table [x index=1, y expr=\thisrow{4}/\thisrow{8}]{\up};
  \addlegendentry{$\vartheta$ ordered}
  \end{semilogxaxis}
\end{tikzpicture}

\else 
\begin{center}
\includegraphics[width = 5.5 cm]{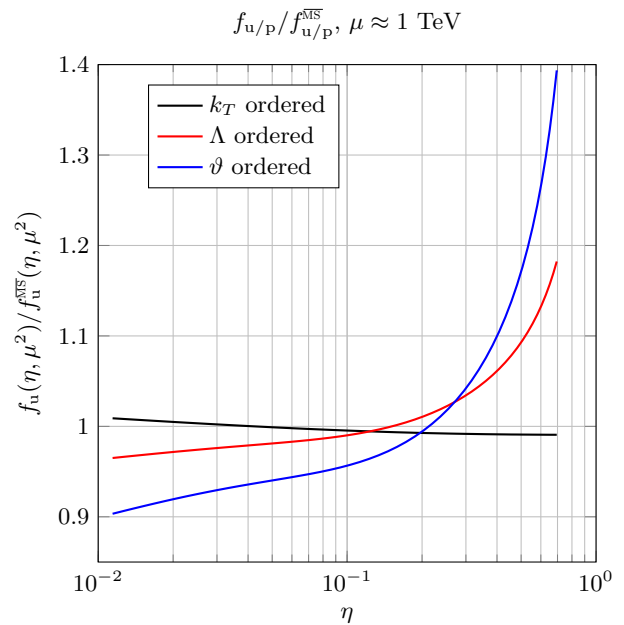}
\end{center}
\fi

\caption{
Ratio of shower oriented up-quark distributions to the $\MSbar$ up quark distribution at $\mu = 1007\ \mathrm{GeV}$: shower oriented distributions for $k_\LT$ ordering, $\Lambda$ ordering, and $\vartheta$ ordering. The rise at large $\eta$ illustrates the PDF part of the summation of threshold logarithms.
}
\label{fig:up1TeVratio}
\end{figure}

In Fig.~\ref{fig:up1TeV}, we plot the up-quark momentum distribution $\eta\,f_{\Lu/\Lp}(\eta,\mu^2)$ as a function of $\eta$ at a large value of the scale, $\mu \approx 1 \TeV$. We show first the $\MSbar$ distribution as a dashed curve. There are quite a lot of up quarks around $\eta \sim 0.1$ because the up quark is a valence quark. The shower oriented distribution for $k_\LT$ ordering is almost the same as the $\MSbar$ distribution. The shower oriented distributions for $\Lambda$ ordering and for $\vartheta$ ordering are somewhat smaller for $\eta < 0.1$. It is difficult to see the differences among ordering choices for $\eta > 0.1$ because the distributions themselves are small. To make these differences visible, we plot in Fig.~\ref{fig:up1TeVratio} the ratios of the three shower oriented PDFs to the $\MSbar$ PDF. We see that for $\Lambda$ and $\vartheta$ ordering, these ratios rise quite steeply as $\eta$ increases. As described in Appendix \ref{sec:thresholdsum}, part of the summation of threshold logarithms is contained in the shower oriented parton distribution functions. This threshold logarithm summation produces the rise at large $\eta$ that we see in Fig.~\ref{fig:up1TeVratio}.

\begin{figure}
\ifusefigs 

\begin{tikzpicture}
  \begin{semilogxaxis}[title = {up-quark distribution, 
    $\mu \approx 50\ \mathrm{GeV}$},
    xlabel={$\eta$}, ylabel={$\eta\,f_{\Lu/\Lp}(\eta,\mu^2)$},
    xmin=0.001, xmax=1,
    legend cell align=left,
    every axis legend/.append style = {
    at={(0.5,0.96)},
    anchor=north west}
    ]
  \pgfplotstableread{PDFPlots/up50.dat}\up
  
  \addplot [black, dashed,restrict x to domain = -6.8:-0.356675] 
  table [x index=1, y expr=\thisrow{8}]{\up};
  \addlegendentry{$\MSbar$}

  \addplot [black, thick,restrict x to domain = -6.8:-0.356675] 
  table [x index=1, y expr=\thisrow{2}]{\up};
  \addlegendentry{$k_T$ ordered}
  
  \addplot [red,thick,restrict x to domain = -6.8:-0.356675] 
  table [x index=1, y expr=\thisrow{3}]{\up};
  \addlegendentry{$\Lambda$ ordered}

  \addplot [blue, thick,restrict x to domain = -6.8:-0.356675] 
  table [x index=1, y expr=\thisrow{4}]{\up};
  \addlegendentry{$\vartheta$ ordered}
  \end{semilogxaxis}
\end{tikzpicture}

\else 
\begin{center}
\includegraphics[width = 5.5 cm]{sample.pdf}
\end{center}
\fi

\caption{Up-quark distributions $\eta\,f_{\Lu/\Lp}(\eta,\mu^2)$ at $\mu = 49.9\ \mathrm{GeV}$: the $\MSbar$ distribution and the shower oriented distributions for $k_\LT$ ordering, $\Lambda$ ordering, and $\vartheta$ ordering.
}
\label{fig:up50GeV}
\end{figure}

\begin{figure}
\ifusefigs 

\begin{tikzpicture}
  \begin{semilogxaxis}[title = {$f_{\Lu/\Lp}/f^\msbar_{\Lu/\Lp}$, 
  $\mu \approx 50\ \mathrm{GeV}$},
    xlabel={$\eta$}, ylabel={$f_{\Lu}(\eta,\mu^2)/f_{\Lu}^{\msbar}(\eta,\mu^2)$},
    xmin=0.001, xmax=1,
    ymin=0.65, ymax= 1.7,
    legend cell align=left,
    every axis legend/.append style = {
    at={(0.1,0.96)},
    anchor=north west}
    ]
  \pgfplotstableread{PDFPlots/up50.dat}\up

  \addplot [black, thick,restrict x to domain = -6.8:-0.356675] 
  table [x index=1, y expr=\thisrow{2}/\thisrow{8}]{\up};
  \addlegendentry{$k_T$ ordered}
  
  \addplot [red,thick,restrict x to domain = -6.8:-0.356675] 
  table [x index=1, y expr=\thisrow{3}/\thisrow{8}]{\up};
  \addlegendentry{$\Lambda$ ordered}

  \addplot [blue, thick,restrict x to domain = -6.8:-0.356675] 
  table [x index=1, y expr=\thisrow{4}/\thisrow{8}]{\up};
  \addlegendentry{$\vartheta$ ordered}
  \end{semilogxaxis}
\end{tikzpicture}

\else 
\begin{center}
\includegraphics[width = 5.5 cm]{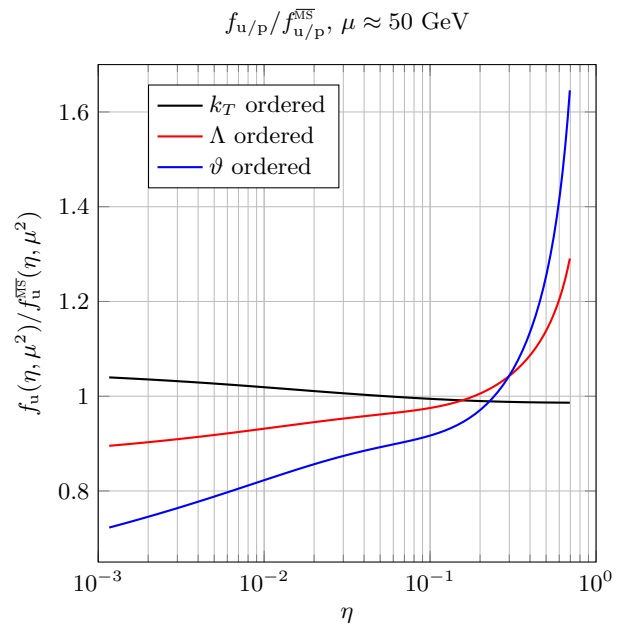}
\end{center}
\fi

\caption{
Ratio of shower oriented up-quark distributions to the $\MSbar$ up quark distribution at $\mu = 49.9\ \mathrm{GeV}$: shower oriented distributions for $k_\LT$ ordering, $\Lambda$ ordering, and $\vartheta$ ordering. 
}
\label{fig:up50GeVratio}
\end{figure}

In Fig.~\ref{fig:up50GeV}, we plot the up-quark momentum distribution at a value of the scale, $\mu \approx 50 \GeV$, that is much larger than $1 \GeV$ but not nearly as large as in Fig.~\ref{fig:up1TeV}. We plot the distributions down to a smaller minimum value of $\eta$, in keeping with the smaller value of $\mu$. We show the $\MSbar$ distribution and the shower oriented distributions for $k_\LT$ ordering, $\Lambda$ ordering, and $\vartheta$ ordering.  In Fig.~\ref{fig:up50GeVratio}, we plot the ratios of the three shower oriented PDFs to the $\MSbar$ PDF. The pattern is similar to what we saw at a 1 TeV scale, but somewhat more pronounced because $\as$ is larger.


\begin{figure}
\ifusefigs 

\begin{tikzpicture}
  \begin{semilogxaxis}[title = {bottom-quark distribution, 
    $\mu \approx 50\ \mathrm{GeV}$},
    xlabel={$\eta$}, ylabel={$\eta\,f_{\Lb/\Lp}(\eta,\mu^2)$},
    xmin=0.001, xmax=1,
    legend cell align=left,
    every axis legend/.append style = {
    at={(0.5,0.96)},
    anchor=north west}
    ]
  \pgfplotstableread{PDFPlots/bottom50.dat}\bottom
  
  \addplot [black, dashed,restrict x to domain = -6.8:-0.356675] 
  table [x index=1, y expr=\thisrow{8}]{\bottom};
  \addlegendentry{$\MSbar$}

  \addplot [black, thick,restrict x to domain = -6.8:-0.356675] 
  table [x index=1, y expr=\thisrow{2}]{\bottom};
  \addlegendentry{$k_T$ ordered}
  
  \addplot [red,thick,restrict x to domain = -6.8:-0.356675] 
  table [x index=1, y expr=\thisrow{3}]{\bottom};
  \addlegendentry{$\Lambda$ ordered}

  \addplot [blue, thick,restrict x to domain = -6.8:-0.356675] 
  table [x index=1, y expr=\thisrow{4}]{\bottom};
  \addlegendentry{$\vartheta$ ordered}
  \end{semilogxaxis}
\end{tikzpicture}

\else 
\begin{center}
\includegraphics[width = 5.5 cm]{sample.pdf}
\end{center}
\fi

\caption{Bottom-quark distributions $\eta\,f_{\Lb/\Lp}(\eta,\mu^2)$ at $\mu = 49.9\ \mathrm{GeV}$: the $\MSbar$ distribution and the shower oriented distributions for $k_\LT$ ordering, $\Lambda$ ordering, and $\vartheta$ ordering.
}
\label{fig:bottom50GeV}
\end{figure}

In Fig.~\ref{fig:bottom50GeV}, we plot the bottom-quark momentum distribution $\eta\,f_{\Lb/\Lp}(\eta,\mu^2)$ at the scale $\mu \approx 50 \GeV$.  As before, we show the $\MSbar$ distribution and the shower oriented distributions for $k_\LT$ ordering, $\Lambda$ ordering, and $\vartheta$ ordering.  The shape of the $\MSbar$ bottom quark distribution is different from that of the up quark distribution because bottom quarks all come from $\Lg \to \Lb \bar{\Lb}$ splittings. However, the relationships among the curves for $\eta < 0.1$ is quite similar to the relationships for up quarks: the distribution $k_\LT$ ordering is slightly larger than the $\MSbar$ distribution, while the distributions for $\Lambda$ ordering and $\vartheta$ ordering are smaller.

\begin{figure}
\ifusefigs 

\begin{tikzpicture}
  \begin{semilogxaxis}[title = {$f_{\Lb/\Lp}/f^\msbar_{\Lb/\Lp}$, 
  $\mu \approx 50\ \mathrm{GeV}$},
    xlabel={$\eta$}, 
    ylabel={$f_{\Lb/\Lp}(\eta,\mu^2)/f_{\Lb/\Lp}^{\msbar}(\eta,\mu^2)$},
    xmin=0.001, xmax=1,
    ymin=0.0, ymax= 1.2,
    legend cell align=left,
    every axis legend/.append style = {
    at={(0.1,0.3)},
    anchor=north west}
    ]
  \pgfplotstableread{PDFPlots/bottom50.dat}\bottom

  \addplot [black, thick,restrict x to domain = -6.8:-0.356675] 
  table [x index=1, y expr=\thisrow{2}/\thisrow{8}]{\bottom};
  \addlegendentry{$k_T$ ordered}
  
  \addplot [red,thick,restrict x to domain = -6.8:-0.74] 
  table [x index=1, y expr=\thisrow{3}/\thisrow{8}]{\bottom};
  \addlegendentry{$\Lambda$ ordered}
  
  \addplot [blue, thick,restrict x to domain = -6.8:-2.4] 
  table [x index=1, y expr=\thisrow{4}/\thisrow{8}]{\bottom};
  \addlegendentry{$\vartheta$ ordered}
  
  \addplot [red,thick,dashed,restrict x to domain = -0.74:-0.356675] 
  table [x index=1, y expr=\thisrow{3}/\thisrow{8}]{\bottom};
  
  \addplot [blue, thick, dashed, restrict x to domain = -2.4:-0.356675] 
   table [x index=1, y expr=\thisrow{4}/\thisrow{8}]{\bottom};

  \end{semilogxaxis}
\end{tikzpicture}

\else 
\begin{center}
\includegraphics[width = 5.5 cm]{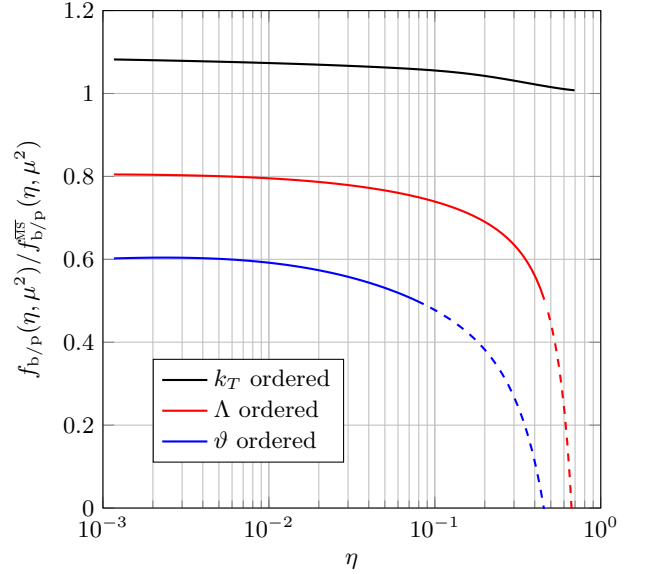}
\end{center}
\fi

\caption{
Ratio of shower oriented bottom-quark distributions to the $\MSbar$ up quark distribution at $\mu = 49.9\ \mathrm{GeV}$: shower oriented distributions for $k_\LT$ ordering, $\Lambda$ ordering, and $\vartheta$ ordering. 
}
\label{fig:bottom50GeVratio}
\end{figure}

In Fig.~\ref{fig:bottom50GeVratio}, we plot the ratios of the three shower oriented PDFs to the $\MSbar$ PDF. Here, we see something quite different from what we saw for up quarks. In the cases of $\Lambda$ ordering and $\vartheta$ ordering, the ratios of shower oriented distributions to the $\MSbar$ distribution do not rise with increasing $\eta$. Instead, they fall. Here, we must face the fact that we are using a first order evolution kernel to go from $k_\LT$ ordering to $\Lambda$ ordering or $\vartheta$ ordering. This is plausibly justified if the change in the parton distribution is reasonably small, say 
\begin{equation}
\label{eq:smallchange}
0.5 <
\frac{f_{\Lb/\Lp}(\eta,\mu^2)}{f_{\Lb/\Lp}^\msbar(\eta,\mu^2)} < 2
\;.
\end{equation}
When the condition (\ref{eq:smallchange}) is violated, we judge that higher order contributions to the evolution kernel are needed. The regions for which the lowest order evaluation appears to be untrustworthy by this criterion are indicated by dashed curves in Fig.~\ref{fig:bottom50GeVratio}.


\subsection{Which PDFs to use}
\label{sec:whichPDFs}

Assume that we base the shower cross section on one of the options in Sec.~\ref{sec:unresolvedregion}: $k_\LT$ ordering, $\Lambda$ ordering, or $\vartheta$ ordering. (Later in this paper, we propose more complicated choices.) Then we use the corresponding definition of shower oriented parton distribution functions as described above in this section. The PDFs appear as the operator $\cF(\muh^2)$ in the cross section formula (\ref{eq:sigmaJshower}).

We still have a choice to make. One possibility is to define the shower oriented PDFs using evolution in the parameter $\lambda$. For this, we begin with the $\MSbar$ PDFs and apply the $P^{(\epsilon)}$ transformation (\ref{eq:Kdef0A}) to obtain the $k_\LT$ ordered PDF. Then we use evolution in $\lambda$: we solve either Eq.~(\ref{eq:dtildefdlambda}) to obtain the $\Lambda$ ordered PDF or Eq.~(\ref{eq:fdlambdaAngular}) to obtain the $\vartheta$ ordered PDF. This produces the PDFs exhibited in the previous subsection. 

Alternatively, we can use $\mu^2$ evolution. We begin with the $\MSbar$ PDFs at the starting scale $m_\perp^2$, adjust these with the $P^{(\epsilon)}$ transformation, and then let the PDFs evolve from $\mu^2 = m_\perp^2$ to higher values of $\mu^2$ using the evolution equation (\ref{eq:evolutionkernel1}) and (\ref{eq:dtildefdlogmusq}) for $\Lambda$ ordering or (\ref{eq:dtildefdlogmusqangle}) and (\ref{eq:evolutionkernel1angle}) for $\vartheta$ ordering or just the ordinary DGLAP equation for $k_\LT$ ordering. 

The $\MSbar$ PDFs are produced using next-to-leading order evolution in $\mu^2$. However, both the $\lambda$ evolution and the $\mu^2$ evolution used to define shower oriented PDFs use just leading order evolution kernels. With this limited accuracy, the results from $\mu^2$ evolution can be noticeably different from the results from $\lambda$ evolution.

For $\mu^2 \gg m_\Lb^2$, we expect the results using $\lambda$ evolution to be the most reliable, except for charm and bottom quarks at very large values of $\eta$.\footnote{When the first inequality in the condition (\ref{eq:smallchange}) is violated, we replace $f_{\Lb/\Lp}(\eta,\mu^2) \to 0.5\,f_{\Lb/\Lp}^\msbar(\eta,\mu^2)$. The distribution functions for $\bar \Lb, \Lc$ and $\bar \Lc$ are treated analogously.} In the operators $\cF(\muh^2)$ and $\cU_\cV(t_\Lf, t_{\scH})$ in Eq.~(\ref{eq:sigmaJshower}) for the cross section, the PDFs are evaluated at or not far from the scale $\mu_\scH$ of the hard scattering. For this reason, we use the $\lambda$-evolution version of the shower oriented PDFs in these operators. In the shower evolution operator $\cU(t_\Lf, t_{\scH})$ in Eq.~(\ref{eq:sigmaJshower}), the PDFs are evaluated at all scales $\mu^2$ down to $m_\perp^2$. At small scales, we find that the PDFs for charm and bottom quarks obtained using $\lambda$ evolution are badly behaved, while the PDFs obtained from $\mu^2$ evolution are well behaved and nicely matched to the evolution of the shower. For this reason, we use the $\mu^2$-evolution version of the shower oriented PDFs in $\cU(t_\Lf, t_{\scH})$.

\section{Multiple scales}
\label{sec:multiplescales}

In Eqs.~(\ref{eq:cDandhatD}) and (\ref{eq:hatDstructure}), we have described the action of $\cF{\cal D}^{[1,0]}_\mathrm{a}\cF^{-1}\sket{\{p,f,c,c'\}_{m}}$ resulting from an initial state splitting using an operator $\hat {\bm D}^\La_{a a'}$ that is a function of $z$ and the momenta and flavors of the partonic states and is an operator on the partonic color state. In general, this operator has both soft and collinear singularities. Our first task in this section is to divide $\hat {\bm D}^\La_{a a'}$ into terms with different singularity structures.

\subsection{Singularity structures}
\label{sec:singularitydecomposition}

Let $\hat {\bm D}^\La_{a \hat a}(z;\{\hat p,\hat f\}_\mpone,\{p,f\}_{m})$ denote the full splitting function in the \textsc{Deductor} algorithm, Eq.~(\ref{eq:hatDstructure}) or (\ref{eq:D10a4}). This is the splitting function before making any approximation with respect to color, even though the \textsc{Deductor} code then treats color approximately. The default color approximation in \textsc{Deductor} is the LC+ approximation \cite{NScolor},\footnote{\textsc{Deductor} can also calculate corrections to the LC+ approximation \cite{NSNewColor}.} in which $\hat{\bm D}^\La_{a a'}$ is replaced by an approximate operator $\hat{\bm D}^{\mathrm{a, LC+}}_{a a'}$. The terms in $\hat{\bm D}^\La_{a a'}$ in which the dipole partner parton is the same as the emitting parton, $k = \La$, have a simple color structure. In the LC+ approximation, these $k = \La$ terms are unchanged. In the terms proportional to the function $W_0$ in Eq.~(\ref{eq:hatDstructure}), which have $k \ne \La$, some contributions are dropped. The terms proportional to $W_0$ inside the integrations are singular in the soft limit, $(1-z) \to 0$ at fixed $\vartheta$, but not in the collinear limit, $\vartheta \to 0$ at fixed $(1-z)$, as we see explicitly in Eq.~(\ref{eq:Wlk-ini-result2}). Thus the difference operator
\begin{equation}
\label{eq:softdef}
\hat{\bm D}^{\mathrm{a, soft}}_{a a'}
= \hat{\bm D}^\La_{a a'}
- \hat{\bm D}^{\mathrm{a, LC+}}_{a a'}
\end{equation}
is singular in the soft limit but not in the collinear limit \cite{NScolor}. These terms are proportional to $\delta_{a a'}$, so $a = a'$. Thus we can use Eq.~(\ref{eq:softdef}) to decompose $\hat{\bm D}^\La_{a a'}$ in the form
\begin{equation}
\label{eq:LCplusandsoft}
\hat{\bm D}^\La_{a a'}
= \hat{\bm D}^{\mathrm{a,LC+}}_{a a'}
+
\hat{\bm D}^{\mathrm{a,soft}}_{a a'}
\;.
\end{equation}
Then $\hat{\bm D}^{\mathrm{a, soft}}_{a a'}$ does not contain a collinear singularity.

This decomposition for $\hat{\bm D}^\La_{a a'}$ induces a corresponding decomposition for the operator $\hat{\bm P}^\mathrm{a,NS}_{a a'}$ that appears in the inclusive infrared finite operator $\cV$, Eqs.~(\ref{eq:cV1integral}) and (\ref{eq:bfVresult2}). The operator $\hat{\bm P}^\mathrm{a,NS}_{a a'}$ is written out in Eq.~(\ref{eq:hatbfPNS}).  We define $\hat{\bm P}^\mathrm{a,LC+}_{a a'}$ to be $\hat{\bm P}^\mathrm{a,NS}_{a a'}$ with the LC+ approximation applied to its color operators. Then we define $\hat{\bm P}^\mathrm{a,soft}_{a a'}$ by
\begin{equation}
\begin{split}
\label{eq:LCplusandsoftP}
\hat {\bm P}^\mathrm{a,NS}_{a a'}(z;\{p\}_{m})
={}&
\hat {\bm P}^\mathrm{a,LC+}_{a a'}(z;\{p\}_{m})
\\& + 
\hat {\bm P}^\mathrm{a,soft}_{a a'}(z;\{p\}_{m})
\;.
\end{split}
\end{equation}
In $\hat {\bm P}^\mathrm{a,NS}_{a a'}(z;\{p\}_{m})$, the $k = \La$ terms are proportional to the unit color operator and remain unchanged between $\hat{\bm P}^\mathrm{a,NS}_{a a'}$ and $\hat{\bm P}^\mathrm{a,LC+}_{a a'}$, so that they do not appear in $\hat {\bm P}^\mathrm{a,soft}_{a a'}$. The terms with $k \ne \La$ are proportional to $\delta_{a a'}$ and have nontrivial color operators $[\bm T_k \cdot \bm T_\La \otimes 1] +[1 \otimes \bm T_k \cdot \bm T_\La]$. Some of these terms are changed in the LC+ approximation and thus contribute to $\hat {\bm P}^\mathrm{a,soft}_{a a'}$. See Eqs.~(\ref{eq:hatbfPNSLCplus}) and (\ref{eq:hatbfPNSsoft}).

In the simplest splitting algorithm, one would use the same unresolved region $U(\vec \mu_\scS)$ for $\hat{\bm D}^{\mathrm{a, LC+}}_{a a'}$ and $\hat{\bm D}^{\mathrm{a, soft}}_{a a'}$ and for $\hat{\bm P}^\mathrm{a,LC+}_{a a'}$ and $\hat{\bm P}^\mathrm{a,soft}_{a a'}$. In this section, we propose an alternative in which we define different unresolved regions for the ``LC+'' operators and the ``soft'' operators based on the differences in their singularity structures. We then use Eq.~(\ref{eq:LCplusandsoft}) and (\ref{eq:LCplusandsoftP}) as the definitions of $\hat{\bm D}^\La_{a a'}$ and $\hat{\bm P}^\mathrm{a,NS}_{a a'}$.

\subsection{Scales for LC+ operators}
\label{sec:LCplusscales}

In our proposed choice for the unresolved region for $\hat{\bm D}^{\mathrm{a,LC+}}_{a a'}$ and $\hat{\bm P}^\mathrm{a,LC+}_{a a'}$, the evolution is described by three scales, 
\begin{equation}
\label{eq:twoscales}
\vec \mu_\scS = (\mu_\scE, \mu_\scC, \mu_{\mi \pi})
\;.
\end{equation}
The third scale, $\mu_{\mi \pi}$, is the scale for the imaginary part of virtual graphs, regarded now as an independent scale. This scale controls the color phase operator, $\mi\pi \cS^{[0,1]}_{\mi\pi}(t)$, as specified in Eq.~(\ref{eq:SiPiResult}), but does not affect real emissions. The scale $\mu_\scC$ will describe part of the evolution of real emissions. The subscript in $\mu_\scC$ denotes one of the scale choices defined previously using $a_\scC(z,\mu_\scC^2)$ with $\mathrm{C} = \ \perp$ for $k_\LT$ as the ordering variable (Eq.~(\ref{eq:aPerpdef})), $\mathrm{C} = \Lambda$ for $\Lambda$ as the ordering variable (Eq.~(\ref{eq:aLambdadef})), or $\mathrm{C} = \angle$ for $\vartheta Q^2$ as the ordering variable (Eq.~(\ref{eq:aAngledef})). We add an additional scale, $\mu_\scE$, that will provide a cut on the energy of an emitted parton.

With scales $\mu_\scE$ and $\mu_\scC$, we define the unresolved region $U(\mu_\scE, \mu_\scC)$ to be used for the operators $\hat{\bm D}^{\mathrm{a, LC+}}_{a a'}$ and $\hat{\bm P}_{a a'}^\mathrm{LC+}$ as follows. We say that $(z,\vartheta) \in U(\mu_\scE, \mu_\scC)$ if $0<z<1$, $0 < \vartheta < 1$ and
\begin{equation}
\label{eq:ylim1}
\vartheta <
\max\left\{
a_\mathrm{cut}(z,\mu_\scE^2,\mu_\scC^2),
a_\perp(z,m_\perp^2(a,\hat a))\right\}
\end{equation}
as in Eq.~(\ref{eq:acutamperp}), where
\begin{equation}
\label{eqYcutdef}
a_\mathrm{cut}(z,\mu_\scE^2,\mu_\scC^2) =
\max\left\{
a_\scE(z,\mu_\scE^2),
a_\scC(z, \mu_\scC^2)\right\}
\;.
\end{equation}
Here we use the definition (\ref{eq:aPerpdef}), (\ref{eq:aLambdadef}), or (\ref{eq:aAngledef}) for $a_\scC(z,\mu_\scC^2)$ and define
\begin{equation}
\label{eq:aEdef}
a_\scE(z,\mu_\scE^2)
=
\begin{cases}
1 & \mathrm{for}\ (1-z)Q^2 < \mu_\scE^2
\\
0 & \mathrm{otherwise}
\end{cases}
\;.
\end{equation}
The inclusion of $a_\scE(z,\mu_\scE^2)$ in Eq.~(\ref{eq:ylim1}) means that a splitting $(z,\vartheta)$ is unresolved when $(1-z) < \mu_\scE^2/Q^2$. Thus $\mu_\textsc{e}$ controls the approach to the soft limit by putting an upper bound on the momentum fraction (and thus the energy, E) of the emitted parton. The detailed specification of the functions that we need in $\hat{\bm D}^\La_{a a'}$ and $\hat{\bm P}^\La_{a a'}$ are given in Appendix \ref{sec:twoscales}.

\begin{figure}
\begin{center}
\ifusefigs 

\begin{tikzpicture}

\newcommand\epsilonperp{0.002}
\newcommand\xmin{0.2}
\newcommand\ymin(0.1)
\newcommand\ymperp{\epsilonperp*(1-x)/x^2}
\newcommand\yLambda{0.1*(1-x)/x}
\newcommand\xymin{0.1/1.1)}
\newcommand\xmperpmin{0.04}

\begin{axis}[
xlabel={$1-z$}, 
ylabel={$\vartheta$},
ymin = 0,
ymax = 1,
xmin = 0,
xmax = 1,
legend cell align=left,
every axis legend/.append style = {
    at={(0.95,0.95)},
    anchor=north east}
]

\addplot [name path=plot1,black,dashed,domain=\xmperpmin:1]
{\ymperp};
\addplot [name path=plot2L,blue,thick,domain=\xymin:\xmin]
{\yLambda};
\addplot [name path=plot2R,blue,thick,domain=\xmin:1]
{\yLambda};
\addplot [blue,thick] coordinates{(\xmin,0) (\xmin,1)};

\addplot [name path=zeroL,black,domain=0:\xmin]{0};
\addplot [name path=zeroR,black,domain=\xmin:1]{0};
\addplot [name path=oneL,black,domain=0:\xmin]{1};
\addplot [name path=oneR,black,domain=\xmin:1]{1};

\addplot[fill opacity=0.2, blue] fill between[of = zeroL and oneL];
\addplot[fill opacity=0.2, blue] fill between[of = zeroR and plot2R];
\addplot[fill opacity=0.2, yellow] fill between[of = plot2R and oneR];

\end{axis}
\end{tikzpicture}

\else 
\begin{center}
\includegraphics[width = 5.5 cm]{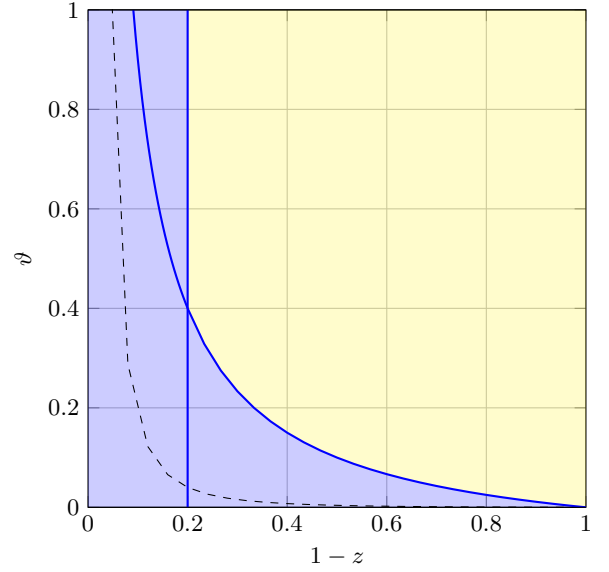}
\end{center}
\fi
\end{center}
\caption{
Resolved and unresolved regions for $\hat{\bm P}_{a a'}^\mathrm{LC+}$ for $\mathrm C = \Lambda$ with $r_\La \mu_\scC^2/Q^2 = 0.1$ and $\mu_\scE^2/Q^2 = 0.2$. The unresolved region is shaded blue while the resolved region is shaded yellow. The curve for an $m_\perp^2(a,a')$ cut with $m_\perp^2(a,a')/Q^2 = 0.002$ is shown as a dashed line, but in this case the $m_\perp^2(a,a')$ cut has no effect.
}
\label{fig:unresolvedSC}
\end{figure}

The unresolved region in the plane of $1-z$ and $\vartheta$ is illustrated in Fig.~\ref{fig:unresolvedSC} for $\Lambda$ ordering, $\mathrm{C} = \Lambda$, with $r_\La \mu_\scC^2/Q^2 = 0.1$ and $\mu_\scE^2/Q^2 = 0.2$. The unresolved region is shaded blue, while the resolved region is shaded yellow. We illustrate a potential $m_\perp^2(a,a')$ cut with $m_\perp^2(a,a')/Q^2 = 0.002$. In this example, the $m_\perp^2(a,a')$ cut has no effect.

\subsection{A special treatment for soft splittings}
\label{sec:softsplittings}

We adopt a different strategy for the soft operators $\hat{\bm D}^{\mathrm{a, soft}}_{a a'}$ and  $\hat{\bm P}_{a a'}^\mathrm{soft}$, which have $(1-z) \to 0$ singularities but no collinear, $\vartheta \to 0$, singularities. We define the unresolved region $U(\mathrm{soft}; \mu_\scE)$ for $\hat{\bm D}^{\mathrm{a, soft}}_{a a'}$ and $\hat{\bm P}_{a a'}^\mathrm{soft}$ by letting $(z,\vartheta) \in U(\mathrm{soft},\mu_\scE)$ if $0<z<1$, $0 < \vartheta < 1$ and 
\begin{equation}
\label{eq:ylim2}
\vartheta <
\max\left\{
a_\scE(z,\mu_\scE^2),
a_\perp(z,m_\perp^2)\right\}
\;,
\end{equation}
Here we have noted that for $\hat{\bm D}^{\mathrm{a, soft}}_{a a'}$ and  $\hat{\bm P}_{a a'}^\mathrm{soft}$, $a = a'$ so $m_\perp^2(a,a') = m_\perp^2$ according to Eq.~(\ref{eq:mperpahata}). This provides a limit on $k_\LT^2$ at a fixed, small, infrared scale $m_\perp^2$. 

\begin{figure}
\begin{center}
\ifusefigs 

\begin{tikzpicture}

\newcommand\epsilonperp{0.002}
\newcommand\xmin{0.2}
\newcommand\ymin(0.1)
\newcommand\ymperp{\epsilonperp*(1-x)/x^2}
\newcommand\yLambda{0.1*(1-x)/x}
\newcommand\xymin{0.1/1.1)}
\newcommand\xmperpmin{0.04}

\begin{axis}[
xlabel={$1-z$}, 
ylabel={$\vartheta$},
ymin = 0,
ymax = 1,
xmin = 0,
xmax = 1,
legend cell align=left,
every axis legend/.append style = {
    at={(0.95,0.95)},
    anchor=north east}
]

\addplot [name path=plot1L,black,dashed,domain=\xmperpmin:\xmin]
{\ymperp};
\addplot [name path=plot1R,black,dashed,domain=\xmin:1]
{\ymperp};
\addplot [blue,thick] coordinates{(\xmin,0) (\xmin,1)};

\addplot [name path=zeroL,black,domain=0:\xmin]{0};
\addplot [name path=zeroR,black,domain=\xmin:1]{0};
\addplot [name path=oneL,black,domain=0:\xmin]{1};
\addplot [name path=oneR,black,domain=\xmin:1]{1};

\addplot[fill opacity=0.2, blue] fill between[of = zeroL and oneL];
\addplot[fill opacity=0.2, blue] fill between[of = zeroR and plot1R];
\addplot[fill opacity=0.2, yellow] fill between[of = plot1R and oneR];

\end{axis}
\end{tikzpicture}

\else 
\begin{center}
\includegraphics[width = 5.5 cm]{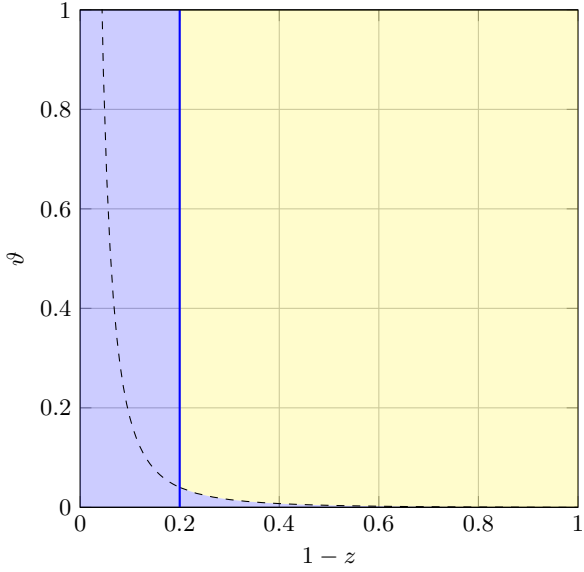}
\end{center}
\fi
\end{center}
\caption{
Resolved and unresolved regions for $\hat{\bm P}_{a a'}^\mathrm{soft}$ for $\mu_\scE^2/Q^2 = 0.2$ and $m_\perp^2/Q^2 = 0.002$ (dashed line).
}
\label{fig:unresolvedSoft}
\end{figure}

The unresolved region for $\hat{\bm D}^{\mathrm{a, soft}}_{a a'}$ and $\hat{\bm P}_{a a'}^\mathrm{soft}$ in the plane of $1-z$ and $\vartheta$ is illustrated in Fig.~\ref{fig:unresolvedSoft} for $\mu_\scE^2/Q^2 = 0.2$ and $m_\perp^2/Q^2 = 0.002$. In the collinear limit, $\vartheta \to 0$ with fixed $z$, $\hat{\bm D}^{\mathrm{a, soft}}_{a a'}$ and $\hat{\bm P}_{a a'}^\mathrm{soft}$ are not singular. For $(1-z) > \mu_\scE^2/Q^2$, the only cut on $\vartheta$ is the fixed infrared cut $k_\LT^2 > m_\perp^2$.

This treatment divides $\cD_l^{[1,0]}$ for $l \in \{\La,\Lb\}$ into separate contributions $\cD_{l,\mathrm{LC+}}^{[1,0]}$ and $\cD_{l,\mathrm{soft}}^{[1,0]}$. For $\cD_{l,\mathrm{LC+}}^{[1,0]}$, the unresolved region is defined by two scale parameters, $\mu_\scE$ and $\mu_\scC$. For $\cD_{l,\mathrm{soft}}^{[1,0]}$, the unresolved region is defined by $\mu_\scE$ only. We do the same thing for final state splittings, $l \in \{1,\dots,m\}$. The details are presented in Ref.~\cite{angleFS}.

\section{A path for the three scale parameters}
\label{sec:pathchoice}

We have defined an unresolved region $(z,\vartheta) \in U(\mu_\scE, \mu_\scC)$ for scale parameters, $(\mu_\scE, \mu_\scC)$, for real emission graphs and a separate unresolved region $(1-\cos\theta) < \mu_{\mi\pi}^2\,f_{\mi\pi}^{lk}(\{p\}_m)$ for the integral that gives the imaginary part of virtual graphs.

We now need to specify a path $\vec \mu(t)$ in the space of $\vec \mu = (\mu_\scE, \mu_\scC, \mu_{\mi\pi})$, with $t$ in the arbitrarily chosen range $0 < t <3$. First, we should specify the endpoints. For the end of the shower, $t \to 3$, we can take $\mu_\scE(3) = 0$, $\mu_\scC(3) = 0$, and $\mu_{\mi\pi}(3) = m_\perp$. For the start of the shower at $t = 0$, for $k_\LT$ or $\Lambda$ ordering, we first choose $\mu_\scC^2(0)$ to be a scale $\muh^2$ appropriate to the hard scattering that initiates the shower. For $\vartheta$ ordering, we choose $\mu_\scC^2(0) = \mu_\scH^2$ with $\mu_\scH^2 = Q_0^2$. For $\mu_\scE(0)$, we need an appropriate matching scale. For this purpose, define $z_\scH$ as the solution of
\begin{equation}
\label{eq:zHdef}
a_\scC(z_\scH,\muh^2) = 1
\;.
\end{equation}
This gives $1-z_\scH$ that is not close to 0. For $\vartheta$ ordering, Eq.~(\ref{eq:zHdef}) is satisfied for any $z_\scH$ in the range $0 < z_\scH < 1$. Thus for $\vartheta$ ordering, we can choose $1 - z_\scH = 1$.

Having chosen $z_\scH$,  we can set
\begin{equation}
\label{eq:muE0}
\mu_\scE^2(0) = (1 - z_\scH) Q_0^2
\;,
\end{equation}
so that $\mu_\scE^2(0)$ is large, of order $Q_0^2$. For $\mu_{\mi\pi}(0)$, we can choose $\mu_{\mi\pi}(0) = \muh$.

\begin{figure}
\begin{center}
\ifusefigs 

\begin{tikzpicture}

\begin{axis}[
xlabel={$\mu_\scE/\mu_\scE(0)$},
ylabel={$\mu_\scC/\mu_\scH$},
zlabel={$\mu_{\mi\pi}/\mu_\scH$},
xmin = 0.0,
xmax = 1.0,
ymin = -0.05,
ymax = 1.0,
zmin = 0.0,
zmax = 1.05,
xtick={0,0.5,1},
ytick={0,0.5,1},
ztick={0,0.5,1},
]

\addplot3 [
scatter,only marks,point meta=explicit symbolic, 
scatter/classes={a={mark=*,black,mark size=2.5pt}
},
] table [meta=label] {
x   y   z   label 
1.0 1.0 1.0 a 
0.0 1.0 1.0 a 
0.0 0.0 1.0 a 
0.0 0.0 0.1 a
};

\addplot3[blue,very thick,postaction={decorate, decoration={markings,
    mark=at position 0.6 with {\arrow[scale=1.0,>=Latex]{>};},}}] 
    coordinates{(1,1,1) (0,1,1)};
\addplot3[red,very thick,postaction={decorate, decoration={markings,
    mark=at position 0.6 with {\arrow[scale=1.0,>=Latex]{>};},}}]
    coordinates{(0,1,1) (0,0,1)};
\addplot3[mygreen,very thick,postaction={decorate, decoration={markings,
    mark=at position 0.6 with {\arrow[scale=1.0,>=Latex]{>};},}}]  coordinates{(0,0,1) (0,0,0.1)};

\node[]at({0.9,1,0.9}){$t = 0$};
\node[]at({0.1,1,0.9}){$t = 1$};
\node[]at({0.1,0.1,0.95}){$t = 2$};
\node[]at({0.15,0.1,0.15}){$t = 3$};

\end{axis}
\end{tikzpicture}

\else 
\begin{center}
\includegraphics[width = 5.5 cm]{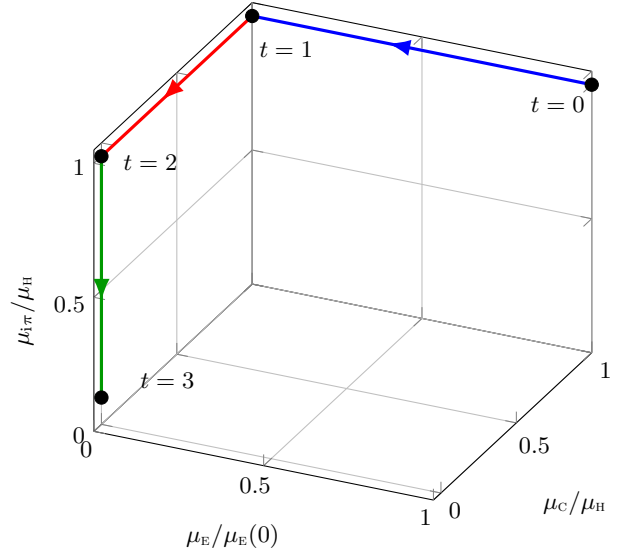}
\end{center}
\fi
\end{center}
\caption{
Evolution path with three segments with $m_\perp= 0.1 \mu_\scH$.
}
\label{fig:evolution-path}
\end{figure}

There are many paths that one might choose that connect $(\mu_\scE(0),\mu_\scC(0), \mu_{\mi\pi}(0))$ and $(\mu_\scE(3),\mu_\scC(3), \mu_{\mi\pi}(3))$. We make a particular choice in which we change each of the three scales one at a time as we change $t$ from 0 to 1, then from 1 to 2, and finally from 2 to 3. The path $\vec \mu(t)$ is illustrated in Fig.~\ref{fig:evolution-path}. We let $\mu_\scE(t)$ decrease from $\mu_\scE(0)$ to 0 as $t$ increases from $0$ to 1. For $1 < t < 3$, $\mu_\scE(t)$ remains at 0. For example, we can take
\begin{equation}
\begin{split}
\mu_\scE(t) 
={}&
\begin{cases}
(1-t) \mu_\scE(0)  & 0 < t < 1\\
0 & 1< t < 3
\end{cases}
\;.
\end{split}
\end{equation}
We let $\mu_\scC(t)$ retain its initial value, $\muh$, for $0<t<1$. Then we let $\mu_\scC(t)$ decrease from $\muh$ to 0 as $t$ increases from $1$ to $2$. For $2 < t < 3$, we let $\mu_\scC(t)$ remain at 0. For example, we can take
\begin{equation}
\begin{split}
\mu_\scC(t)
=
\begin{cases}
\muh & 0 < t < 1\\
(2-t) \muh & 1 < t < 2\\
0 & 2 < t < 3
\end{cases}
\;.
\end{split}
\end{equation}
Finally, we let $\mu_{\mi\pi}(t)$ remain at $\muh$ for $0 < t < 2$. Then we let $\mu_{\mi\pi}(t)$ decrease from $\muh$ to $m_\perp$ as $t$ increases from $2$ to $3$. For example, we can take
\begin{equation}
\begin{split}
\mu_{\mi\pi}(t) 
=
\begin{cases}
\muh & 0 < t < 2\\
(t-2)m_\perp + (3-t)\muh & 2 < t < 3\\
\end{cases}
\;.
\end{split}
\end{equation}
%

Let us consider the three operators $\cU(1,0)$, $\cU(2,1)$, and $\cU(3,2)$ corresponding to the three segments of this path. In each case, $\cU$ is determined by  $d\cD^{[1,0]}(t)/dt$ or $d\,\mathrm{Im}\,\cD^{[0,1]}(t)/dt$ on that path segment, according to Eqs.~(\ref{eq:cS1decomposition1}), (\ref{eq:cS10}), and (\ref{eq:SiPi}).

On the first segment, $0 < t < 1$, $\mu_{\mi\pi}$ is fixed so there is no contribution from $\mathrm{Im}\,\cD^{[0,1]}(t)$. There is a contribution from $\cD^{[1,0]}(t)$, which contains terms proportional to $\hat{\bm D}^{\mathrm{a, LC+}}_{a a'}$ and $\hat{\bm D}^{\mathrm{a, soft}}_{a a'}$, according to Eqs.~(\ref{eq:cDandhatD}) and (\ref{eq:LCplusandsoft}). On this segment, $\mu_\scE$ decreases. Since $\hat{\bm D}^{\mathrm{a, soft}}_{a a'}$ depends on $\mu_\scE$, there is a contribution from $\hat{\bm D}^{\mathrm{a, soft}}_{a a'}$. The scale $\mu_\scC$ is fixed at $\mu_\scC(0) = \mu_\scH$. 

Is there a contribution from $\hat{\bm D}^{\mathrm{a, LC+}}_{a a'}$? The unresolved region $U(\mu_\scE, \mu_\scC)$ for $\hat{\bm D}^{\mathrm{a, LC+}}_{a a'}$ is $U(\mu_\scE(t), \mu_\scH)$ for $0 < t < 1$. With a little analysis in Appendix \ref{sec:cUmuEmuC}, we find that for $0 < t < 1$, $(z,\vartheta) \in U(\mu_\scE(t), \mu_\scH)$ if $\vartheta < a_\mathrm{total}(z,\mu_\scE^2(t))$, where
\begin{align}
\label{eq:atotaldef}
a_\mathrm{total}&(z,\mu_\scE^2(t)) 
\\ \notag ={}&
\min\!\Big[ 1,
\\ \notag &\
\max\!\left\{
a_\scE(z,\mu_\scE^2(t)),
a_\scC(z, \muh^2),
a_\perp(z,m_\perp^2(a,\hat a))\right\}\!
\Big]
.
\end{align}
We find in Eq.~(\ref{eq:atotalresult}) that, for $0 < t < 1$,
\begin{equation}
\label{eq:atotal}
a_\mathrm{total}(z,\mu_\scE^2(t))
= 
\begin{cases}
1 & 1-z < 1 - z_\scH \\
a_\mathrm{total}(z,\mu_\scE^2(0)) & 
1 - z_\scH < 1-z
\end{cases}
\;.
\end{equation}
This is independent of $t$. Thus when we construct the generator $\cS^{[1,0]}(t)$ of shower evolution by differentiating $\cD^{[1,0]}(t)$ with respect to $t$, $\hat{\bm D}^{\mathrm{a, soft}}_{a a'}$ contributes but $\hat{\bm D}^{\mathrm{a, LC+}}_{a a'}$ does not.

On the second segment, $1 < t < 2$, $\mu_{\mi\pi}$ is fixed, so there is no contribution from $\mathrm{Im}\,\cD^{[0,1]}(t)$. Also, $\mu_\scE(t)$ is fixed at $\mu_\scE(t) = 0$. The unresolved region for $\hat{\bm D}^{\mathrm{a, soft}}_{a a'}$ depends only on $\mu_\scE$ according to Eq.~(\ref{eq:ylim2}), so there is no contribution from $\hat{\bm D}^{\mathrm{a, soft}}_{a a'}$. On the second segment, $\mu_\scC^2$ decreases from $\mu_\scH^2$ to 0. In $\hat{\bm D}^{\mathrm{a, LC+}}_{a a'}$, the unresolved region is determined by $\mu_\scC^2(t)$ with $\mu_\scE^2$ set to zero. Thus $\hat{\bm D}^{\mathrm{a, LC+}}_{a a'}$ {\em does} contribute to $\cU(2,1)$.  With one notable qualification, this gives us a shower based on $k_\LT$ ordering ($\LC = \ \perp$), $\Lambda$ ordering ($\LC = \Lambda$), or $\vartheta$ ordering ($\LC = \angle$) from Sec.~\ref{sec:simpleordering}. The qualification is that $\hat{\bm D}^{\mathrm{a, LC+}}_{a a'}$ uses the LC+ approximation for color.

On the third segment, $2 < t < 3$, $\mu_\scE^2$ and $\mu_\scC^2$ are fixed, so there is no contribution from $\cD^{[1,0]}(t)$. Since $\mu_{\mi\pi}^2$ decreases from $\muh^2$ to $m_\perp^2$, there is a contribution from $\mathrm{Im}\,\cD^{[0,1]}(t)$. Thus there is a color phase factor $\mi\pi\cS^{[0,1]}_{\mi\pi}(t)$.

With this choice of shower scales and path, the shower evolution from the hard scale to the scale at which the shower stops has the form from Eq.~(\ref{eq:cU}),
\begin{equation}
\cU(3,0)
= \cU(3,2)\, \cU(2,1)\,
\cU(1,0)
\;.
\end{equation}
We thus arrive at what seems to us quite a surprising structure. The operator $\cU(2,1)$ is the most important of these evolution operators. It represents a full first order shower from the hard scale $\muh^2$ to zero using $k_\LT$, $\Lambda$, or $\vartheta$ ordering and using the LC+ approximation for color. The operator $\cU(1,0)$ provides an exponentiated correction to the LC+ color approximation. It is generated by an operator that produces soft gluon emissions and exchanges using the difference between color matrices for full color and the corresponding color matrices for LC+ color. The remaining operator, $\cU(3,2)$ produces an exponentiation of the color phase from the imaginary part of virtual graphs, generated by $S_{\mi\pi}(t)$.

We have chosen a path $(\mu_\scE(t), \mu_\scC(t), \mu_{\mi\pi}(t))$ such that $\mu_\scE^2(t) \to 0$ with fixed $\mu_\scC^2(t)$ for as $t$ increases, then $\mu_\scC^2(t) \to 0$. Thus it is $\mu_\scC^2(t)$ that controls  the resolved region for large $t$. Then in the general definition Eq.~(\ref{eq:ylimdef0}),
\begin{equation}
\mu_\mathrm{lim}(\vec \mu_\scS) = \mu_\scC
\;.
\end{equation}
Then also
\begin{equation}
a_\mathrm{lim}(z,\mu_\scC^2(t)) = a_\scC(z,\mu_\scC^2(t))
\;,
\end{equation}
as in Eq.~(\ref{eq:ylimdef0}).

\section{The cross section}
\label{sec:crosssection}

We now describe the components of an infrared safe cross section when we use three scale parameters $\vec \mu = (\mu_\scE, \mu_\scC, \mu_{\mi\pi})$ with the special path described in the previous section in which $\mu_\scE$ changes first, then $\mu_\scC$, then $\mu_{\mi\pi}$. The scale $\mu_\scC$ can correspond to any of $k_\LT$, $\Lambda$, or $\vartheta$ ordering. We consider any infrared safe cross section, but concentrate on cross sections for which the beginning hard scattering process is either Drell-Yan muon pair production or the hard scattering of two partons. The measured cross section could involve more than just the beginning hard process. For instance, one could measure the transverse momentum of the muon pair in the Drell-Yan process.

The parton shower representation for the cross section is described in some detail in Ref.~\cite{NSAllOrder} and is stated in Eq.~(\ref{eq:sigmaJshower}). If we decompose  $\cU(3, 0)$ as $\cU(3, 2)\, \cU(2, 1)\,\cU(1, 0)$ and decompose $\cU_\cV(3, 0)$ as $\cU_\cV(3, 2)\, \cU_\cV(2, 1)\, \cU_\cV(1, 0)$, we have
\begin{equation}
\begin{split}
\label{eq:sigmaJshowerthreescale}
\sigma ={}& \sbra{1}
\cO_J\,
\cU(3, 2)\, \cU(2, 1)\,\cU(1, 0)
\\&\times
\cU_\cV(3, 2)\, \cU_\cV(2, 1)\, \cU_\cV(1, 0)
\cF(\muh^2)
\sket{\rho_\scH}
\;.
\end{split}
\end{equation}
At the end of the shower, $\cO_J$ makes the desired measurement on the many-parton state. Then $\sbra{1}$ represents an inclusive sum over the parton state variables. At the start of the shower, $\sket{\rho_\scH}$ is the parton statistical state at the hard interaction. It can be computed beyond the leading perturbative order if we use infrared subtractions that are matched to the shower \cite{NSAllOrder}. Since this matching is not yet implemented in \textsc{Deductor}, we concentrate in this section on the case that $\sket{\rho_\scH}$ is calculated at lowest order. The operator $\cF(\muh^2)$ supplies the parton luminosity factor needed to form a cross section. It uses the PDFs that match the organization of the parton shower according to the ordering represented by $\mu_\scC$, calculated using evolution in the parameter $\lambda$ as specified in Sec.~\ref{sec:PDFs}. If we use $k_\LT$ ordering, the shower PDFs are very close the $\MSbar$ PDFs. If we use $\Lambda$ or $\vartheta$ ordering, the shower PDFs can differ substantially from the $\MSbar$ PDFs. This change is part of the summation of threshold logarithms \cite{NSThresholdII} that is part of the shower algorithm represented in Eq.~(\ref{eq:sigmaJshowerthreescale}). (We present a simple example in the following section and an analysis of the structure of the summation of threshold logarithms according to the shower algorithm in Appendix \ref{sec:thresholdsum}.)

The next operator in Eq.~(\ref{eq:sigmaJshowerthreescale}) is $\cU_\cV(3, 0)$, written as three separate factors. This operator is also part of the summation of threshold logarithms. It uses shower oriented PDFs calculated using evolution in the parameter $\lambda$ as specified in Sec.~\ref{sec:PDFs}. In general, $\cU_\cV(3, 0)$ leaves the number of partons and their momenta and flavors unchanged. The operator $\cU_\cV(t, t')$ is the ordered exponential of $\cS_\cV(t)$, as given in Eq.~(\ref{eq:cUcV}). The generator $\cS_\cV(t)$ at lowest order is given by the first order version of Eq.~(\ref{eq:cScV})
\begin{equation}
\label{eq:cScV1}
\cS_\cV^{[1]}(t)
= 
-  
\frac{d}{d t}
\cV^{[1]}(\mur(t),\vec\mu_\scS(t))
.
\end{equation}
This first order operator is described in Appendix.~\ref{sec:cScV}. The operator $\cV^{[1]}$ is associated with real emissions, derived using Eq.~(\ref{eq:cVdef}) from $\cD^{[1,0]}$ and the real part of virtual exchanges, $\mathrm{Re}\,\cD^{[0,1]}$. There is no contribution from the imaginary part of virtual exchanges, $\mathrm{Im}\,\cD^{[0,1]}$, because $\sbra{1}\mathrm{Im}\,\cD^{[0,1]} = 0$. Additionally, only initial state emissions contribute:
\begin{equation}
\cS_\cV^{[1]}(t)
= \cS_{\cV,\La}^{[1]}(t) + \cS_{\cV,\Lb}^{[1]}(t)
\;.
\end{equation}
Emissions from final state partons do not contribute to $\cS_\cV^{[1]}(t)$ because virtual exchanges cancel real emissions \cite{NSThresholdII, angleFS}. To calculate $\cS_{\cV,\La}^{[1]}(t)$, we differentiate $\cV^{[1]}_\mathrm{a}$ according to Eq.~(\ref{eq:cScV1}). Here $\cV^{[1]}_\mathrm{a}$ is obtained by combining Eq.~(\ref{eq:cV1integral}) with Eq.~(\ref{eq:bfVresult2}) and (\ref{eq:LCplusandsoftP}):
\begin{equation}
\begin{split}
\label{eq:cV1integralnet}
\cV^{[1]}_\mathrm{a}(&\mur, \mu_\scE, \mu_\scC)
\sket{\{p,f,c,c'\}_{m}}
\\={}&
\sket{\{p,f\}_{m}}
\frac{\as(\mur^2)}{2\pi}\,
\sum_{a'}
\int_0^1\!\frac{dz}{z}\,
\frac{
f_{a'/A}(\eta_{\La}/z,\mur^{2})}
{f_{a/A}(\eta_{\La},\mur^{2})}
\\&\times
\big[{\bm P}^\mathrm{a,LC+}_{a a'}(z; \{p\}_m; \mu_\scC)
\\&\quad +
{\bm P}^\mathrm{a,soft}_{a a'}(z; \{p\}_m; \mu_\scE) \big]
\sket{\{c,c'\}_m}
\;.
\end{split}
\end{equation}

Consider the three operators $\cU_\cV(1,0)$, $\cU_\cV(2,1)$, and $\cU_\cV(3,2)$ corresponding to the three segments of this path. In each case, $\cU_\cV$ is determined by $d\,\cV^{[1]}(t)/dt$ on that path segment, according to Eq.~(\ref{eq:cScV1}).

On the third segment, $2 < t < 3$, $\mu_\scE$ and $\mu_\scC$ are fixed, so there is no contribution from $\cV^{[1]}(t)$. That is, $\cU_\cV(3,2) = 1$.

On the second segment, $1 < t < 2$, $\mu_\scE$ is fixed at $\mu_\scE = 0$ while $\mu_\scC^2$ varies from $\mu_\scH^2$ to 0. Thus ${\bm P}^\mathrm{a,soft}_{a a'}$ does not contribute. Only ${\bm P}^\mathrm{a,LC+}_{a a'}$ contributes. That is, $\cU_\cV(2, 1)$, is $\cU_\cV(2, 1)$ as it would be calculated using the LC+ approximation. The LC+ approximation has the property that the color basis vectors used in the shower algorithm are eigenvectors of $\cU_\cV(2, 1)$. Thus $\cU_\cV(2, 1)$ applied to a parton basis state $\sket{\{p,f,c,c'\}_m}$ gives just a numerical factor. 

On the first segment, $0 < t < 1$, $\mu_\scC$ is fixed at $\mu_\scC^2 = \mu_\scH^2$ while $\mu_\scE^2$ varies from $\mu_\scE^2(0)$ to $0$. As we saw in Eq.~(\ref{eq:atotal}), the unresolved region that applies in ${\bm P}^\mathrm{a,LC+}_{a a'}$ is independent of $t$ on the first path segment, so that ${\bm P}^\mathrm{a,LC+}_{a a'}$ does not contribute. Thus we are left with ${\bm P}^\mathrm{a,soft}_{a a'}$, which provides a correction in $\cU_\cV(1, 0)$ to the LC+ approximation used in $\cU_\cV(2, 1)$. The operator $\cV$ leaves the number of partons and their momenta and flavors unchanged, but it can change the parton color state. The operator ${\bm P}^\mathrm{a,soft}_{a a'}$, in fact, contains nontrivial color operators. Thus $\cU_\cV(1, 0)$ is, in general, a nontrivial operator on the color space for the partons involved in the hard process, as described in $\isket{\rho_\scH}$. Fortunately from the point of view of computation, this color space is finite dimensional. Thus it should be possible to calculate $\cU_\cV(1, 0)\isket{\rho_\scH}$ numerically. Even more fortunately, in the color space for the $q \bar q$ final state in the lowest order Drell-Yan process, the LC+ approximation is exact, so that ${\bm P}^\mathrm{a,soft}_{a a'}$ acting in this space vanishes. That is, for the lowest order Drell-Yan process, $\cU_\cV(1, 0)\isket{\rho_\scH} = \isket{\rho_\scH}$.

Let us now consider the three operators $\cU(1,0)$, $\cU(2,1)$, and $\cU(3,2)$ corresponding to the probability conserving shower evolution on the three segments of the path. 

On the first segment, $0 < t < 1$, because of Eq.~(\ref{eq:atotal}), ${\bm D}^{\mathrm{a, LC+}}_{a a'}$ does not contribute to $\cU(1,0)$. Only  ${\bm D}^{\mathrm{a, soft}}_{a a'}$ contributes. This operator contains non-trivial color operator and it creates more partons each time it acts. That is, the dimensionality of the color space grows each time $\cS^{[1,0]}$ acts. This means that, at least as far as we know, one cannot, in general, calculate the action of $\cU(1,0)$ exactly. However, we expect that one can expand $\cU(1,0)$ in powers of ${\bm D}^{\mathrm{a, soft}}_{a a'}$ up to order $[{\bm D}^{\mathrm{a, soft}}_{a a'}]^n$ and calculate these contributions numerically for any $n$ that we choose, limited by the available computer power available. It remains a future research project to implement this scheme.\footnote{In fact, we have done this in Refs.~\cite{NSNewColor, GapColor, angleFS} in the more difficult case of a shower formulation in which appearances of ${\bm D}^{\mathrm{a, soft}}_{a a'}$ are interleaved with ordinary LC+ splittings. We found that the contributions from ${\bm D}^{\mathrm{a, soft}}_{a a'}$ are practical to compute and can be substantial in the case of the rapidity gap fraction in hadron-hadron collisions but are small in other cases that we have examined.} 

There is one case in which $\cU(1,0)$ can be computed exactly. In the Drell-Yan process at lowest order, one starts with $q\bar q$ color states. These color states form a one-dimensional space and on this space the LC+ approximation is exact. Thus ${\bm D}^{\mathrm{a, soft}}_{a a'}$ acting on the single state in this space vanishes. That is, $\cU(1,0) = 1$ when acting on this space. We can simply ignore $\cU(1,0)$.

On the second segment, $1 < t < 2$, $\cU(2,1)$ creates a full shower using first order splitting functions derived from ${\bm D}^{\mathrm{a, LC+}}_{a a'}$. The soft splitting functions from ${\bm D}^{\mathrm{a, soft}}_{a a'}$ do not contribute because $\mu_\scE^2$ is fixed. This gives us a first order shower based on the LC+ approximation from $\muh^2$ to zero (with infrared cutoffs based on $m_\perp^2(a,\hat a)$).

On the third segment, $2 < t < 3$, neither $\mu_\scE^2$ nor $\mu_\scC^2$ varies, so there is no contribution from the splittings in $\cD^{[1,0]}$. The only contribution is from the color phase operator $S_{\mi\pi}(t)$, which leaves the number of partons and their momenta and flavors unchanged. The result depends on what we measure with the operator $\cO_J$ in Eq.~(\ref{eq:sigmaJshowerthreescale}). The usual case is that $\cO_J$ measures only parton momenta and flavors, but not colors. Then $\cO_J$ commutes with $S_{\mi\pi}(t)$, so that
\begin{equation}
\begin{split}
\sigma ={}& \sbra{1}
\cU(3, 2)\, \cO_J\, \cU(2, 1)\,\cU(1, 0)
\\&\times
\cU_\cV(2, 1)\, \cU_\cV(1, 0)
\cF(\muh^2)
\sket{\rho_\scH}
\;.
\end{split}
\end{equation}
However, $\sbra{1}S_{\mi\pi}(t) = 0$ so that $\sbra{1}\cU(3, 2) = \sbra{1}$. This leaves
\begin{equation}
\begin{split}
\label{eq:sigmaJshowerthreescaleend}
\sigma ={}& \sbra{1}
\cO_J\, \cU(2, 1)\,\cU(1, 0)
\\&\times
\cU_\cV(2, 1)\, \cU_\cV(1, 0)
\cF(\muh^2)
\sket{\rho_\scH}
\;.
\end{split}
\end{equation}
That is, the effect of $\cU(3, 2)$ cancels and there is no need to calculate $\cU(3, 2)$.

There is a possible exception to this. One may want to add a nonperturbative model of hadronization to the perturbative parton shower. For instance, one can start with a perturbative \textsc{Deductor} event, add a nonperturbative underlying event, and then pass the event to \textsc{Pythia} for hadronization according to the Lund string model, as in Ref.~\cite{NSThresholdII}. In this case, the result depends on the color configuration in the \textsc{Deductor} event, so that the effect of $\cU(3,2)$ does not automatically cancel.

We can summarize these results for the three segment path. First, for the third segment we can use $\cU_\cV(3,2) = 1$ and, as long as the measurement operator $\cO_J$ is blind to parton color, $\cU(3,2) \to 1$. For the second segment, we need both $\cU_\cV(2,1)$ and $\cU(2,1)$, but these operators are straightforward to calculate with the current \textsc{Deductor} code because they use the LC+ approximation. For the first segment, if we consider the lowest order Drell-Yan process then $\cU_\cV(1,0) \to 1$ and $\cU(1,0) \to 1$, leaving
\begin{equation}
\begin{split}
\label{eq:sigmaJDY}
\sigma ={}& \sbra{1}
\cO_J\, \cU(2, 1)\, \cU_\cV(2, 1)\, 
\cF(\muh^2)
\sket{\rho_\scH}
\;.
\end{split}
\end{equation}

Alternatively, if we consider parton-parton scattering as the hard process, then $\cU_\cV(1,0)$ should be fairly simple to calculate by exponentiating a finite dimensional matrix. We expect that one can calculate shower evolution operator $\cU(1,0)$ perturbatively in powers of ${\bm D}^{\mathrm{a, soft}}_{a a'}$. We leave implementation of the calculation of $\cU_\cV(1,0)$ and $\cU(1,0)$ to future research.

\section{The Drell-Yan cross section}
\label{sec:DrellYan}

We can test how the methods of this paper work by calculating the cross section for $p + p \to \mu  + \bar \mu + X$. We use the three segment evolution path presented in this paper. We start with the expression (\ref{eq:sigmaJshowerthreescaleend}) for the cross section. Here the starting statistical state $\isket{\rho_\scH}$ is obtained from the Born matrix elements. It would be preferable to correct this to NLO, but the required matching is not available in the present version of \textsc{Deductor} and, with matching, the treatment of $\cU_\cV(1,0)$ and $\cU(1,0)$ would be nontrivial.

We first look at the cross section $d\sigma/dM_{\mu\bar\mu}$, in which only the mass $M_{\mu\bar\mu}$ of the muon pair is measured. In this case the measurement operator $\cO_J$ commutes with the shower operator $\cU(2, 1)$ and we can use $\sbra{1}\cU(2, 1) = \sbra{1}$ to eliminate the shower operator, leaving
\begin{equation}
\begin{split}
\label{eq:sigmaJDYsimple}
\sigma ={}& \sbra{1}
\cO_J\, \cU_\cV(2, 1)\, 
\cF(\muh^2)
\sket{\rho_\scH}
\;.
\end{split}
\end{equation}
There are three versions of each of the operators $\cU_\cV(2, 1)$ and $\cF(\muh^2)$, one version for each of $k_\LT$ ordering, $\Lambda$ ordering, and $\vartheta$ ordering.

\begin{figure}
\ifusefigs 

\begin{tikzpicture}
\begin{axis}[title = {Drell-Yan cross sections, ratios to Born},
   xlabel={$M_{\mu \bar\mu}\,\mathrm{[TeV]}$}, 
   ylabel={$[d\sigma/dM_{\mu\bar\mu}]/[d\sigma(\mathrm{Born})/dM_{\mu\bar\mu}]$},
   xmin=0, xmax=7000,
   ymin=0.8, ymax=1.9,
   legend cell align=left,
   every axis legend/.append style = {
    at={(0.05,0.96)},
    anchor=north west},
    x coord trafo/.code={
    \pgflibraryfpuifactive{
    \pgfmathparse{(#1)/(1000)}
  }{
     \pgfkeys{/pgf/fpu=true}
     \pgfmathparse{(#1)/(1000)}
     \pgfkeys{/pgf/fpu=false}
   }
 },
 xminorgrids=false,
 yminorgrids=false,
 minor x tick num=4,
 minor y tick num=4
]

\pgfplotstableread{  
4.000000e+02      2.191017e-06   2.220462e-06       2.191260e-06  2.069995e-06       2.190946e-06   1.794160e-06
5.000000e+02      7.797486e-07   7.870209e-07	    7.790379e-07  7.446310e-07	     7.786794e-07   6.573485e-07
6.000000e+02      3.311029e-07   3.331625e-07	    3.310510e-07  3.194073e-07	     3.310196e-07   2.860640e-07
7.000000e+02      1.585479e-07   1.591469e-07	    1.584944e-07  1.541262e-07	     1.585697e-07   1.397061e-07
8.000000e+02      8.270352e-08   8.285214e-08	    8.276382e-08  8.103627e-08	     8.277817e-08   7.414596e-08
9.000000e+02      4.608996e-08   4.609679e-08	    4.604916e-08  4.536641e-08	     4.604679e-08   4.184565e-08
1.000000e+03      2.695872e-08   2.692526e-08	    2.700060e-08  2.674979e-08	     2.695380e-08   2.481582e-08
1.100000e+03      1.643589e-08   1.639595e-08	    1.644130e-08  1.637376e-08	     1.644148e-08   1.531785e-08
1.200000e+03      1.035536e-08   1.031957e-08	    1.035697e-08  1.036463e-08	     1.035670e-08   9.756196e-09
1.300000e+03      6.700737e-09   6.671572e-09	    6.696019e-09  6.732063e-09	     6.698026e-09   6.374801e-09
1.400000e+03      4.443443e-09   4.420593e-09	    4.440641e-09  4.484108e-09	     4.441710e-09   4.268859e-09
1.500000e+03      3.001922e-09   2.984386e-09	    3.004885e-09  3.047039e-09	     3.005190e-09   2.915457e-09
1.600000e+03      2.059943e-09   2.046619e-09	    2.063247e-09  2.100712e-09	     2.063072e-09   2.019509e-09
1.700000e+03      1.440295e-09   1.430176e-09	    1.438770e-09  1.470638e-09	     1.440689e-09   1.422302e-09
1.800000e+03      1.017927e-09   1.010255e-09	    1.016137e-09  1.042613e-09	     1.018625e-09   1.013989e-09
1.900000e+03      7.269504e-10   7.211366e-10	    7.275751e-10  7.493184e-10	     7.263281e-10   7.288986e-10
2.000000e+03      5.250435e-10   5.206241e-10	    5.245839e-10  5.422280e-10	     5.251857e-10   5.311719e-10
2.100000e+03      3.811724e-10   3.778175e-10	    3.819387e-10  3.961815e-10	     3.817608e-10   3.890687e-10
2.200000e+03      2.811215e-10   2.785485e-10	    2.806817e-10  2.921722e-10	     2.805270e-10   2.880717e-10
2.300000e+03      2.078817e-10   2.059114e-10	    2.069244e-10  2.161415e-10	     2.074244e-10   2.145894e-10
2.400000e+03      1.543774e-10   1.528679e-10	    1.543485e-10  1.617752e-10	     1.543071e-10   1.607970e-10
2.500000e+03      1.155175e-10   1.143559e-10	    1.156200e-10  1.215897e-10	     1.156127e-10   1.213432e-10
2.600000e+03      8.722967e-11   8.633044e-11	    8.712793e-11  9.193467e-11	     8.698804e-11   9.195295e-11
2.700000e+03      6.588875e-11   6.519342e-11	    6.571750e-11  6.957339e-11	     6.581724e-11   7.006141e-11
2.800000e+03      5.009129e-11   4.955162e-11	    4.992488e-11  5.302611e-11	     5.006314e-11   5.367012e-11
2.900000e+03      3.818156e-11   3.776202e-11	    3.820100e-11  4.070578e-11	     3.814458e-11   4.117286e-11
3.000000e+03      2.921843e-11   2.889166e-11	    2.923854e-11  3.125683e-11	     2.920072e-11   3.174047e-11
3.100000e+03      2.242258e-11   2.216769e-11	    2.244082e-11  2.406784e-11	     2.242748e-11   2.454438e-11
3.200000e+03      1.724585e-11   1.704679e-11	    1.732924e-11  1.864618e-11	     1.726772e-11   1.902725e-11
3.300000e+03      1.333980e-11   1.318364e-11	    1.334076e-11  1.440031e-11	     1.336687e-11   1.483055e-11
3.400000e+03      1.035680e-11   1.023391e-11	    1.032283e-11  1.117832e-11	     1.032228e-11   1.153172e-11
3.500000e+03      8.022947e-12   7.926562e-12	    8.022006e-12  8.715333e-12	     8.010294e-12   9.010402e-12
3.600000e+03      6.224465e-12   6.148838e-12	    6.229376e-12  6.789293e-12	     6.227140e-12   7.052267e-12
3.700000e+03      4.848838e-12   4.789281e-12	    4.853730e-12  5.307016e-12	     4.847411e-12   5.527825e-12
3.800000e+03      3.779115e-12   3.732227e-12	    3.796787e-12  4.164859e-12	     3.782909e-12   4.343555e-12
3.900000e+03      2.957895e-12   2.920844e-12	    2.953977e-12  3.250843e-12	     2.948543e-12   3.409134e-12
4.000000e+03      2.307406e-12   2.278237e-12	    2.308561e-12  2.548968e-12	     2.314794e-12   2.694905e-12
4.100000e+03      1.806107e-12   1.783082e-12	    1.810170e-12  2.005119e-12	     1.810707e-12   2.122721e-12
4.200000e+03      1.418764e-12   1.400531e-12	    1.414359e-12  1.571872e-12	     1.415128e-12   1.670901e-12
4.300000e+03      1.109710e-12   1.095341e-12	    1.107407e-12  1.234879e-12	     1.109550e-12   1.319174e-12
4.400000e+03      8.703761e-13   8.590250e-13	    8.705505e-13  9.739692e-13	     8.699656e-13   1.041741e-12
4.500000e+03      6.876758e-13   6.786460e-13	    6.840206e-13  7.678321e-13	     6.809290e-13   8.212174e-13
4.600000e+03      5.360135e-13   5.289308e-13	    5.365306e-13  6.043161e-13	     5.367446e-13   6.519219e-13
4.700000e+03      4.221931e-13   4.165791e-13	    4.204561e-13  4.751892e-13	     4.222236e-13   5.165633e-13
4.800000e+03      3.319392e-13   3.275000e-13	    3.306815e-13  3.749984e-13	     3.300568e-13   4.068130e-13
4.900000e+03      2.604578e-13   2.569556e-13	    2.595772e-13  2.953938e-13	     2.595180e-13   3.221973e-13
5.000000e+03      2.041638e-13   2.014040e-13	    2.034364e-13  2.323076e-13	     2.046202e-13   2.559119e-13
5.100000e+03      1.608321e-13   1.586471e-13	    1.602628e-13  1.836353e-13	     1.600199e-13   2.016486e-13
5.200000e+03      1.260691e-13   1.243483e-13	    1.255238e-13  1.443403e-13	     1.255240e-13   1.593950e-13
5.300000e+03      9.865987e-14   9.730732e-14	    9.841450e-14  1.135735e-13	     9.868088e-14   1.262527e-13
5.400000e+03      7.693104e-14   7.587153e-14	    7.721420e-14  8.942845e-14	     7.720038e-14   9.954062e-14
5.500000e+03      6.042456e-14   5.958922e-14	    6.051051e-14  7.033438e-14	     6.039710e-14   7.848667e-14
5.600000e+03      4.736882e-14   4.671144e-14	    4.731500e-14  5.520337e-14	     4.745162e-14   6.215154e-14
5.700000e+03      3.692358e-14   3.640920e-14	    3.698978e-14  4.331552e-14	     3.695764e-14   4.879164e-14
5.800000e+03      2.881120e-14   2.840844e-14	    2.889021e-14  3.395850e-14	     2.904378e-14   3.865124e-14
5.900000e+03      2.267931e-14   2.236118e-14	    2.263517e-14  2.670637e-14	     2.258305e-14   3.029903e-14
6.000000e+03      1.763613e-14   1.738799e-14	    1.760125e-14  2.084546e-14	     1.757054e-14   2.376851e-14
6.100000e+03      1.373627e-14   1.354239e-14	    1.364374e-14  1.622141e-14	     1.366393e-14   1.864118e-14
6.200000e+03      1.061664e-14   1.046636e-14	    1.066984e-14  1.273552e-14	     1.065372e-14   1.465576e-14
6.300000e+03      8.245576e-15   8.128550e-15	    8.258069e-15  9.896271e-15	     8.255789e-15   1.145621e-14
6.400000e+03      6.369307e-15   6.278661e-15	    6.398206e-15  7.698285e-15	     6.431983e-15   9.001666e-15
6.500000e+03      4.932181e-15   4.861813e-15	    4.945718e-15  5.975106e-15	     4.972082e-15   7.019762e-15
6.600000e+03      3.823079e-15   3.768402e-15	    3.817989e-15  4.631779e-15	     3.833934e-15   5.462688e-15
6.700000e+03      2.944312e-15   2.902107e-15	    2.928576e-15  3.567689e-15	     2.935466e-15   4.221075e-15
6.800000e+03      2.255577e-15   2.223179e-15	    2.275810e-15  2.783969e-15	     2.263916e-15   3.284202e-15
}\xsecPDF

\pgfplotstableread{  
4.000000e+02     2.189466e-06    2.000980e-06       2.190141e-06   2.177705e-06       2.191425e-06   2.499178e-06
5.000000e+02     7.783851e-07    7.332003e-07	    7.784605e-07   7.840040e-07	      7.781054e-07   8.827349e-07
6.000000e+02     3.312769e-07    3.190046e-07	    3.310072e-07   3.367958e-07	      3.309288e-07   3.739738e-07
7.000000e+02     1.583526e-07    1.554138e-07	    1.584980e-07   1.626536e-07	      1.582085e-07   1.782412e-07
8.000000e+02     8.275517e-08    8.262493e-08	    8.274166e-08   8.554722e-08	      8.278089e-08   9.303365e-08
9.000000e+02     4.607068e-08    4.667941e-08	    4.596607e-08   4.784071e-08	      4.607997e-08   5.168263e-08
1.000000e+03     2.698659e-08    2.771498e-08	    2.699850e-08   2.826881e-08	      2.700098e-08   3.023209e-08
1.100000e+03     1.640233e-08    1.705331e-08	    1.644610e-08   1.731637e-08	      1.641203e-08   1.834856e-08
1.200000e+03     1.036768e-08    1.089857e-08	    1.035416e-08   1.095861e-08	      1.034288e-08   1.154817e-08
1.300000e+03     6.704490e-09    7.123076e-09	    6.690137e-09   7.114969e-09	      6.699277e-09   7.471249e-09
1.400000e+03     4.444718e-09    4.769577e-09	    4.440788e-09   4.744647e-09	      4.427736e-09   4.932842e-09
1.500000e+03     3.005448e-09    3.255423e-09	    3.002671e-09   3.222360e-09	      3.000495e-09   3.339698e-09
1.600000e+03     2.068874e-09    2.261216e-09	    2.063010e-09   2.223390e-09	      2.065738e-09   2.297309e-09
1.700000e+03     1.439785e-09    1.587397e-09	    1.439888e-09   1.558301e-09	      1.434824e-09   1.594440e-09
1.800000e+03     1.019516e-09    1.133701e-09	    1.016799e-09   1.104719e-09	      1.017478e-09   1.129863e-09
1.900000e+03     7.268853e-10    8.148925e-10	    7.269018e-10   7.929049e-10	      7.265143e-10   8.062282e-10
2.000000e+03     5.245925e-10    5.923457e-10	    5.242137e-10   5.739557e-10	      5.244519e-10   5.816372e-10
2.100000e+03     3.808448e-10    4.337651e-10	    3.815356e-10   4.193211e-10	      3.828446e-10   4.243491e-10
2.200000e+03     2.800167e-10    3.213293e-10	    2.802430e-10   3.091275e-10	      2.806330e-10   3.108908e-10
2.300000e+03     2.085333e-10    2.410789e-10	    2.075809e-10   2.298282e-10	      2.075633e-10   2.298253e-10
2.400000e+03     1.546606e-10    1.801282e-10	    1.547108e-10   1.719023e-10	      1.542563e-10   1.707192e-10
2.500000e+03     1.157459e-10    1.357908e-10	    1.153677e-10   1.286432e-10	      1.159008e-10   1.282129e-10
2.600000e+03     8.689421e-11    1.026820e-10	    8.737404e-11   9.776340e-11	      8.726320e-11   9.649064e-11
2.700000e+03     6.568456e-11    7.815983e-11	    6.587533e-11   7.396838e-11	      6.572942e-11   7.265045e-11
2.800000e+03     4.991923e-11    5.982315e-11	    5.021001e-11   5.657500e-11	      4.999464e-11   5.523685e-11
2.900000e+03     3.812268e-11    4.599961e-11	    3.828299e-11   4.328712e-11	      3.822608e-11   4.221835e-11
3.000000e+03     2.915935e-11    3.543372e-11	    2.921222e-11   3.314719e-11	      2.922950e-11   3.227058e-11
3.100000e+03     2.246138e-11    2.748565e-11	    2.254765e-11   2.567404e-11	      2.251626e-11   2.485009e-11
3.200000e+03     1.730440e-11    2.132311e-11	    1.729466e-11   1.976038e-11	      1.727066e-11   1.905447e-11
3.300000e+03     1.337708e-11    1.659906e-11	    1.338656e-11   1.534855e-11	      1.336968e-11   1.474575e-11
3.400000e+03     1.035903e-11    1.294079e-11	    1.031369e-11   1.186610e-11	      1.035974e-11   1.142241e-11
3.500000e+03     7.997709e-12    1.006076e-11	    8.008220e-12   9.245041e-12	      8.022854e-12   8.843197e-12
3.600000e+03     6.263454e-12    7.933849e-12	    6.227218e-12   7.215065e-12	      6.219500e-12   6.853427e-12
3.700000e+03     4.864194e-12    6.211677e-12	    4.838378e-12   5.625961e-12	      4.856322e-12   5.349790e-12
3.800000e+03     3.785890e-12    4.862411e-12	    3.779667e-12   4.410195e-12	      3.789003e-12   4.172879e-12
3.900000e+03     2.933279e-12    3.794269e-12	    2.954735e-12   3.459992e-12	      2.962628e-12   3.261901e-12
4.000000e+03     2.319288e-12    3.017416e-12	    2.318195e-12   2.724436e-12	      2.316194e-12   2.549492e-12
4.100000e+03     1.807064e-12    2.370514e-12	    1.804908e-12   2.128953e-12	      1.799586e-12   1.980363e-12
4.200000e+03     1.424106e-12    1.881449e-12	    1.419613e-12   1.680432e-12	      1.418741e-12   1.560869e-12
4.300000e+03     1.105262e-12    1.470834e-12	    1.110909e-12   1.319923e-12	      1.115107e-12   1.226530e-12
4.400000e+03     8.731638e-13    1.170256e-12	    8.715535e-13   1.039456e-12	      8.773888e-13   9.648267e-13
4.500000e+03     6.881708e-13    9.289594e-13	    6.855950e-13   8.205929e-13	      6.784666e-13   7.459104e-13
4.600000e+03     5.377173e-13    7.311267e-13	    5.353125e-13   6.431933e-13	      5.395311e-13   5.930336e-13
4.700000e+03     4.228676e-13    5.792656e-13	    4.217539e-13   5.087282e-13	      4.207591e-13   4.623836e-13
4.800000e+03     3.301799e-13    4.556159e-13	    3.315232e-13   4.014072e-13	      3.311535e-13   3.638363e-13
4.900000e+03     2.594884e-13    3.607979e-13	    2.589334e-13   3.147809e-13	      2.597538e-13   2.853309e-13
5.000000e+03     2.040371e-13    2.858572e-13	    2.043668e-13   2.493923e-13	      2.019840e-13   2.218269e-13
5.100000e+03     1.591917e-13    2.248186e-13	    1.592265e-13   1.951100e-13	      1.593208e-13   1.749385e-13
5.200000e+03     1.250131e-13    1.778971e-13	    1.254140e-13   1.542753e-13	      1.254612e-13   1.377329e-13
5.300000e+03     9.855061e-14    1.413177e-13	    9.845580e-14   1.216344e-13	      9.888234e-14   1.085330e-13
5.400000e+03     7.758885e-14    1.121363e-13	    7.695079e-14   9.545026e-14	      7.732013e-14   8.485114e-14
5.500000e+03     6.094215e-14    8.878642e-14	    6.025718e-14   7.506287e-14	      6.033411e-14   6.619832e-14
5.600000e+03     4.757475e-14    6.986174e-14	    4.731466e-14   5.919321e-14	      4.742335e-14   5.202364e-14
5.700000e+03     3.717608e-14    5.504329e-14	    3.699041e-14   4.647523e-14	      3.749950e-14   4.112977e-14
5.800000e+03     2.888195e-14    4.312024e-14	    2.874780e-14   3.627560e-14	      2.872273e-14   3.149801e-14
5.900000e+03     2.278981e-14    3.429862e-14	    2.246767e-14   2.847302e-14	      2.248696e-14   2.465545e-14
6.000000e+03     1.749182e-14    2.655736e-14	    1.758995e-14   2.239262e-14	      1.750111e-14   1.918560e-14
6.100000e+03     1.381270e-14    2.114570e-14	    1.373232e-14   1.755921e-14	      1.380713e-14   1.513359e-14
6.200000e+03     1.062610e-14    1.641142e-14	    1.061228e-14   1.363397e-14	      1.062495e-14   1.164390e-14
6.300000e+03     8.327320e-15    1.292976e-14	    8.213400e-15   1.060180e-14	      8.417644e-15   9.223398e-15
6.400000e+03     6.419365e-15    1.009075e-14	    6.368001e-15   8.258811e-15	      6.397636e-15   7.008957e-15
6.500000e+03     4.923877e-15    7.812551e-15	    4.944607e-15   6.444035e-15	      4.923765e-15   5.393430e-15
6.600000e+03     3.791837e-15    6.070447e-15	    3.804050e-15   4.982862e-15	      3.809856e-15   4.172644e-15
6.700000e+03     2.919122e-15    4.717725e-15	    2.952781e-15   3.886836e-15	      2.946576e-15   3.226683e-15
6.800000e+03     2.264606e-15    3.694132e-15	    2.268526e-15   3.001017e-15	      2.253590e-15   2.467469e-15
}\xsecUV

\addplot[blue,thick] table [x={0},y expr=\thisrow{2}/\thisrow{1}]{\xsecPDF};
\addlegendentry{$k_\LT$, $\cU_\cV = 1$}

\addplot[red,thick] table [x={0},y expr=\thisrow{4}/\thisrow{3}]{\xsecPDF};
\addlegendentry{$\Lambda$, $\cU_\cV = 1$}

\addplot[darkgreen,thick] table [x={0},y expr=\thisrow{6}/\thisrow{5}]{\xsecPDF};
\addlegendentry{$\vartheta$, $\cU_\cV = 1$}

\addplot[blue,thick,dashed] table [x={0},y expr=\thisrow{2}/\thisrow{1}]{\xsecUV};
\addlegendentry{$k_\LT$, PDF = $\MSbar$}

\addplot[red,thick,dashed] table [x={0},y expr=\thisrow{4}/\thisrow{3}]{\xsecUV};
\addlegendentry{$\Lambda$, PDF = $\MSbar$}

\addplot[darkgreen,thick,dashed] table [x={0},y expr=\thisrow{6}/\thisrow{5}]{\xsecUV};
\addlegendentry{$\vartheta$, PDF = $\MSbar$}

\addplot[black, domain=400:6800]{1};

\end{axis}
\end{tikzpicture}

\else 
\begin{center}
\includegraphics[width = 5.5 cm]{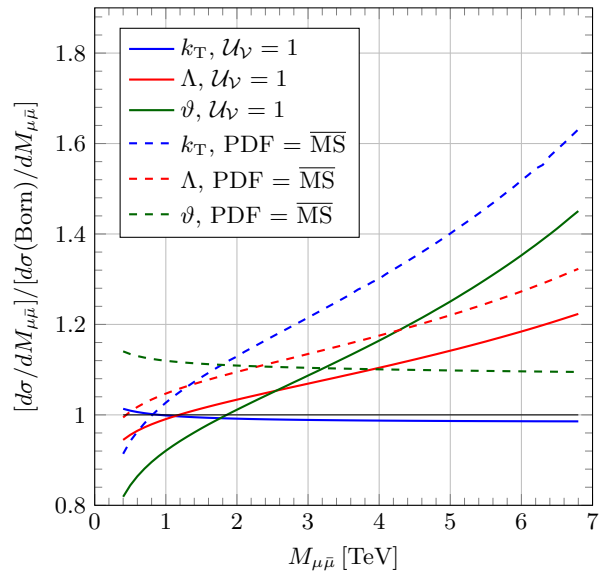}
\end{center}
\fi

\caption{
Ratio of the Drell-Yan cross section $d\sigma/dM_{\mu\bar\mu}$ to the Born cross section as a function of the dimuon mass $M_{\mu\bar\mu}$ for $\sqrt s = 13 \TeV$. In the full result, there is a PDF factor and a $\cU_\cV$ factor for each of $k_\LT$ ordering, $\Lambda$ ordering and $\vartheta$ ordering. Here we show partial results in which we include the proper PDF factors but set $\cU_\cV = 1$ (solid curves) and then in which we use $\MSbar$ PDFs instead of the proper shower oriented PDFs but use the proper factors of $\cU_\cV$ (dashed curves).
}
\label{fig:DYparts}
\end{figure}

In Fig.~\ref{fig:DYparts}, we show the separate effect of the operators $\cU_\cV(2, 1)$ and $\cF(\muh^2)$. First, we replace $\cU_\cV(2, 1)$ by 1 and use the $k_\LT$-ordered, $\Lambda$-ordered, and $\vartheta$-ordered versions of the shower oriented PDFs in the operator $\cF(\muh^2)$. These PDFs are determined from the standard CT14 \cite{CT14} $\MSbar$ PDFs using the $\lambda$-evolution equation (\ref{eq:dtildefdlambda}). We find the cross sections shown as solid curves in Fig.~\ref{fig:DYparts}. Evidently, the choice of ordering variable makes a substantial difference in the results. Second, we fix the PDFs in $\cF(\muh^2)$ to be the $\MSbar$ PDFs and use the full operators $\cU_\cV(2, 1)$ corresponding to the three choices of ordering variable. We find the cross sections shown as dashed curves in Fig.~\ref{fig:DYparts}. Again, the choice of ordering variable makes a substantial difference in the results.


\begin{figure}
\ifusefigs 

\begin{tikzpicture}
\begin{axis}[title = {Drell-Yan cross sections, ratios to Born},
   xlabel={$M_{\mu \bar\mu}\,\mathrm{[TeV]}$}, 
   ylabel={$[d\sigma/dM_{\mu\bar\mu}]/[d\sigma(\mathrm{Born})/dM_{\mu\bar\mu}]$},
   xmin=0, xmax=7000,
   ymin=0.8, ymax=1.9,
   legend cell align=left,
   every axis legend/.append style = {
    at={(0.1,0.96)},
    anchor=north west},
    x coord trafo/.code={
    \pgflibraryfpuifactive{
    \pgfmathparse{(#1)/(1000)}
  }{
     \pgfkeys{/pgf/fpu=true}
     \pgfmathparse{(#1)/(1000)}
     \pgfkeys{/pgf/fpu=false}
   }
 },
 xminorgrids=false,
 yminorgrids=false,
 minor x tick num=4,
 minor y tick num=4
]

\pgfplotstableread{  
4.000000e+02      2.189684e-06    2.049081e-06       2.191229e-06   2.041736e-06       2.188739e-06   2.051380e-06
5.000000e+02      7.800841e-07    7.467776e-07 	     7.784180e-07   7.441785e-07       7.796974e-07   7.488113e-07
6.000000e+02      3.306830e-07    3.224543e-07 	     3.315930e-07   3.235085e-07       3.308391e-07   3.237946e-07
7.000000e+02      1.586108e-07    1.571160e-07 	     1.585166e-07   1.572878e-07       1.585229e-07   1.576266e-07
8.000000e+02      8.281662e-08    8.317731e-08 	     8.270349e-08   8.327982e-08       8.257607e-08   8.325149e-08
9.000000e+02      4.608082e-08    4.685515e-08 	     4.605663e-08   4.698213e-08       4.600536e-08   4.696041e-08
1.000000e+03      2.701449e-08    2.778338e-08 	     2.692303e-08   2.779519e-08       2.697129e-08   2.783320e-08
1.100000e+03      1.641889e-08    1.706268e-08 	     1.643611e-08   1.715421e-08       1.645173e-08   1.715480e-08
1.200000e+03      1.037443e-08    1.088862e-08 	     1.037047e-08   1.093430e-08       1.036029e-08   1.090669e-08
1.300000e+03      6.695903e-09    7.093912e-09 	     6.715459e-09   7.149087e-09       6.711834e-09   7.130790e-09
1.400000e+03      4.449239e-09    4.756075e-09 	     4.437644e-09   4.767081e-09       4.448731e-09   4.766736e-09
1.500000e+03      2.983655e-09    3.217705e-09 	     3.008638e-09   3.260248e-09       2.993789e-09   3.234793e-09
1.600000e+03      2.052961e-09    2.232151e-09 	     2.057098e-09   2.247845e-09       2.054150e-09   2.237638e-09
1.700000e+03      1.440279e-09    1.578334e-09 	     1.439003e-09   1.584905e-09       1.437300e-09   1.577691e-09
1.800000e+03      1.018510e-09    1.124662e-09 	     1.012297e-09   1.124022e-09       1.018551e-09   1.126490e-09
1.900000e+03      7.301910e-10    8.124250e-10 	     7.265232e-10   8.128825e-10       7.267005e-10   8.097071e-10
2.000000e+03      5.244125e-10    5.878963e-10 	     5.262752e-10   5.931642e-10       5.238415e-10   5.878821e-10
2.100000e+03      3.827496e-10    4.322592e-10 	     3.813298e-10   4.329587e-10       3.816485e-10   4.313000e-10
2.200000e+03      2.801627e-10    3.186246e-10 	     2.808757e-10   3.212651e-10       2.807119e-10   3.194956e-10
2.300000e+03      2.072993e-10    2.374262e-10 	     2.078056e-10   2.393303e-10       2.077393e-10   2.380330e-10
2.400000e+03      1.543673e-10    1.780224e-10 	     1.546489e-10   1.793824e-10       1.544466e-10   1.782035e-10
2.500000e+03      1.153352e-10    1.339646e-10 	     1.157849e-10   1.352372e-10       1.155962e-10   1.342537e-10
2.600000e+03      8.678429e-11    1.014905e-10 	     8.722839e-11   1.025886e-10       8.754453e-11   1.023562e-10
2.700000e+03      6.578431e-11    7.746243e-11 	     6.616053e-11   7.832497e-11       6.574602e-11   7.739461e-11
2.800000e+03      4.992910e-11    5.916462e-11 	     5.014097e-11   5.975750e-11       5.009766e-11   5.934802e-11
2.900000e+03      3.800430e-11    4.535154e-11 	     3.799175e-11   4.559486e-11       3.814904e-11   4.549222e-11
3.000000e+03      2.896024e-11    3.479307e-11 	     2.914454e-11   3.520551e-11       2.929559e-11   3.515895e-11
3.100000e+03      2.242746e-11    2.712116e-11 	     2.243226e-11   2.728151e-11       2.247875e-11   2.716090e-11
3.200000e+03      1.738437e-11    2.116643e-11 	     1.732625e-11   2.120547e-11       1.731028e-11   2.105102e-11
3.300000e+03      1.335438e-11    1.637160e-11 	     1.327992e-11   1.636257e-11       1.337604e-11   1.637340e-11
3.400000e+03      1.036952e-11    1.279581e-11 	     1.030390e-11   1.277974e-11       1.032500e-11   1.272197e-11
3.500000e+03      7.940583e-12    9.867655e-12 	     8.060941e-12   1.006403e-11       8.024300e-12   9.951317e-12
3.600000e+03      6.289123e-12    7.864945e-12 	     6.184295e-12   7.773653e-12       6.207954e-12   7.749120e-12
3.700000e+03      4.862054e-12    6.120530e-12 	     4.834284e-12   6.119074e-12       4.853433e-12   6.098252e-12
3.800000e+03      3.812056e-12    4.832151e-12 	     3.783573e-12   4.821284e-12       3.784202e-12   4.785582e-12
3.900000e+03      2.971137e-12    3.792000e-12 	     2.947637e-12   3.781966e-12       2.961231e-12   3.769899e-12
4.000000e+03      2.309893e-12    2.968583e-12 	     2.302092e-12   2.972920e-12       2.313805e-12   2.966259e-12
4.100000e+03      1.809252e-12    2.341222e-12 	     1.810562e-12   2.353960e-12       1.803939e-12   2.328051e-12
4.200000e+03      1.412103e-12    1.840071e-12 	     1.413831e-12   1.851152e-12       1.416554e-12   1.840350e-12
4.300000e+03      1.117451e-12    1.466515e-12 	     1.113689e-12   1.467979e-12       1.103646e-12   1.443549e-12
4.400000e+03      8.732750e-13    1.154017e-12 	     8.748057e-13   1.161393e-12       8.734407e-13   1.150145e-12
4.500000e+03      6.866504e-13    9.140458e-13 	     6.843855e-13   9.149726e-13       6.870025e-13   9.108572e-13
4.600000e+03      5.389157e-13    7.225957e-13 	     5.400789e-13   7.271921e-13       5.354400e-13   7.150986e-13
4.700000e+03      4.230835e-13    5.711807e-13 	     4.209185e-13   5.709344e-13       4.226462e-13   5.683549e-13
4.800000e+03      3.312451e-13    4.505631e-13 	     3.289435e-13   4.494906e-13       3.301339e-13   4.471385e-13
4.900000e+03      2.600330e-13    3.562014e-13 	     2.604154e-13   3.583715e-13       2.598634e-13   3.544095e-13
5.000000e+03      2.039598e-13    2.815887e-13 	     2.037053e-13   2.824623e-13       2.045421e-13   2.810531e-13
5.100000e+03      1.601401e-13    2.227730e-13 	     1.598184e-13   2.231960e-13       1.595551e-13   2.208114e-13
5.200000e+03      1.261153e-13    1.768028e-13 	     1.254011e-13   1.764917e-13       1.252277e-13   1.746155e-13
5.300000e+03      9.913585e-14    1.400548e-13 	     9.911853e-14   1.405759e-13       9.865346e-14   1.384996e-13
5.400000e+03      7.743129e-14    1.101859e-13 	     7.665507e-14   1.095450e-13       7.671158e-14   1.085948e-13
5.500000e+03      6.059744e-14    8.694622e-14 	     6.036892e-14   8.694014e-14       6.046960e-14   8.620105e-14
5.600000e+03      4.742779e-14    6.858580e-14 	     4.753077e-14   6.898802e-14       4.719264e-14   6.780940e-14
5.700000e+03      3.707970e-14    5.405580e-14 	     3.699917e-14   5.411693e-14       3.712509e-14   5.376586e-14
5.800000e+03      2.892034e-14    4.249559e-14 	     2.899595e-14   4.276012e-14       2.860555e-14   4.175903e-14
5.900000e+03      2.243574e-14    3.325773e-14 	     2.278283e-14   3.387785e-14       2.235697e-14   3.289352e-14
6.000000e+03      1.762006e-14    2.632870e-14 	     1.779922e-14   2.667957e-14       1.744818e-14   2.588035e-14
6.100000e+03      1.349604e-14    2.034610e-14 	     1.357928e-14   2.053906e-14       1.365802e-14   2.042086e-14
6.200000e+03      1.059276e-14    1.609503e-14 	     1.061758e-14   1.618441e-14       1.057349e-14   1.593908e-14
6.300000e+03      8.218662e-15    1.259972e-14 	     8.357304e-15   1.285127e-14       8.296436e-15   1.261413e-14
6.400000e+03      6.360884e-15    9.841594e-15 	     6.326364e-15   9.813281e-15       6.424927e-15   9.851877e-15
6.500000e+03      4.869169e-15    7.602120e-15 	     4.905199e-15   7.678218e-15       4.922784e-15   7.614206e-15
6.600000e+03      3.802033e-15    5.989622e-15 	     3.805073e-15   6.006463e-15       3.841470e-15   5.992157e-15
6.700000e+03      2.984434e-15    4.743694e-15 	     2.949698e-15   4.699437e-15       2.938362e-15   4.624683e-15
6.800000e+03      2.269336e-15    3.642430e-15 	     2.243696e-15   3.609742e-15       2.289636e-15   3.637977e-15
}\xsec

\pgfplotstableread{  
500     1.18525
600     1.18698
700     1.18933
800     1.19248
900     1.19664
1000    1.20195
1100    1.20828
1200    1.21515
1300    1.22166
1400    1.22691
1500    1.23053
1600    1.2376
1700    1.24471
1800    1.25159
1900    1.25803
2000    1.26395
2100    1.27095
2200    1.27785
2300    1.2846
2400    1.29118
2500    1.29766
2600    1.30542
2700    1.31341
2800    1.32154
2900    1.32977
3000    1.33807
3100    1.34681
3200    1.35563
3300    1.36447
3400    1.37329
3500    1.38204
3600    1.39071
3700    1.39932
3800    1.40788
3900    1.41641
4000    1.42496
4100    1.43383
4200    1.44285
4300    1.45205
4400    1.46148
4500    1.47123
4600    1.4822
4700    1.49356
4800    1.50521
4900    1.51703
5000    1.52891
5100    1.53999
5200    1.551
5300    1.56199
5400    1.57303
5500    1.58419
5600    1.5957
5700    1.60751
5800    1.61968
5900    1.63231
6000    1.64552
6100    1.66037
6200    1.67586
6300    1.69184
6400    1.70817
6500    1.72471
6600    1.74127
6700    1.75767
6800    1.77374
}\bnx

\addplot[blue,thick] table [x={0},y expr=\thisrow{2}/\thisrow{1}]{\xsec};
\addlegendentry{$k_\LT$ ordered}

\addplot[red,thick] table [x={0},y expr=\thisrow{4}/\thisrow{3}]{\xsec};
\addlegendentry{$\Lambda$ ordered}

\addplot[darkgreen,thick] table [x={0},y expr=\thisrow{6}/\thisrow{5}]{\xsec};
\addlegendentry{$\vartheta$ ordered}


\addplot[purple,dashed,thick] table [x={0},y={1}]{\bnx};
\addlegendentry{analytic (BNX)}
\addplot[black,thick, dashed, domain=500:6800]{1.14563 + 0.0000460973*x + 1.198e-9*x*x};
\addlegendentry{NLO}

\addplot[black, domain=400:6800]{1};

\end{axis}
\end{tikzpicture}

\else 
\begin{center}
\includegraphics[width = 5.5 cm]{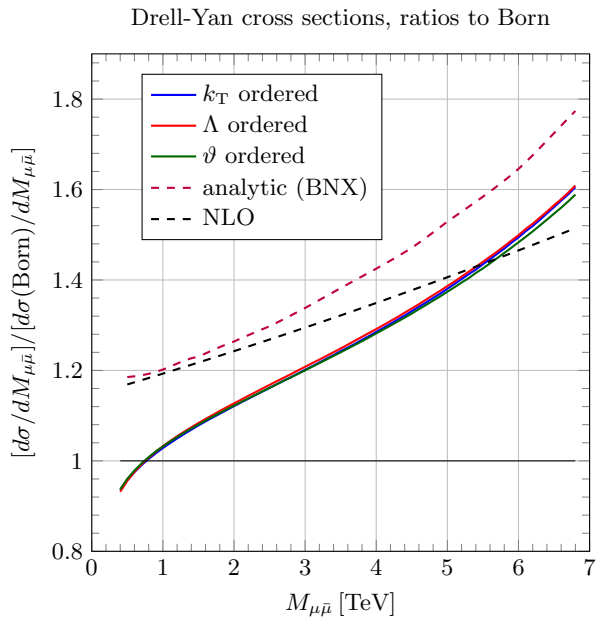}
\end{center}
\fi

\caption{
Ratio of the Drell-Yan cross section $d\sigma/dM_{\mu\bar\mu}$ to the Born cross section as a function of the dimuon mass $M_{\mu\bar\mu}$ for $\sqrt s = 13 \TeV$.
The two dashed curves shown for comparison are from Fig.~3 of Ref.~\cite{NSThreshold}. The curve labeled analytic (BNX) is the analytic summation of threshold logarithms of Becher, Neubert, and Xu, comparable to the NNLO curve of Fig.~8 of Ref.~\cite{BNX}. The curve labeled  NLO is obtained from a perturbative calculation using MCFM \cite{MCFM}.
}
\label{fig:DY}
\end{figure}

In Fig.~\ref{fig:DY}, we calculate the cross section using the product $\cU_\cV(2, 1)\, \cF(\muh^2)$ of operators for each of $k_\LT$ ordering, $\Lambda$ ordering, and $\vartheta$ ordering. We find the cross sections shown as solid curves in Fig.~\ref{fig:DY}. Now we see that the choice of ordering variable makes almost no difference in the results. We also see that the effect of summing threshold logarithms using $\cU_\cV(2, 1)\, \cF(\muh^2)$ is quite substantial. This is not a surprise because threshold logarithms are known to be quite important for processes that involve partons with large momentum fractions. In Fig.~\ref{fig:DY}, we show as a dashed red curve the result of the analytic summation of threshold logarithms by Becher, Neubert, and Xu \cite{BNX}. We see that the shower version of the summation of threshold logarithms matches the analytic version quite nicely except that the shower result needs an approximately constant correction factor of about 1.15 to equal the analytic result. This factor would plausibly be supplied by matching the shower result to a perturbative NLO calculation. In Fig.~\ref{fig:DY}, we also show the perturbative NLO result obtained from MCFM \cite{MCFM}. This fixed order result lacks the summation of threshold logarithms, but it does show the approximately 15\% correction to the shower result at $M_{\mu\bar\mu} \approx 1 \TeV$.


In Appendix \ref{sec:thresholdsum}, we use a simple leading logarithm approximation to analyze how the operators $\cU_\cV(2, 1)$ and $\cF(\muh^2)$ combine to sum threshold logarithms. This analysis shows why the curves in Fig.~\ref{fig:DYparts} have the qualitative behavior that we see in the figure. This analysis also shows why, at least in the leading approximation, the three curves in Fig.~\ref{fig:DY} match.

\begin{figure}
\ifusefigs 

\begin{tikzpicture}
\begin{axis}[title = {Drell-Yan $p_\LT$ distribution},
   xlabel={$p_T\,\mathrm{[GeV]}$}, 
   ylabel={$(1/\sigma)\,d\sigma/dp_\LT$},
   xmin=0, xmax=100,
   ymin=0, ymax=0.035,
   legend cell align=left,
   every axis legend/.append style = {
    at={(0.5,0.96)},
    anchor=north west},
  xminorgrids=false,
 yminorgrids=false,
 minor x tick num=4,
 minor y tick num=4
]

\pgfplotstableread{  
 5.000000e-01      3.608538e-03        5.159409e-03      5.993437e-03     0.00270587	   
 1.500000e+00      1.037223e-02        1.319205e-02      1.661020e-02	  0.00883755	   
 2.500000e+00      1.517866e-02        1.882987e-02      2.374168e-02	  0.0135983	   
 3.500000e+00      1.874817e-02        2.282949e-02      2.867429e-02	  0.0171181	   
 4.500000e+00      2.120059e-02        2.500150e-02      3.127449e-02	  0.0196568	   
 5.500000e+00      2.275161e-02        2.606824e-02      3.243657e-02	  0.020964	   
 6.500000e+00      2.354026e-02        2.632319e-02      3.253476e-02	  0.0216509	   
 7.500000e+00      2.382566e-02        2.613951e-02      3.200743e-02	  0.0219991	   
 8.500000e+00      2.379616e-02        2.558389e-02      3.115156e-02	  0.0219673	   
 9.500000e+00      2.339149e-02        2.487268e-02      3.003149e-02	  0.0217916	   
 1.050000e+01      2.297048e-02        2.408859e-02      2.881717e-02	  0.0215672	   
 1.150000e+01      2.243012e-02        2.324895e-02      2.754605e-02	  0.0212597	   
 1.250000e+01      2.194099e-02        2.239193e-02      2.631840e-02	  0.0207951	   
 1.350000e+01      2.124122e-02        2.156070e-02      2.508691e-02	  0.0203295	   
 1.450000e+01      2.068729e-02        2.070508e-02      2.391927e-02	  0.019868	   
 1.550000e+01      2.000140e-02        1.994061e-02      2.275003e-02	  0.0193716	   
 1.650000e+01      1.948334e-02        1.922863e-02      2.165973e-02	  0.0188878	   
 1.750000e+01      1.882125e-02        1.847849e-02      2.064646e-02	  0.0184465	   
 1.850000e+01      1.828800e-02        1.780842e-02      1.971260e-02	  0.0180028	   
 1.950000e+01      1.771243e-02        1.720355e-02      1.880238e-02	  0.0174863	   
 2.050000e+01      1.721348e-02        1.658299e-02      1.795177e-02	  0.0170324	   
 2.150000e+01      1.666170e-02        1.602903e-02      1.716466e-02	  0.0165773	   
 2.250000e+01      1.619106e-02        1.552791e-02      1.638517e-02	  0.0161545	   
 2.350000e+01      1.571341e-02        1.499414e-02      1.568680e-02	  0.0157219	   
 2.450000e+01      1.522658e-02        1.453852e-02      1.501729e-02	  0.0152628	   
 2.550000e+01      1.474560e-02        1.405216e-02      1.435828e-02	  0.0149284	   
 2.650000e+01      1.432853e-02        1.362687e-02      1.374689e-02	  0.014526	   
 2.750000e+01      1.392268e-02        1.320063e-02      1.324902e-02	  0.0140715	   
 2.850000e+01      1.358199e-02        1.285052e-02      1.269459e-02	  0.0137186	   
 2.950000e+01      1.317604e-02        1.245677e-02      1.220917e-02	  0.0133133	   
 3.050000e+01      1.282583e-02        1.217626e-02      1.174511e-02	  0.01303	   
 3.150000e+01      1.245626e-02        1.172722e-02      1.128085e-02	  0.0127002	   
 3.250000e+01      1.217467e-02        1.148801e-02      1.088162e-02	  0.0124343	   
 3.350000e+01      1.184768e-02        1.118665e-02      1.048738e-02	  0.0120487	   
 3.450000e+01      1.147063e-02        1.086733e-02      1.010251e-02	  0.01176	   
 3.550000e+01      1.114901e-02        1.060211e-02      9.770912e-03	  0.0115668	   
 3.650000e+01      1.094286e-02        1.030723e-02      9.430432e-03	  0.0111793	   
 3.750000e+01      1.063063e-02        1.003367e-02      9.114583e-03	  0.0109402	   
 3.850000e+01      1.037563e-02        9.798613e-03      8.802232e-03	  0.0106992	   
 3.950000e+01      1.012198e-02        9.573526e-03      8.505101e-03	  0.010531	   
 4.050000e+01      9.834222e-03        9.312959e-03      8.225288e-03	  0.0102152	   
 4.150000e+01      9.675787e-03        9.107283e-03      7.958656e-03	  0.0100404	   
 4.250000e+01      9.381252e-03        8.909184e-03      7.716242e-03	  0.00976311	   
 4.350000e+01      9.218533e-03        8.685979e-03      7.473056e-03	  0.00954145	   
 4.450000e+01      9.004817e-03        8.548052e-03      7.246891e-03	  0.00933789	   
 4.550000e+01      8.791252e-03        8.335552e-03      7.043547e-03	  0.00912951	   
 4.650000e+01      8.568385e-03        8.110689e-03      6.813541e-03	  0.00898488	   
 4.750000e+01      8.338907e-03        7.923975e-03      6.632639e-03	  0.00876441	   
 4.850000e+01      8.194671e-03        7.779405e-03      6.427235e-03	  0.00861545	   
 4.950000e+01      7.994489e-03        7.579581e-03      6.259172e-03	  0.00843629	   
 5.050000e+01      7.865554e-03        7.425366e-03      6.073818e-03	  0.0082177	   
 5.150000e+01      7.730697e-03        7.320455e-03      5.914760e-03	  0.00804609	   
 5.250000e+01      7.553569e-03        7.142517e-03      5.744034e-03	  0.00791573	   
 5.350000e+01      7.383137e-03        7.004977e-03      5.612375e-03	  0.00778268	   
 5.450000e+01      7.183083e-03        6.865859e-03      5.461211e-03	  0.00757076	   
 5.550000e+01      7.055458e-03        6.709990e-03      5.326959e-03	  0.00743518	   
 5.650000e+01      6.898779e-03        6.586343e-03      5.174656e-03	  0.00724082	   
 5.750000e+01      6.764048e-03        6.434199e-03      5.021133e-03	  0.00718896	   
 5.850000e+01      6.678704e-03        6.313319e-03      4.916268e-03	  0.0069986	   
 5.950000e+01      6.540321e-03        6.198981e-03      4.778329e-03	  0.00684893	   
 6.050000e+01      6.412039e-03        6.121204e-03      4.669593e-03	  0.006699	   
 6.150000e+01      6.271730e-03        5.974333e-03      4.561150e-03	  0.00662059	   
 6.250000e+01      6.158184e-03        5.865251e-03      4.443803e-03	  0.00652552	   
 6.350000e+01      6.055603e-03        5.765458e-03      4.342665e-03	  0.00637303	   
 6.450000e+01      5.897536e-03        5.629380e-03      4.232612e-03	  0.0062582	   
 6.550000e+01      5.818361e-03        5.548980e-03      4.132524e-03	  0.00615157	   
 6.650000e+01      5.690637e-03        5.416913e-03      4.047067e-03	  0.00601693	   
 6.750000e+01      5.557502e-03        5.361925e-03      3.953966e-03	  0.00594694	   
 6.850000e+01      5.503056e-03        5.233874e-03      3.873128e-03	  0.00588354	   
 6.950000e+01      5.359011e-03        5.163188e-03      3.777866e-03	  0.005734	   
 7.050000e+01      5.295265e-03        5.055957e-03      3.692267e-03	  0.00561756	   
 7.150000e+01      5.207838e-03        4.956197e-03      3.601191e-03	  0.00555492	   
 7.250000e+01      5.089202e-03        4.896211e-03      3.522442e-03	  0.00549924	   
 7.350000e+01      5.004038e-03        4.818237e-03      3.461540e-03	  0.00539484	   
 7.450000e+01      4.903603e-03        4.764831e-03      3.377591e-03	  0.0052409	   
 7.550000e+01      4.808507e-03        4.694646e-03      3.324688e-03	  0.00515962	   
 7.650000e+01      4.782471e-03        4.586651e-03      3.243253e-03	  0.00514174	   
 7.750000e+01      4.679160e-03        4.497535e-03      3.183323e-03	  0.00503421	   
 7.850000e+01      4.615345e-03        4.421724e-03      3.120905e-03	  0.00491944	   
 7.950000e+01      4.533951e-03        4.358433e-03      3.043714e-03	  0.00487378	   
 8.050000e+01      4.419641e-03        4.300549e-03      2.989715e-03	  0.0047481	   
 8.150000e+01      4.343090e-03        4.199656e-03      2.932281e-03	  0.00469404	   
 8.250000e+01      4.290260e-03        4.103014e-03      2.860674e-03	  0.00465083	   
 8.350000e+01      4.248665e-03        4.094806e-03      2.807841e-03	  0.00454712	   
 8.450000e+01      4.202261e-03        4.026359e-03      2.754675e-03	  0.00450064	   
 8.550000e+01      4.114004e-03        3.961274e-03      2.705295e-03	  0.00444006	   
 8.650000e+01      4.066742e-03        3.871484e-03      2.664815e-03	  0.00432044	   
 8.750000e+01      3.958262e-03        3.829982e-03      2.597668e-03	  0.00427321	   
 8.850000e+01      3.893179e-03        3.783719e-03      2.559170e-03	  0.00418244	   
 8.950000e+01      3.855212e-03        3.717476e-03      2.514480e-03	  0.00419085	   
 9.050000e+01      3.815278e-03        3.693101e-03      2.465751e-03	  0.00401486	   
 9.150000e+01      3.761199e-03        3.624291e-03      2.435592e-03	  0.0040562	   
 9.250000e+01      3.674321e-03        3.565077e-03      2.368196e-03	  0.0039918	   
 9.350000e+01      3.625202e-03        3.474281e-03      2.336604e-03	  0.00386719	   
 9.450000e+01      3.558207e-03        3.457994e-03      2.305637e-03	  0.00386745	   
 9.550000e+01      3.508405e-03        3.416475e-03      2.252529e-03	  0.0039057	   
 9.650000e+01      3.439462e-03        3.356372e-03      2.214871e-03	  0.00367034	   
 9.750000e+01      3.423029e-03        3.320083e-03      2.182058e-03	  0.00361572	   
 9.850000e+01      3.353358e-03        3.273860e-03      2.137659e-03	  0.00359091	   
 9.950000e+01      3.314241e-03        3.227430e-03      2.105578e-03	  0.0036522        
 }\xsec

\addplot[blue,thick] table [x={0},y={1}]{\xsec};
\addlegendentry{$k_\LT$ ordered}

\addplot[red,thick] table [x={0},y={2}]{\xsec};
\addlegendentry{$\Lambda$ ordered}

\addplot[darkgreen,thick] table [x={0},y={3}]{\xsec};
\addlegendentry{$\vartheta$ ordered}

\addplot [black,semithick, dashed] table [x={0},y={4}]{\xsec};
\addlegendentry{\textsc{ResBos}}

\end{axis}
\end{tikzpicture}

\else 
\begin{center}
\includegraphics[width = 5.5 cm]{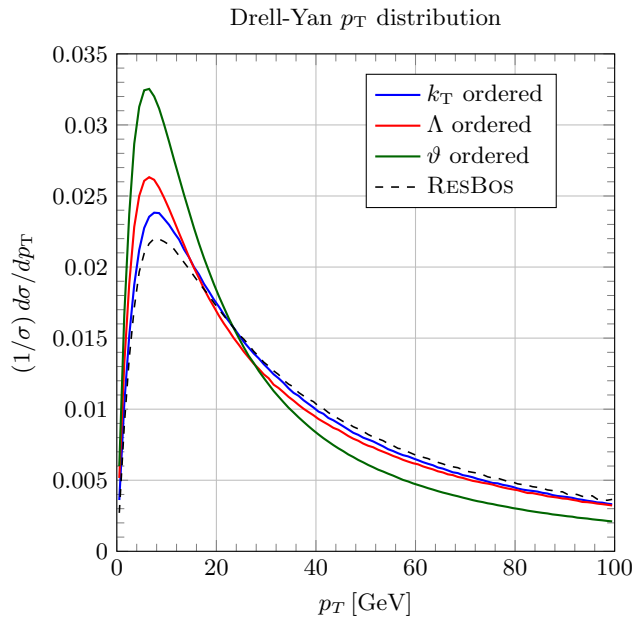}
\end{center}
\fi

\caption{
The normalized Drell-Yan transverse momentum distribution, $(1/\sigma) d\sigma/dp_\LT$ for the LHC at 13 TeV. Here $p_\LT$ is the transverse momentum of the $\mu \bar\mu$ pair, $d\sigma/dp_\LT$ is $d\sigma/(dM_{\mu\bar \mu} dp_\LT)$ integrated over $2 \TeV < M_{\mu\bar \mu} < 2.1 \TeV$ and $\sigma$ is this cross section integrated over $0 < p_\LT < 100 \GeV$, so that the area under the curve is 1. We show curves for $k_\LT$ ordering, $\Lambda$ ordering, and $\vartheta$ ordering of the shower. We also show the corresponding result obtained using ResBos \cite{ResBos1,ResBos} as a dashed black curve.
}
\label{fig:DYpT}
\end{figure}

We now look at the cross section $d\sigma/(dM_{\mu\bar\mu}\,dp_\LT$), in which both the mass and the transverse momentum $p_\LT$ of the muon pair are measured. Now the shower operator $\cU(2, 1)$ in Eq.~(\ref{eq:sigmaJDY}) does not commute with the measurement operator $\cO_J$, so that the parton shower has an effect. In Fig.~\ref{fig:DYpT}, we show $d\sigma/(dM_{\mu\bar\mu}\,dp_\LT)$ integrated in $M_{\mu\bar\mu}$ over the range $2 \TeV < M_{\mu\bar\mu} < 2.1 \TeV$ and normalized by this cross section integrated in $p_\LT$ over $0 < p_\LT < 100 \GeV$. The shower approximately sums logarithms of $p_\LT/M_{\mu\bar\mu}$. We compare to an analytic summation of these logarithms \cite{CSSpT} using the program \textsc{ResBos} \cite{ResBos1,ResBos}. The \textsc{ResBos} result includes some nonperturbative broadening that is not present in the shower result.

Fig.~\ref{fig:DYpT}, is comparable to Fig.~4 of Ref.~\cite{NSThreshold}, but now the definitions for the shower algorithm are a little different and now we show results for $k_\LT$ ordering, $\Lambda$ ordering, and $\vartheta$ ordering of the shower. 

We see that the distributions for $k_\LT$ ordering and $\Lambda$ ordering agree with each other pretty closely. The \textsc{ResBos} result is, as expected, a bit broader than these. The result with $\vartheta$ ordering is qualitatively similar to the $k_\LT$ and $\Lambda$ ordering results, but noticeably narrower.


\section{Summary and comments}
\label{sec:Conclusions}

In a previous paper \cite{angleFS}, we showed how, in the case of electron-positron annihilation, one can formulate a first order parton shower using more than one shower scale, $\vec \mu = (\mu_1, \mu_2, \dots)$. Then one has the freedom to specify an evolution path $\vec\mu(t)$. This can lead to a simpler end result for cross sections than one would get with, say, $k_\LT$ evolution. In this paper, we extend the use of multiscale evolution to showers with initial state partons. This paper covers many issues, so we briefly summarize the main points here.

A parton shower with initial state partons in inherently more complicated than a shower with just final state partons. One needs to factor parton distribution functions out of the cross section and one needs to connect the evolution of the shower with the evolution of the parton distribution functions. This has two consequences. 

First, the parton distribution functions used internally in the shower should not be the same as the $\MSbar$ parton distribution functions used for calculations in fixed order perturbation theory. We have analyzed this in earlier papers \cite{NSThreshold, NSAllOrder, NSThresholdII, ISevolve}, but in Sec.~\ref{sec:PDFs} of this paper, we have presented a more precise derivation. We also present a direct relation between the shower oriented PDFs and the $\MSbar$ PDFs that is better suited  than the relationship used in Refs.~\cite{NSThreshold, NSAllOrder, NSThresholdII, ISevolve} when the scale arguments of the PDFs are large.

A second consequence of the connection between shower evolution and PDF evolution is that, in addition to the shower evolution operator $\cU$, a second operator, $\cU_\cV$ appears in the shower cross section. The operator $\cU_\cV$ creates a summation of threshold logarithms. This operator was included in Refs.~\cite{NSThreshold, NSAllOrder, NSThresholdII, ISevolve}, but in this paper in Sec.~\ref{sec:cU} we have presented what we believe is a clearer derivation. We also present an improved expression for $\cU_\cV$ and the infrared sensitive operator $\cD$ on which the derivation is based. In particular, we have simplified the definition of the part of $\cD$ at first order that comes from virtual parton exchanges. This is achieved by introducing what we call the momentum sum rule in Sec.~\ref{sec:RecD01}.

With initial state partons included, the first order virtual exchange operator $\cD^{[0,1]}$ acquires an imaginary part. Sec.~\ref{sec:ImcD01} of this paper provides a simpler definition of $\mathrm{Im}\,\cD^{[0,1]}$ than appears in Refs.~\cite{NScolor, NSColoriPi}.

Using this structure for initial state showers, we consider a shower with multiple shower scales. We use three shower scales, defined in Sec.~\ref{sec:multiplescales}. The first is $\mu_\scC$, which stands for one of $\mu_\perp$, $\mu_\Lambda$, or $\mu_\angle$, corresponding to $k_\LT$ ordering, $\Lambda$ ordering (based on virtuality), or angular ordering. The scale $\mu_\scC$ determines the boundary of a region in the space of splitting variables $(z,\vartheta)$ that is defined to represent splittings that are unresolved at scale $\mu_\scC$. With $k_\LT$ ordering or $\Lambda$ ordering, a splitting is unresolved if it is either too soft ($(1-z)$ too small) or too collinear ($\vartheta$ too small.) The second scale is $\mu_\scE$, for which a splitting is unresolved if it is too soft, $(1-z) < \mu_\scE^2/Q^2$. There are some terms in the shower splitting functions that have a soft singularity but not a collinear singularity. We take advantage of having the scale $\mu_\scE$ by eliminating the cuts based on $\mu_\scC$ for these terms. We introduce a third scale $\mu_{\mi\pi}$ that defines an unresolved region for $\mathrm{Im}\,\cD^{[0,1]}$ but not for other terms in $\cD^{[1]}$.

With three scales available, we define a simple, three-component path in the space of scales that takes us from large scales to small scales. This choice gives a simpler result for a calculated cross section than one would have with a single scale $\mu_\scC$. First, as long as the measurement applied at the end of the shower is not sensitive to the color state of the partons, the imaginary part of the splitting function has no effect. Second, one can choose to use the LC+ approximation for color on the second segment of the path. Then there is a term in the splitting functions that is proportional to the difference between full color and the LC+ approximation. This term has a soft singularity by no collinear singularity. This difference term appears by itself on the first segment of the path. In the case of the lowest-order Drell-Yan process, the shower operator, $\cU(1,0)$, is just the unit operator and the threshold operator $\cU_\cV(1,0)$ is also just the unit operator. 

This gives us a very simple result for the cross section. One can simply use the result for one scale $\mu_\scC$ in which we ignore the imaginary part of virtual exchanges and in which we use the LC+ approximation for color. In Sec.~\ref{sec:DrellYan} we exhibit results using this approach for the Drell-Yan cross section $d\sigma/dM_{\mu\bar\mu}$  and for the $p_\LT$ distribution of the produced muon pair. These results are obtained for $k_\LT$ ordering, $\Lambda$ ordering, and $\vartheta$ ordering. The results for $d\sigma/dM_{\mu\bar\mu}$ are remarkably independent of the ordering choice and the results for the $p_\LT$ distribution show a noticeable dependence on the ordering choice in the case of $\vartheta$ ordering.

Having summarized the main points of the paper, we can offer some comments. 

We view a parton shower as something that is, in principle, defined at all perturbative orders \cite{NSAllOrder}. In particular, the infrared sensitive operator $\cD$ should be thought of as having an expansion in powers of $\as$: $\cD = 1 + \cD^{[1]} + \cdots$. Then, in the formulation of Ref.~\cite{NSAllOrder}, the generator $\cS(t)$ of parton splittings is determined and has an expansion in powers of $\as$ beginning at order $\as^1$. In a first order shower, we use the first order term. In fact, it is an important open problem to define a prescription in which the second order contributions to $\cD$ and thus $\cS(t)$ can be calculated.

With a first order shower, there is considerable freedom to define $\cD^{[1]}$ and thus $\cS^{[1]}(t)$. For instance, one can use a single scale that defines either $k_\LT$ ordering or angular ordering. The resulting first order splitting functions are not the same. However, as analyzed in Ref.~\cite{angleFS}, the contribution to a cross section resulting from this difference starts at order $\as^2$. If we were able to work with order $\as^2$ splitting functions, $\cS^{[2]}$, then the differences in calculated cross sections between prescriptions would be order $\as^3$ and therefore would be less significant. In addition, if we were able to use splitting functions $\cS^{[2]}$, we would have a sensible way to judge the merits of two prescriptions for a first order shower: we would prefer a prescription at first order that makes the second order splitting functions $\cS^{[2]}$ numerically small in applications.

In this paper, we have used the option to use multiple shower scales $\vec \mu_\scS$ in a way that makes the first order shower simpler. The difference between the resulting shower and a more standard one scale shower would then be generated by higher order contributions to the shower splitting functions.  We can hope, but cannot now prove, that the first order shower thus defined will not have large contributions to its splitting functions when extended to second order.

We expect the $\cD^{[2]}$ operator in a second order shower to be quite complicated. In order to deal with the complications, it may be helpful to use multiple splitting scales along the lines used in this paper.

\acknowledgments{ 
This work was supported in part by the United States Department of Energy under grant DE-SC0011640. The work benefited from access to the University of Oregon high performance computer cluster, Talapas.  We thank Thomas Becher for advice about threshold summation for the Drell-Yan cross section and Frank Petriello for advice about the perturbative Drell-Yan cross section.
}
\bigskip
\appendix

\section{Initial state splittings in \textsc{Deductor}}
\label{sec:ISsplittings}

In this appendix, we describe the details of initial state real emissions from parton ``a'' in \textsc{Deductor}. We begin with the kinematic variables. An initial state parton with momentum $p_\La$ splits in backward evolution into a new initial state parton with momentum $\hat p_\La$ and a final state parton labeled $m+1$ with momentum $\hat p_{m+1}$. All partons are treated as massless. \textsc{Deductor} uses splitting variables consisting of a dimensionless virtuality $y$, a momentum fraction $z$, and an azimuthal angle $\phi$. Before the splitting, the initial state parton momenta are
\begin{equation}
\begin{split}
p_\La ={}& \eta_\La p_\LA
\;,
\\
p_\Lb ={}& \eta_\Lb p_\LB
\;.
\end{split}
\end{equation}
Here $p_\LA$ and $p_\LB$ are the hadron momenta, approximated as being massless. After the splitting, the new momenta are
\begin{align}
\label{eq:ISmomenta}
\notag
\hat p_\La ={}& \frac{1}{z}\,p_\La
\;,
\\ \notag
\hat p_{m+1} ={}& \frac{1 - z - yz}{z}\, p_\La
+ z y\, p_\Lb
+ k_\perp
\;,
\\
\hat p_\Lb ={}& p_\Lb
\;.
\end{align}
Here $k_\perp \cdot p_\La = k_\perp \cdot p_\Lb = 0$ and $\phi$ is the azimuthal angle of $k_\perp$. This gives us the virtuality variable $y$:
\begin{equation}
2 \hat p_\La \cdot \hat p_{m+1} = y\, Q^2
\;,
\end{equation}
where $Q^2 = 2 p_\La \cdot p_\Lb$. Using $\hat p_{m+1}^2 = 0$, we find
\begin{equation}
- k_\perp^2 = (1- z - yz) y\,Q^2
\;.
\end{equation}

Since $\hat p_\La - \hat p_{m+1} \ne p_\La$ when $y > 0$, we need to take some momentum from the final state partons in order to conserve momentum. \textsc{Deductor} takes the needed momentum from {\em all} of the final state partons before the splitting using
\begin{equation}
\hat p_l^\mu = \Lambda^\mu_\nu p_l^\nu
\;,
\hskip 1 cm l\in \{1,\dots,m\}
\end{equation}
for a suitably chosen Lorentz transformation $\Lambda^\mu_\nu$. Then momentum is conserved provided that
\begin{equation}
\Lambda^\mu_\nu (p_\La^\nu + p_\Lb^\nu)
= \hat p^\mu_\La - \hat p^\mu_{m+1} + p^\mu_\Lb
\;.
\end{equation}
It is possible to find a Lorentz transformation with this property because the momenta in Eq.~(\ref{eq:ISmomenta}) obey
\begin{equation}
(\hat p_\La - \hat p_{m+1} + p_\Lb)^2
= (p_\La + p_\Lb)^2
\;.
\end{equation}
The choice made in \textsc{Deductor} is as follows \cite{NSZpT}. If 
\begin{equation}
p = \alpha\, p_\La + \beta\, p_\Lb + p_\perp
\end{equation}
with $p_\perp \cdot p_\La = p_\perp \cdot p_\Lb = 0$, then $\hat p^\mu = \Lambda^\mu_\nu p^\nu$ is
\begin{equation}
\begin{split}
\hat p ={}& (1+y) \alpha\, p_\La
\\& +
\frac{1}{1+y}\,\left[
\beta - \frac{2 \bm p_\perp \cdot \bm k_\perp}{Q^2}
+ \alpha\,\frac{\bm k_\perp^2}{Q^2}
\right] p_\Lb
\\ & +
p_\perp - \alpha k_\perp
\;.
\end{split}
\end{equation}

We now turn to the initial state real emission operator $\cD^{[1,0]}_\La(\mur^2,\mu_\scC^2)$. Here $\mur^2$ is the renormalization scale and $\mu_\scC^2$ is a single shower scale corresponding to $k_\LT$ ordering, $\Lambda$ ordering, or $\vartheta$ ordering as specified in Sec.~\ref{sec:simpleordering}. We apply $\cD^{[1,0]}_\La(\mur^2,\mu_\scC^2)$ with its accompanying PDF factors to an $m$-parton state and write the result in the form
\begin{widetext}
\begin{equation}
\begin{split}
\label{eq:Dhatdef}
\cF_{\!\La}(\mur^2)
{\cal D}^{[1,0]}_\mathrm{a}&(\mur^2,\mu_\scC^2)
\cF_{\!\La}^{-1}(\mur^2)\sket{\{p,f,c,c'\}_{m}}
\\
={}& 
\int\!d\{\hat p,\hat f\}_\mpone\
\sket{\{\hat p,\hat f\}_{\mpone}} 
\frac{\as(\mur^2)}{2\pi}
\sum_{\hat a}\int_0^1\!\frac{dz}{z}\
\frac{
f_{\hat a/A}(\eta_{\La}/z,\mur^{2})}
{f_{a/A}(\eta_{\La},\mur^{2})}
\hat {\bm D}^\La_{a \hat a}(z;\{\hat p,\hat f\}_\mpone,\{p,f\}_{m};\epsilon)
\sket{\{c,c'\}_{m}}
\;.
\end{split}
\end{equation}
Here ${\cal D}^{[1,0]}_\mathrm{a}$ adds one new parton and we integrate over the momenta and flavors $\{\hat p,\hat f\}_\mpone$ of the partons after the splitting. There are dimensionally regulated singularities, so this integration is in $4 - 2 \epsilon$ dimensions for each momentum. Then $\hat {\bm D}^\La_{a \hat a}$ is a function of the momenta and flavors before and after the splitting and is an operator that maps the $m$-parton color space to the $\mpone$ parton color space. The index $a$ in $\hat {\bm D}^\La_{a \hat a}$ is fixed. It is the flavor of the incoming parton ``a.'' The operator $\hat {\bm D}^\La_{a \hat a}$ also depends on the two scales $\mur^2$ and $\mu_\scC^2$, but we have not made that dependence explicit in the notation.

The operator $\cF = \cF_{\!\La} \cF_{\!\Lb}$ provides the parton distribution functions needed to make a cross section, Eq.~(6.3) of Ref.~\cite{NSThreshold},
\begin{equation}
\label{eq:Fdef}
\cF_{\!\La}(\mur^2)\sket{\{p,f,c',c\}_{m}} = 
\frac{f_{a/A}(\eta_{\La},\mur^2)}
{n_\Lc(a) n_\Ls(a)\,2\eta_{\La}[p_\LA\!\cdot\!p_\LB]^{1/2}}\
\sket{\{p,f,c',c\}_{m}} 
\;.
\end{equation}
Here $n_c(a)$ is the number of colors for flavor $a$ and $n_\Ls(a)$ is the number of spins, 2 for a quark and $2(1 - \epsilon)$ for a gluon.

The operator $\hat {\bm D}^\La_{\hat a a}$ is
\begin{equation}
\begin{split}
\label{eq:D10a4}
\hat {\bm D}^\La_{a \hat a}(z;&\{\hat p,\hat f\}_\mpone,\{p,f\}_{m};\epsilon)
\\={}&
\frac{z^\epsilon}{(1-z)^{2\epsilon}}
\left(\frac{\mur^2}{Q^2}\right)^{\!\epsilon}
\frac{(4\pi)^\epsilon}{\Gamma(1-\epsilon)}\,
\int_0^1\!\frac{d\vartheta}{\vartheta}\
[\vartheta(1-\vartheta)]^{-\epsilon}
\int\!\frac{d^{1-2\epsilon}\phi}{S(2-2\epsilon)}
\\&\times
\delta\big(\{\hat p, \hat f\}_\mpone - R_\La(z,\vartheta,\phi, 
 \hat a;\{p,f\}_{m})\big)\,
\Theta\big((z,\vartheta) \in U(\mu_\scC^2)\big)\,
\\&\times
\sum_{k}\frac{1}{2}\bigg[
\theta(k = \La)\,
\frac{1}{N(a ,\hat a)}\, \hat P_{a \hat a}(z, \vartheta,\epsilon)
-
\theta(k\ne \La)\,
\delta_{a \hat a}\,
\frac{2}{1-z}\, 
W_0\!\left(\xi_{\La k}, z, \vartheta, \phi - \phi_k\right)
\bigg]
\\&\times
\left\{
t^\dagger_\La(f_\La \to \hat f_\La\!+\!\hat f_\mpone)\otimes
t_k(f_k \to \hat f_k \!+\! \hat f_\mpone)
+
t^\dagger_k(f_k \to \hat f_k \!+\! \hat f_\mpone)\otimes
t_\La(f_\La \to \hat f_\La \!+\! \hat f_\mpone)
\right\}
\;.
\end{split}
\end{equation}
\end{widetext}
The operator $\hat {\bm D}^\La_{a \hat a}$ depends on the splitting variable $z$ and on $\hat a$, which is the flavor of the incoming parton after the splitting (in the sense of backward evolution). We use $\hat a$ as a splitting variable that specifies the flavor content of the splitting. For instance, $\hat a = a$ corresponds to gluon emission from the incoming line, so that parton $\mpone$ is a gluon. If $a$ is a quark flavor and $\hat a = \Lg$, then parton $\mpone$ is the corresponding flavor of antiquark. The right-hand side of Eq.~(\ref{eq:D10a4}) begins with integrations over the other two splitting variables $\vartheta$ and $\phi$, with appropriate dependence on the dimensional regulation parameter $\epsilon$. The variable $\phi$ is a unit vector in the $2 - 2 \epsilon$ dimensional transverse momentum space and represents the azimuthal angle of $\hat p_\mpone$. There is a sum over the index $k \in \{\La,\Lb,1,\dots,m\}$ of a dipole partner parton for a splitting. Then $\phi_k$ is the azimuthal angle of $\hat p_k$ if $k \notin \{\La,\Lb\}$. The integration over $\phi$ is an integration over a unit sphere that is a $1 - 2\epsilon$ dimensional surface. The function $S(2 - 2\epsilon)$ is the surface area of this sphere, so that
\begin{equation}
\int\!\frac{d^{1-2\epsilon}\phi}{S(2-2\epsilon)}\ 1 = 1
\;.
\end{equation}
Integration over $\phi$ of a function of $\phi - \phi_k$ is a shorthand notation for integration over a unit vector $\phi$ in a coordinate system with the unit vector $\phi_k$ equal to $(1,0,\dots)$. After the integrations, there is a delta function that sets $\{\hat p, \hat f\}_\mpone$ to the momenta and flavors obtained from a splitting with variables $(z,\vartheta,\phi, \hat a)$ applied to partons with momenta and flavors $\{p,f\}_{m}$ according to \textsc{Deductor} conventions.

The idea of the singular operator ${\cal D}^{[1,0]}$ is that it integrates over splittings that are arbitrarily close to the soft and collinear limits, but with an ultraviolet cutoff that depends on a scale parameter $\mu_\scC^2$. The region of $(z,\vartheta)$ allowed by the cutoff is called the unresolved region and is denoted by $U(\mu_\scC^2)$. The subscript $\scC$ denotes the  definition of the unresolved region, as described in Sec.~\ref{sec:simpleordering}. We therefore insert a theta function that specifies that $(z,\vartheta)$ lies in the unresolved region.

There is a special feature that applies when the hard scattering process is parton-parton scattering, $\La + \Lb \to 1 + 2$, with transverse momentum $P_\perp$  defined by $p_1 = \alpha\, p_\La + \beta\, p_\Lb + P_\perp$. Here the partons are massless and $P_\perp\cdot p_\La = P_\perp\cdot p_\Lb = 0$. In a parton shower, the initial state partons split in backwards evolution into other final state partons besides partons 1 and 2 from the hard scattering, leaving at the end of the shower new initial state partons $\La'$ and $\Lb'$ with larger momentum fractions than the partons that created the hard scattering. In this situation, the scattering identified as the hard scattering should be the hardest of all the scatterings. That is, the $k_\LT^2$ of each initial state emission should be no greater than $|P_\perp^2|$.\footnote{This was argued in some detail in Ref.~\cite{ShowerTime} with a slightly different choice of variables.} For this reason, the unresolved region $U(\mu_\scC^2)$ in Eq.~(\ref{eq:D10a4}) should be a subregion of $0<z<1$, $0 < \vartheta < 1$, and $k_\LT^2 < |P_\perp^2|$. That is, we define $\Theta((z,\vartheta) \in U(\mu_\scC^2))$ in Eq.~(\ref{eq:D10a4}) by 
\begin{equation}
\begin{split}
\Theta\big((z,\vartheta) \in U(&\mu_\scC^2)\big) 
\\={}& 
\theta(\vartheta < a_\scC(z,\mu_\scC^2))\,
\theta(0<\vartheta < 1)
\\&\times
\theta(\vartheta < a_\perp(z,|P_\perp^2|))
\;.
\end{split}
\end{equation}
Then the resolved region consists of all points $(z,\vartheta)$ with $0<z<1$, $0 < \vartheta < 1$, and $k_\LT^2 < |P_\perp^2|$ that are not in the unresolved region. Only splittings with $k_\LT^2 < |P_\perp^2|$ can then be generated in the shower. When the hard process is Drell-Yan muon pair production, this restriction does not apply. When the hard scattering is parton-parton scattering, if the shower is $k_\LT$ ordered, this restriction does not matter, but in the case of $\Lambda$ or $\vartheta$ ordering, this restriction does matter. In order to avoid extra clutter, we omit effects of the factor $\theta(\vartheta < a_\perp(z,|P_\perp^2|))$ in our formulas that follow.

In the next factor in Eq.~(\ref{eq:D10a4}), there is a sum over dipole partner partons $k$. In the first term, the partner parton is the same as the emitting parton, $k = \La$. This term contains a color factor $N(a, \hat a)$ defined by 
\begin{align}
\label{eq:naahat}
\notag
N(q, \Lg) ={}& T_\LR \;,
\\ \notag
N(\Lg, q) ={}& C_\LF \;,
\\  \notag
N(q, q) ={}& C_\LF \;,
\\
N(\Lg, \Lg) ={}& C_\LA \;,
\end{align}
\vskip 1 em
\noindent
where $q$ is any quark or antiquark flavor. Then there is a splitting function $\hat P_{a \hat a}(z,\vartheta,\epsilon)$. For the case that $\hat a \ne a$, this function is related to the function $\overline w_{\La\La}(\{\hat p,\hat f\}_\mpone)$ that appears in Eq.~(5.7) of Ref.~\cite{NScolor} by
\begin{align}
\label{eq:PaadefI}
\frac{\as(\mur^2)}{2\pi}\,\frac{1}{\vartheta}\,&
\frac{\hat P_{a \hat a}(z,\vartheta,\epsilon)}{N(a, \hat a)} 
\\ \notag = {}&
\frac{Q^2}{16 \pi^2}\,
\frac{n_\Ls(a)}{n_\Ls(\hat a)}\,(1-z)\,
\overline w_{\La\La}(\{\hat p,\hat f\}_\mpone)
\;.
\end{align}
Here $n_\Ls(a)$ is the number of spin states for a parton of flavor $a$ as in Eq.~(\ref{eq:Fdef}). We calculate $\overline w_{\La\La}(\{\hat p,\hat f\}_\mpone)$ and similar functions in the cases below from the definitions in Ref.~\cite{NSI} so that the sums over parton spins are performed in $4 - 2 \epsilon$ dimensions. For the case that $\hat a = a$, $\hat P_{a a}(z,\vartheta,\epsilon)$ is related to the functions $\overline w_{\La\La}(\{\hat p,\hat f\}_\mpone)$ and $\overline w_{\La\La}^{\rm eikonal}(\{\hat p,\hat f\}_\mpone)$ that appear in Eq.~(5.7) of Ref.~\cite{NScolor} by
\begin{widetext}
\begin{equation}
\begin{split}
\label{eq:PaadefII}
\frac{\as(\mur^2)}{2\pi}\,\frac{1}{\vartheta}\, 
\frac{\hat P_{a a}(z,\vartheta,\epsilon)}{N(a, a)}
={}& 
\frac{Q^2}{16 \pi^2}
\left((1-z)\,\overline w_{\La\La}(\{\hat p,\hat f\}_\mpone)
- (1-z)\,\overline w_{\La\La}^{\rm eikonal}(\{\hat p,\hat f\}_\mpone)
+ \frac{8\pi\as(\mur^2)}{Q^2}\, \frac{2z}{(1-z)\vartheta}\right)
\;.
\end{split}
\end{equation}

The functions $\hat P_{a \hat a}(z, \vartheta, \epsilon)$ are rather simple. We simplify the notation by defining 
\begin{equation}
\begin{split}
\label{eq:Paanotation}
\hat P_{a \hat a}(z,\vartheta,0) ={}& \hat P_{a \hat a}(z,\vartheta) 
\;,
\\
\hat P_{a \hat a}(z,0,0) ={}& \hat P_{a \hat a}(z)
\;,
\\
\hat P_{a \hat a}(z,0,\epsilon)
= {}&
\hat P_{a \hat a}(z)
- \epsilon \hat P_{a \hat a}^{(\epsilon)}(z)
+ \cO(\epsilon^2)
\;.
\end{split}
\end{equation}
Then we find
\begin{equation}
\begin{split} 
\label{eq:hatP}
\hat P_{qq}(z,\vartheta) ={}& 
C_F\left(
\frac{1+z^2}{1-z}
- z y
\right)
\;,
\\ 
\hat P_{gg}(z,\vartheta) ={}& 
2C_A\left(
\frac{z[1 + yz(1+y)]}{(1-z)(1+y)}
+ \frac{1-z(1+y)}{z (1+y)^2}
+  z [1-z(1+y)]
\right)
\;,
\\ 
\hat P_{q\Lg}(z,\vartheta) ={}& T_R\big(1 - 2\,z\,(1+y)\,[1-z(1+y)]\big)
\;, 
\\ 
\hat P_{\Lg q}(z,\vartheta) ={}&C_F\left(z + 2\,\frac{1-z(1+y)}{z\,(1+y)^2}\right)
\;,
\end{split}
\end{equation}
where $y = (1-z)\vartheta /z$ and $q$ is any quark or antiquark flavor. 
\end{widetext}

The contributions proportional to $\epsilon$ (at $y = 0$) are
\begin{equation}
\begin{split}
\label{eq:Pepsilon}
\hat P_{q q}^{(\epsilon)}(z) ={}& C_\LF (1-z)
\;,
\\ 
\hat P_{\Lg \Lg}^{(\epsilon)}(z) ={}& 0
\;,
\\ 
\hat P_{q \Lg}^{(\epsilon)}(z) ={}& T_\LR\,2z(1-z)
\;,
\\
\hat P_{\Lg q}^{(\epsilon)}(z) ={}& C_\LF\,z
\;.
\end{split}
\end{equation}
Finally, at $y = 0$ and $\epsilon = 0$ we have
\begin{equation}
\begin{split}
\label{eq:hatPzeroy}
\hat P_{qq}(z)
={}& C_\LF\, \frac{1+z^2}{1-z}
\;,
\\ 
\hat P_{\Lg \Lg}(z)
={}& 2C_\LA
\left(
\frac{z}{1-z} + \frac{1-z}{z} + z(1-z)
\right) 
\;,
\\
\hat P_{q \Lg}(z)
={}& T_\LR\big(1-2 z(1-z)\big) 
\;,
\\
\hat P_{\Lg q}(z)
={}& C_\LF
\left(
z + 2\,\frac{1-z}{z}
\right) 
\;.
\end{split}
\end{equation}
That is, these are the familiar unregulated DGLAP kernels.

Next, in Eq.~(\ref{eq:D10a4}) is a term proportional to a function $W_0\!\left(\xi_{\La k}, z, \vartheta, \phi - \phi_k\right)$. This term comes from interference between emission of a gluon from parton ``a'' and emission from dipole partner parton $k$ with $k \ne \La$.  The variable $\xi_{\La k}$ is $(1 - \cos\theta_{\La, k})/2$ where $\theta_{\La, k}$ is the angle between $p_k$ and $p_\La$ as measured in the rest frame of $Q$,
\begin{equation}
\xi_{\La k} = \frac{p_\La \cdot p_k\,Q^2}{2\, p_\La \cdot Q\,p_k\cdot Q}
= \frac{p_\La \cdot p_k}{p_k\cdot Q}
\;.
\end{equation}

The function $W_0$ is related to the functions $\overline w_{\La k}^\mathrm{dipole}(\{\hat p\}_\mpone)$ and $A'_{\La k}(\{\hat p\}_{m+1})$ that appear in Eq.~(5.7) of Ref.~\cite{NScolor} by
\begin{align}
\label{eq:Wlk0-ini-definition}
\notag
\frac{\as(\mur^2)}{2\pi}\,&\frac{1}{\vartheta}\,
\frac{2}{1-z}\, 
W_0\!\left(\xi_{\La k}, z, \vartheta, \phi - \phi_k\right) 
\\ \notag ={}&
\frac{Q^2}{16 \pi^2}
\Bigg[
(1-z)\,A'_{\La k}(\{\hat p\}_{m+1})\, 
\overline w_{\La k}^\mathrm{dipole}(\{\hat p\}_{m+1})
\\&
- \frac{8\pi\as(\mur^2)}{Q^2}\, \frac{2z}{(1-z)\vartheta}
\Bigg]
\;.
\end{align}
The functions $\overline w_{\La k}^\mathrm{dipole}(\{\hat p\}_{m+1})$ and $A'_{\La k}(\{\hat p\}_{m+1})$ are (from \cite{NScolor}, Eq.~(5.3) and \cite{NSspin}, Eq.~(7.12))
\begin{equation}
\begin{split}
\overline w_{\La k}^\mathrm{dipole}(\{\hat p\}_{m+1}) ={}& 
4\pi\as(\mur^2)\,\frac{2 \hat p_k\cdot \hat p_\La}
{\hat p_\mpone\cdot \hat p_k\, \hat p_\mpone\cdot \hat p_\La}
\;,
\\
A'_{\La k}(\{\hat p\}_{m+1}) ={}& 
\frac{\hat p_\mpone\cdot \hat p_k\ \hat p_\La \cdot \hat Q}
{\hat p_\mpone\cdot \hat p_k\ \hat p_\La \cdot \hat Q
+ \hat p_\mpone\cdot \hat p_\La\ \hat p_k \cdot \hat Q}
\;.
\end{split}
\end{equation}
The function $(1-z)\,A'_{\La k} \overline w_{\La k}^\mathrm{dipole}$ is singular both in the collinear limit, $\vartheta \to 0$ at fixed $(1-z)$, and in the soft limit, $(1-z) \to 0$ at fixed $\vartheta$. In the collinear limit, $(1-z)\,A'_{\La k} \overline w_{\La k}^\mathrm{dipole}$ is proportional to $z/[(1-z)\vartheta]$. In Eq.~(\ref{eq:Wlk0-ini-definition}), we have subtracted this singular behavior, leaving a function $W_0$ that vanishes in the collinear limit, although it remains finite in the soft limit.

The explicit expression for $W_0$ is
\begin{widetext}
\begin{equation}
\begin{split}
\label{eq:W0}
W_0({}&\xi,z,\vartheta,\phi) = \frac{z^2+z(1-z)\vartheta}{1+\vartheta(1-z)}
\left[\frac{z-(1+(1-z)\vartheta)(1-\xi)
\left(z (1-\vartheta) + \vartheta\right)}
{z - \big(1+(1-z)\vartheta\big)\left[ (1-\xi)(z(1-\vartheta) - \vartheta)
+ 2\cos\phi\, \sqrt{z \vartheta(1-\vartheta)}\sqrt{\xi(1-\xi)}\right]}
- 1\right]
\;.
\end{split}
\end{equation}

In our study of the inclusive splitting probability, we will need the azimuthal average of $W_0$ at $\epsilon = 0$,
\begin{equation}
\label{eq:Waveragedef}
W\!\left(\xi_{\La k}, z, \vartheta\right)
= \int_{-\pi}^\pi\! \frac{d\phi}{2\pi}\
W_0\!\left(\xi_{\La k}, z, \vartheta, \phi-\phi_k\right)
\;.
\end{equation}
It has a reasonably simple form:
\begin{equation}
\begin{split}
\label{eq:Wlk-ini-result1}
W(\xi, z, \vartheta) = {}& \frac{z^2+z(1-z)\vartheta}{1+\vartheta(1-z)}
\left[
\frac{1 - \delta}
{\sqrt{(1-\delta)^2 + 4 \vartheta^2 \delta}}
-1
\right]
\;,
\end{split}
\end{equation}
where
\begin{equation}
\delta = (1+(1-z)\vartheta)\left(1+\frac{1-z}{z}\,\vartheta\right)(1-\xi)
\;.
\end{equation}
We can rewrite this in a form that shows that $W(\xi_{\La k}, z, \vartheta)$ is proportional to $\vartheta^2$ as $\vartheta \to 0$ with fixed $z$:
\begin{equation}
\begin{split}
\label{eq:Wlk-ini-result2}
W(\xi, z, \vartheta) = {}& -\frac{z^2+z(1-z)\vartheta}{1+\vartheta(1-z)}
\frac{4 \vartheta^2 \delta}
{\sqrt{(1-\delta)^2 + 4 \vartheta^2 \delta}\,
\left[(1-\delta) + \sqrt{(1-\delta)^2 + 4 \vartheta^2 \delta}\right]}
\;.
\end{split}
\end{equation}
\end{widetext}
We will also need $W$ at $z = 1$:
\begin{equation}
\begin{split}
\label{eq:Wlk-ini-result3}
W(\xi, 1, \vartheta) = {}& 
\frac{\xi}
{\sqrt{\xi^2 + 4 \vartheta^2 (1-\xi)}}
-1
\;.
\end{split}
\end{equation}

Finally in Eq.~(\ref{eq:D10a4}) there is a factor with color operators. The operator  $t^\dagger_\La(f_\La \to \hat f_\La\!+\!\hat f_\mpone)$, acting on the ket color state $\ket{\{c\}_m}$, gives the new color state $\ket{\{\hat c\}_\mpone}$ that one gets after emitting the new parton $\mpone$ from parton ``a'' with flavor $f_\La$. This operator is described in some detail in Ref.~\cite{NSI}. Similarly,   $t_k(f_k \to \hat f_k\!+\!\hat f_\mpone)$, acting on the bra color state $\bra{\{c'\}_m}$, gives the new color state $\bra{\{\hat c'\}_\mpone}$ that one gets after emitting the new parton $\mpone$ from parton $k$ with flavor $f_k$.

When parton $\mpone$ is a gluon, the color operators obey the identity
\begin{equation}
\sum_{k} t_k(f_k \to f_k\!+\!\Lg)
= 0
\;.
\end{equation}
This identity arises from the fact that the parton color state is an overall color singlet, so that attaching a color generator matrix $\bm T_k^c$ to all of the parton lines $k$ in the state, including $k = \La$, gives zero. We have used this identity to add the same term, proportional to $z/[(1-z)\vartheta]$, to both the $k = a$ term and the $k \ne a$ terms in Eq.~(\ref{eq:D10a4}). We have added this term in both places in order to move the soft$\times$collinear singularity from the $k \ne a$ terms to the $k = a$ term. After this change, the $k \ne \La$ terms, proportional to $W_0$, have a soft singularity but not a collinear singularity.

We now turn to the splitting operator. The shower operator $\cU(t_2,t_1)$ is the ordered exponential of the integral over shower time $t$ of a splitting operator $\cS(t)$ according to Eq.~(\ref{eq:cUfromcS}). Consider the first order contribution to the splitting operator for splitting of initial state parton ``a,'' $\cS_\La^{[1]}(t)$. This operator consists of a real emission part, $\cS^{[1,0]}_\La(t)$, and a virtual exchange part $\cS^{[0,1]}_\La(t)$. The real emission part is determined from the derivative of $\cD^{[1,0]}_\La$ with respect to the shower time using Eq.~(\ref{eq:cS}) specialized to first order, which gives Eq.~(\ref{eq:cS10}). For the contribution from initial state parton ``a,'' this is
\begin{equation}
\begin{split}
\label{eq:cS10a}
\cS^{[1,0]}_\La(t) ={}& - 
\cF_\La(\mur^2(t))
\sum_i
\\&\times \frac{d\mu_{\scS,i}^2(t)}{dt}
\frac{\partial \cD^{[1,0]}_\La(\mur(t),\vec\mu_\scS(t))}{\partial \mu_{\scS,i}^2}
\\&\times
\cF^{-1}_\La(\mur^2(t))
\;.
\end{split}
\end{equation}

We differentiate $\cD^{[1,0]}_\La$ as given in Eq.~(\ref{eq:Dhatdef}). In Eq.~(\ref{eq:Dhatdef}), the ratio of PDFs corresponds to the PDF operators $\cF_{\!\La}(\mur^2) \cdots \cF_{\!\La}^{-1}(\mur^2)$ rather than $\cD^{[1,0]}_\La$, so we do not differentiate the PDF factor. There is a factor of $\as(\mur^2)$ that is part of $\cD^{[1,0]}_\La$, but the derivative of $\as$ with respect to its scale argument is of order $\as^2$ and we want only the first order contribution to $\cS_\La(t)$, so we do not differentiate the $\as$ factor. This leaves the derivative of $\hat {\bm D}^\La_{a \hat a}$, which is given in Eq.~(\ref{eq:D10a4}). The dependence on $\vec\mu_\scS(t)$ is contained in the factor $\Theta((z,\vartheta) \in U(\mu_\scC^2))$, so it is only the theta functions contained in this factor that we should differentiate, producing delta functions that fix the integration variable $\vartheta$ in $\hat {\bm D}^\La_{a \hat a}$. Then the dimensional regularization is not needed, so we can set $\epsilon = 0$. These considerations give one a straightforward calculation of $\cS^{[1,0]}(t)$. We omit further details.

The virtual exchange part, $\cS^{[0,1]}_\La(t)$, of $\cS^{[1]}_\La(t)$ leaves the number of partons and their momenta and flavors unchanged, but can change the color state of the partons. It has both a real part and an imaginary part. The imaginary part, $ \mi\pi\cS^{[0,1]}_{\mi\pi}(t)$, is analyzed in Sec.~\ref{sec:ImcD01} and given in Eq.~(\ref{eq:SiPiResult}). The real part of $\cS^{[0,1]}_\La(t)$ can be determined rather simply because the definition of $\cS(t)$ is arranged so that the shower operator $\cU(t_2,t_1)$ preserves probability. This implies that $\sbra{1} \cS_\La^{[1]}(t) = 0$.  Since $\sbra{1}\, \mi\pi\cS^{[0,1]}_{\mi\pi}(t) = 0$, if we define $\cS^{[0,1]}$ by $\cS^{[1]} = \cS^{[1,0]} + \cS^{[0,1]}$, this gives us
\begin{equation}
\sbra{1}\,\mathrm{Re}\,\cS_\La^{[0,1]}(t) = - \sbra{1}\,\cS_\La^{[1,0]}(t)
\;.
\end{equation}
This requirement, together with a convention on the color content of $\mathrm{Re}\,\cS_\La^{[0,1]}(t)$, is enough to determine $\mathrm{Re}\,\cS_\La^{[0,1]}(t)$ from $\cS_\La^{[1,0]}(t)$, as described in some detail in Secs.~IV and XVI.C of Ref.~\cite{NSLogSum}. We write the relation in the form defined in Ref.~\cite{NSLogSum},
\begin{equation}
\mathrm{Re}\,\cS_\La^{[0,1]}(t) = -\P{\cS_\La^{[1,0]}(t)}
\;.
\end{equation}
The mapping $\cS_\La^{[1,0]}(t) \to \iP{\cS_\La^{[1,0]}(t)}$ is straightforward. We start with ${\cal D}^{[1,0]}_\mathrm{a}$ with scales $\{\mur(t),\mu_\scC(t)\}$ as given by the combination of Eqs.~(\ref{eq:Dhatdef}) and (\ref{eq:D10a4}). We multiply by $\sbra{1}$ to form the inclusive probability and use $\sbrax{1_{p,f}}\sket{\{p,f\}_{m+1}} = 1$. Then the delta function in Eq.~(\ref{eq:D10a4}) eliminates the integration over $\{p,f\}_{m+1}$ in Eq.~(\ref{eq:Dhatdef}), leaving integrations over the splitting variables $\{z,\vartheta,\phi\}$. We differentiate the theta function $\Theta\big((z,\vartheta) \in U(\mu_\scC^2)\big)$ with respect to the shower time $t$ in order to obtain $\cS_\La^{[1,0]}$ from ${\cal D}^{[1,0]}_\mathrm{a}$ according to Eq.~(\ref{eq:cS10}). Then the dimensional regularization is no longer needed, so we set $\epsilon = 0$. Finally, the color operator $t_\La^\dagger \otimes t_k$ for a real emission is mapped into the operator $1 \otimes t_k\,t_\La^\dagger$ for the corresponding virtual exchange and $t_k^\dagger \otimes t_\La$ is mapped to  $t_\La t_k^\dagger \otimes 1$. The result is

\begin{widetext}
\begin{equation}
\begin{split}
\label{eq:D01A}
\P{\cS_\La^{[1,0]}(t)}&
\sket{\{p,f,c,c'\}_{m}}
\\
={}& 
-\frac{\as(\mur^2(t))}{2\pi}
\sum_{\hat a}\int_0^1\!\frac{dz}{z}\
\frac{
f_{\hat a/A}(\eta_{\La}/z,\mur^{2}(t)}
{f_{a/A}(\eta_{\La},\mur^{2}(t))}
\int_0^1\!\frac{d\vartheta}{\vartheta}\
\int\!\frac{d\phi}{2\pi}\
\frac{d}{dt}
\Theta\big((z,\vartheta) \in U(\mu_\scC^2(t))\big)\,
\\&\times
\sum_{k}\frac{1}{2}\bigg[
\theta(k = \La)\,
\frac{1}{N(a ,\hat a)}\, \hat P_{a \hat a}(z, \vartheta)
-
\theta(k\ne \La)\,
\delta_{a \hat a}\,
\frac{2}{1-z}\, 
W_0\!\left(\xi_{\La k}, z, \vartheta, \phi - \phi_k\right)
\bigg]
\\&\times
\left\{
1\otimes
t_k(f_k \to \hat f_k \!+\! \hat f_\mpone)
t^\dagger_\La(f_\La \to \hat f_\La\!+\!\hat f_\mpone)
+
t_\La(f_\La \to \hat f_\La \!+\! \hat f_\mpone)
t^\dagger_k(f_k \to \hat f_k \!+\! \hat f_\mpone)\otimes 1
\right\}
\\&\times
\sket{\{p,f,c,c'\}_{m}}
\;.
\end{split}
\end{equation}
This is the operator that appears in Eq.~(\ref{eq:cS1decomposition1}).
\end{widetext}

\section{Pole structure}
\label{sec:poles}

Inclusive splitting at first order is given in Eq.~(\ref{eq:1D10and01}) by an operator ${\bm P}^\La_{a a'}(z; \{p\}_m)$ that operates on the parton color space. This operator has two parts. The first, $\hat{\bm P}^\La_{a a'}(z; \{p\}_m)$ comes from real emission graphs, while the second, $\bm \Gamma_{a'}(\{p\}_m)$ comes from virtual exchange graphs:
\begin{equation}
\begin{split}
\label{eq:bmGammadefbis}
{\bm P}^\La_{a a'}(z; \{p\}_m)
={}&
\hat{\bm P}^\La_{a a'}(z; \{p\}_m)
\\&
+ \delta_{a a'}\,\delta(1-z)\,
\bm \Gamma_{a'}(\{p\}_m)
\;.
\end{split}
\end{equation}
In ${\bm P}^\La_{a a'}(z; \{p\}_m)$, there are infrared singularities in the integration over splitting variables $y,z,\phi$. The singularities are regulated by working in $4 - 2 \epsilon$ dimensions, so that integrals over $z$ of $\hat{\bm P}^\La_{a a'}(z; \{p\}_m)$ have poles $1/\epsilon$ and $1/\epsilon^2$. The integrals over $y,z,\phi$ are confined to the region of unresolved splittings set by a scale or scales $\vec \mu_\scS$.

For the virtual graphs that contribute to $\bm \Gamma_{a'}(\{p\}_m)$, there is some freedom available in defining a region of unresolved loop momenta that corresponds to the unresolved region for real emission graphs. However, the pole structure of $\bm \Gamma_{a'}(\{p\}_m)$ is independent of the definition of an unresolved region and is given by
\begin{align}
\label{eq:Gammapoles}
\notag
\Big[ \bm \Gamma_{a'}(\{p\}_m)& \Big]_\mathrm{poles}
\\={}&
-\left[\sum_a \int_0^1\! dz\ z\, \hat{\bm P}^\La_{a a'}(z; \{p\}_m)
\right]_\mathrm{poles}
.
\end{align}
This was stated in the main text as Eq.~(\ref{eq:polestructure}). The purpose of this appendix is to establish this result. 

The method of proof is straightforward. We know based on collinear factorization in QCD that the infrared finite operator $\cV$ defined in Eq.~(\ref{eq:cVdef}) has no poles. Each pole in $\bm \Gamma_{a'}(\{p\}_m)$ would contribute a corresponding pole to $\cV$, so the $\bm \Gamma_{a'}(\{p\}_m)$ poles must cancel poles from other contributions to $\cV$. Thus we can adopt the algebraically simple strategy of assuming Eq.~(\ref{eq:Gammapoles}) and showing that then
\begin{equation}
\label{eq:cVpoles}
\big[ \cV \big]_\mathrm{poles} = 0
\;.
\end{equation}
We omit an explicit demonstration that the structure of the coefficients of the poles requires that Eq.~(\ref{eq:Gammapoles}) is the {\em unique} solution to Eq.~(\ref{eq:cVpoles}). 

We assume Eq.~(\ref{eq:Gammapoles}) for the pole part of $\bm \Gamma_{a'}(\{p\}_m)$ and extend this to the finite part as $\epsilon \to 0$ by defining
\begin{equation}
\begin{split}
\label{eq:GammaMSR}
\bm \Gamma_{a'}(\{p\}_m) 
={}&
- \sum_a \int_0^1\! dz\ z\, \hat{\bm P}^\La_{a a'}(z; \{p\}_m)
\;.
\end{split}
\end{equation}
We then show that this leads to Eq.~(\ref{eq:cVpoles}). We call Eq.~(\ref{eq:GammaMSR}) the momentum-sum-rule (MSR) ansatz. Of course, Eq.~(\ref{eq:GammaMSR}) could be supplemented by adding to $\bm \Gamma_{a'}(\{p\}_m)$ any operator that has no poles.

In Eq.~(\ref{eq:bmGammadefbis}), the inclusive probability $\hat {\bm P}^\La_{a a'}$ for real emissions is related to $\cD^{[1,0]}$ in Eq.~(\ref{eq:1D10}). Its structure is given in Eq.~(\ref{eq:1PC}). We can rewrite this equation by changing the integration variable from $y$ to $\vartheta = z y/(1-z)$. We write the result in terms of three functions, $A^\mathrm{(R)}_{a a'}(z)$, $B^\mathrm{(R)}_{k, a'}(z)$, and $C^\mathrm{(R)}_{k, a'}(z;\xi_{\La k})$, where the superscript R denotes real emissions,
\begin{widetext}
\begin{equation}
\begin{split}
\label{eq:1PCmod}
\hat {\bm P}^\La_{a a'}(z;\{p\}_{m})
={}&
A^\mathrm{(R)}_{a a'}(z) [1 \otimes 1]
+
B^\mathrm{(R)}_{a a'}(z) [1 \otimes 1]
\\&
- \delta_{a a'}\,
\left[\frac{(1-z)^2}{z}\right]^{-\epsilon}
\frac{2}{1-z}\, 
\sum_{k\ne \La}\,
C^\mathrm{(R)}_{k, a'}(z;\xi_{\La k})\
\frac{1}{2}\Big\{
[\bm T_k \cdot \bm T_\La \otimes 1]
+[1 \otimes \bm T_k \cdot \bm T_\La]
\Big\}
\;.
\end{split}
\end{equation}
Here
\begin{equation}
\begin{split}
\label{eq:1PAterm}
A^\mathrm{(R)}_{a a'}(z)
={}&
\left(\frac{\mur^2}{Q^2}\right)^{\!\epsilon}
\frac{(4\pi)^\epsilon}{\Gamma(1-\epsilon)}\,
\left[\frac{(1-z)^2}{z}\right]^{-\epsilon}\,
\int_0^1\!\frac{d\vartheta}{\vartheta}\
\left[\vartheta (1-\vartheta)\right]^{-\epsilon}
\hat P_{a a'}(z)
\;,
\end{split}
\end{equation}
where $\hat P_{a a'}(z)$ is the familiar DGLAP kernel, equal to $\hat P_{a a'}(z,\vartheta,\epsilon)$ at $\vartheta = \epsilon = 0$. Next,
\begin{equation}
\begin{split}
\label{eq:1PBterm}
B^\mathrm{(R)}_{a a'}(z)
={}&
\left(\frac{\mur^2}{Q^2}\right)^{\!\epsilon}
\frac{(4\pi)^\epsilon}{\Gamma(1-\epsilon)}\,
\left[\frac{(1-z)^2}{z}\right]^{-\epsilon}\,
\int_0^1\!\frac{d\vartheta}{\vartheta}\
\left[\vartheta (1-\vartheta)\right]^{-\epsilon}
\left[\Theta\big((z,\vartheta) \in U(\vec \mu_\scS)\big)\,
\hat P_{a a'}(z,\vartheta,\epsilon)
- \hat P_{a a'}(z)\right]
\;.
\end{split}
\end{equation}
Finally,
\begin{equation}
\begin{split}
\label{eq:1PCterm}
C^\mathrm{(R)}_{k, a'}(z;\xi_{\La k}) 
={}& \left(\frac{\mur^2}{Q^2}\right)^{\!\epsilon}
\frac{(4\pi)^\epsilon}{\Gamma(1-\epsilon)}\,
\int_0^1\!\frac{d\vartheta}{\vartheta}\
\left[\vartheta (1-\vartheta)\right]^{-\epsilon}
\int\!\frac{d^{1-2\epsilon}\phi}{S(2-2\epsilon)}\
\Theta\big((z,\vartheta) \in U(\vec \mu_\scS)\big)\,
W_0\!\left(\xi_{\La k}, z, \vartheta, \phi - \phi_k\right)
\;.
\end{split}
\end{equation}
\end{widetext}

For the virtual exchange operator $\bm \Gamma_{a'}(\{p\}_m)$, we can define functions $A^\mathrm{(V)}_{\hat a}$, $B^\mathrm{(V)}_{\hat a}$, and $C^\mathrm{(V)}_k(\xi_{\La k})$, with superscripts V for virtual exchanges:
\begin{equation}
\begin{split}
\bm \Gamma_{a'}(\{&p\}_m) 
\\ ={}& A^\mathrm{(V)}_{a'} [1 \otimes 1]
+
B^\mathrm{(V)}_{a'} [1 \otimes 1]
\\& + \sum_{k\ne \La}\, C^\mathrm{(V)}_{k, a'}(\xi_{\La k})
\frac{1}{2}\Big\{
[\bm T_k \cdot \bm T_\La \otimes 1]
+[1 \otimes \bm T_k \cdot \bm T_\La]
\Big\}
.
\end{split}
\end{equation}
Here a virtual gluon exchange between the initial state parton ``a'' and another parton $k$ is represented by $C^\mathrm{(V)}_{k, a'}(\xi_{\La k})$. It has the same color structure as the $C^\mathrm{(R)}_{k, a'}(z;\xi_{\La k})$ term in $\hat{\bm P}^\La_{a a'}$.

We start our investigation with the third term in Eq.~(\ref{eq:1PCmod}) for $\hat {\bm P}^\La_{a a'}$, proportional to the function $C^\mathrm{(R)}_{k, a'}(z;\xi_{\La k})$. This term has a distinctive color structure, $\left\{ [\bm T_k \cdot \bm T_\La \otimes 1] + [1 \otimes \bm T_k \cdot \bm T_\La] \right\}$.  This term in $\hat {\bm P}^\La_{a a'}$ does not have a collinear, $\vartheta \to 0$, singularity because $W_0$ vanishes as $\vartheta \to 0$, as one can see directly from the expression (\ref{eq:W0}) for $W_0$. However it has a soft gluon singularity, as seen in the factor $1/(1-z)$. This singularity produces a pole $1/\epsilon$ after we integrate over $z$. For any smooth function $h(z)$ we have
\begin{equation}
\begin{split}
\label{eq:realBintegral}
\int_0^1\!dz\ h(z) \bigg[&\frac{(1-z)^2}{z}\bigg]^{-\epsilon}
\frac{2 C^\mathrm{(R)}_{k, a'}(z;\xi_{\La k})}{1-z}
\\
={}& - \frac{h(1)\, C^\mathrm{(R)}_{k, a'}(1;\xi_{\La k})}{\epsilon} 
+ \cO(\epsilon^0)
\;.
\end{split}
\end{equation}
The corresponding contribution to virtual exchange integrated against a smooth function $h(z)$ is
\begin{equation}
\label{eq:virtualBintegral}
\int_0^1\!dz\ h(z)\, \delta(1-z)\, C^\mathrm{(V)}_{k, a'}(\xi_{\La k}) = 
h(1)\, C^\mathrm{(V)}_{k, a'}(\xi_{\La k})
\;.
\end{equation}
In order for the two $1/\epsilon$ poles in Eqs.~(\ref{eq:realBintegral}) and (\ref{eq:virtualBintegral}) to cancel, we need
\begin{equation}
\label{eq:virtualpoleneeded}
C^\mathrm{(V)}_{k, a'}(\xi_{\La k}) = 
\frac{C^\mathrm{(R)}_{k, a'}(1; \xi_{\La k})}{\epsilon}
+ \cO(\epsilon^0)
\;.
\end{equation}

In order to obtain this relation, we use the definition given by the momentum sum rule:
\begin{equation}
\begin{split}
C^\mathrm{(V)}_{k, a'}(\xi_{\La k}) ={}& 
-\int_0^1\!d\bar z\
\left[\frac{(1-\bar z)^2}{\bar z}\right]^{-\epsilon} 
\frac{2\bar z}{1-\bar z}
\\&\times
C^\mathrm{(R)}_{k, a'}(\bar z; \xi_{\La k})
\;.
\end{split}
\end{equation}
Then we indeed satisfy Eq.~(\ref{eq:virtualpoleneeded}).

Of course, there are other ways to satisfy Eq.~(\ref{eq:virtualpoleneeded}). Indeed,  $C^\mathrm{(V)}_{k, a'}$ is expressed as a dimensionally regulated integral over a $4 - 2 \epsilon$ dimensional loop momentum $q$. We need to cut out the ultraviolet contributions to this integral using the shower scales $\vec \mu_\scS$ used for the real emission integrations. However we do this, the infrared pole $1/\epsilon$ will match between the real and virtual graphs. We have seen that if we simply define the virtual integral in terms of the corresponding real integral by using the momentum sum rule, the $1/\epsilon$ contributions will cancel as required. Then the momentum sum rule ansatz fixes the finite part of the virtual contribution as a function of $\vec \mu_\scS$.

We can now consider the second term, proportional to $B^\mathrm{(R)}_{a a'}(z)$. There is a factor
\begin{equation}
\Theta\big((z,\vartheta) \in U(\vec \mu_\scS)\big) = 1 
- \Theta\big((z,\vartheta) \notin U(\vec \mu_\scS)\big)
\;.
\end{equation}
The term $\Theta\big((z,\vartheta) \notin U(\vec \mu_\scS)\big)$ cannot contribute any $1/\epsilon$ poles because in the resolved region, $(z,\vartheta)$ can never reach the singular surfaces $\vartheta = 0$ or $(1-z) = 0$. This leaves $\hat P_{a a'}(z,\vartheta,\epsilon) - \hat P_{a a'}(z)$. The contributions to $\hat P_{a a'}(z,\vartheta,\epsilon)$ proportional to $\epsilon$ are completely nonsingular, as we see in Eq.~(\ref{eq:Pepsilon}), so these contributions cannot contribute any poles. The contributions proportional to $y = (1-z) \vartheta/z$ contain a factor $\vartheta$, which cancels the collinear singularity, and a factor $(1-z)$, which cancels the soft singularity. 

Thus the integral of $B^\mathrm{(R)}_{a a'}(z)$ over $z$ has no $1/\epsilon$ poles. There is a corresponding virtual contribution, $B^\mathrm{(V)}_{a a'}(z)$, but if we define it from $B^\mathrm{(R)}_{a a'}(z)$ according to the momentum sum rule, it will contain no poles also. We conclude that $B_{a a'}(z)$ is free of $1/\epsilon$ poles.

Finally, we consider the first term in Eq.~(\ref{eq:1PCmod}), proportional to $A^\mathrm{(R)}_{a a'}(z)$. We can perform the $\vartheta$ integration to give
\begin{equation}
\begin{split}
\label{eq:ARresult}
A^\mathrm{(R)}_{a a'}(z) ={}&
-\left(\frac{\mur^2}{Q^2}\right)^{\!\epsilon}
\frac{(4\pi)^\epsilon}{\Gamma(1-\epsilon)}\,
\left[\frac{1}{\epsilon} + \cO(\epsilon)\right]
Q^\mathrm{(R)}_{a a'}(z)
\;,
\end{split}
\end{equation}
where
\begin{equation}
Q^\mathrm{(R)}_{a a'}(z) = \left[\frac{(1-z)^2}{z}\right]^{-\epsilon}\,
\hat P_{a a'}(z)
\;.
\end{equation}

Caution is needed because $\hat P_{a a'}(z)$ has a $1/(1-z)$ singularity:
\begin{equation}
\hat P_{a a'}(z) = \delta_{a a'}\,\frac{2 C_a}{1-z} + \hat P_{a a'}^\mathrm{reg}(z)
\;,
\end{equation}
where $\hat P_{a a'}^\mathrm{reg}(z)$ is finite for $z \to 1$. Treating $Q^\mathrm{(R)}_{a a'}(z)$ as a distribution by integrating against a test function $h(z)$ that vanishes for $z \to 0$, we obtain
\begin{equation}
\begin{split}
\label{eq:QRresult}
Q^\mathrm{(R)}_{a a'}(z) ={}& 
-2C_a\,\delta_{a a'} \left[\frac{1}{2\epsilon} + 1\right]
\delta(1-z)
\\ &
+\delta_{a a'}\frac{1}{z}\left[\frac{2 C_a \,z}{1-z}\right]_+
+\hat P_{a a'}^\mathrm{reg}(z)
+\cO(\epsilon)
\;.
\end{split}
\end{equation}
Here we have used the dimensional regularization factor $[(1-z)^2/z]^{-\epsilon}$ in the singular term. Then instead of an unregulated $1/(1-z)$ singularity, we obtain a distribution that is well defined as long as $\epsilon \ne 0$, with $\delta(1-z)$ and $[z/(1-z)]_+$ terms. Overall, counting the $1/\epsilon$ in Eq.~(\ref{eq:ARresult}), we have $1/\epsilon^2$ and $1/\epsilon$ infrared singularities in $A^\mathrm{(R)}_{a a'}(z)$ when $\epsilon \to 0$.

We can now add the contributions from virtual graphs. We can use the momentum sum rule as an ansatz to define these contributions. Then we can check whether this ansatz works to remove all $1/\epsilon^n$ poles in a physical cross section. The momentum-sum-rule ansatz is to replace $Q^\mathrm{(R)}_{a a'}(z)$ by
\begin{equation}
Q_{a a'}(z) = Q^\mathrm{(R)}_{a a'}(z) + Q^\mathrm{(V)}_{a a'}(z)
\;,
\end{equation}
with
\begin{equation}
Q^\mathrm{(V)}_{a a'}(z)
= - \delta_{a a'}\delta(1-z) 
\sum_c \int_0^1\!d\bar z\ \bar z\,Q^\mathrm{(R)}_{c a'}(\bar z)
\;.
\end{equation}
This gives
\begin{equation}
\begin{split}
\label{eq:Qresult}
Q_{a a'}(z) ={}& 
\delta_{a a'}\frac{1}{z}\left[\frac{2 C_a \,z}{1-z}\right]_+
+\hat P_{a a'}^\mathrm{reg}(z)
\\&
- \delta_{a a'}\delta(1-z)
\sum_c \int_0^1\!d\bar z\ \bar z\,
\hat P_{c a'}^\mathrm{reg}(\bar z)
\\& +\cO(\epsilon)
\;.
\end{split}
\end{equation}
Note that the contribution proportional to $\delta(1-z)$ with a $1/\epsilon$ singularity has cancelled. Since $P_{a a'}(z)$ obeys the momentum sum rule, we have
\begin{equation}
\label{eq:QisP}
Q_{a a'}(z) = P_{a a'}(z) + \cO(\epsilon)
\;.
\end{equation}

We can now define $A_{a a'}(z) = A^\mathrm{(R)}_{a a'}(z) + A^\mathrm{(V)}_{a a'}(z)$ with the result that
\begin{equation}
\label{eq:Aresult}
A_{a a'}(z) = 
-\left(\frac{\mur^2}{Q^2}\right)^{\!\epsilon}
\frac{(4\pi)^\epsilon}{\Gamma(1-\epsilon)}\,
\left[\frac{1}{\epsilon} + \cO(\epsilon)\right]
P_{a a'}(z)
\;.
\end{equation}
Of course, we could also add a term proportional to $\delta(1-z)$ with no poles to $A^\mathrm{(V)}_{a a'}(z)$. The construction presented here determines only the pole structure of the contributions from virtual graphs to ${\bm P}^\La_{a a'}(z; \{p\}_m)$.

We can use Eq.~(\ref{eq:Aresult}) in Eq.~(\ref{eq:bmVstartbis}). For $\cV$ in Eq.~(\ref{eq:bmVstartbis}), counting the factorization subtraction, we obtain
\begin{equation}
\begin{split}
\frac{1}{\epsilon}\frac{(4\pi)^\epsilon}{\Gamma(1-\epsilon)}\,
P_{a a'}(z) &
+ A_{a a'}(z)
\\={}& 
- \log\left(\frac{\mur^2}{Q^2}\right)P_{a a'}(z)
\;.
\end{split}
\end{equation}
The $1/\epsilon$ singularity cancels from $\cV$.

We conclude that as long as the contribution to ${\bm P}^\La_{a a'}(z; \{p\}_m)$ from virtual graphs is defined using the momentum sum rule, we obtain
\begin{equation}
\label{eq:polestructurebis}
\left[\sum_a \int_0^1\! dz\ z\, {\bm P}^\La_{a a'}(z; \{p\}_m)
\right]_\mathrm{poles} = 0
\;,
\end{equation}
as required for the infrared finite operator $\cV$ in Eq.~(\ref{eq:cVdef}) to be free of infrared poles in the dimensionally regulated theory. This is the result reported in Eq.~(\ref{eq:polestructure}).

\section{Structure of $\cV$ at first order}
\label{sec:cV}

In this appendix, we examine the structure of the operator $\cV$ at first order in $\as$. We start with Eq.~(\ref{eq:cVdef}),
\begin{equation}
\begin{split}
\label{eq:cVdefbis}
\sbra{1}\cV(\mur^2,\vec \mu_\scS) ={}& 
\sbra{1}
\big[\cF(\mur^2)\circ\cK(\mur^2)\circ\cZ_F(\mur^2)\big]
\\&\times
\cD(\mur^2,\vec \mu_\scS)\cF^{-1}(\mur^2)
\;.
\end{split}
\end{equation}
The operators $\cK$, $\cZ_F$, and $\cD$ are simply unit operators at order $\as^0$. Thus at first order, the part of $\cV$ that applies to hadron A becomes
\begin{equation}
\begin{split}
\label{eq:cV1A}
\sbra{1}\cV^{[1]}_\La &(\mur^2,\vec \mu_\scS) 
\\={}& 
\sbra{1}
\big[\cF_{\!\La}(\mur^2)\circ\cK_\La^{[1]}(\mur^2)\big]
\cF_{\!\La}^{-1}(\mur^2)
\\&
+ \sbra{1}
\big[\cF_{\!\La}(\mur^2)\circ\cZ_{F,\La}^{[1]}(\mur^2)\big]
\cF^{-1}(\mur^2)
\\&
+ \sbra{1}\cF_{\!\La}(\mur^2)
\cD^{[1]}_\La(\mur^2,\vec \mu_\scS)\cF_{\!\La}^{-1}(\mur^2)
\;.
\end{split}
\end{equation}
The $\circ$ symbols in the first two terms stand for convolutions, as in Eq.~(63) of Ref.~\cite{NSAllOrder}, so that 
\begin{equation}
\begin{split}
\label{eq:cKform}
\big[\cF_{\!\La}(\mur^2)&\circ\cK^{[1]}_\La(\mur^2)\big]
\cF_{\!\La}^{-1}(\mur^2)
\\
={}& \sum_{a'}\int_0^1\!\frac{dz}{z}\,
\frac{f_{a'/A}(\eta_\La/z, \mur^2)}{f_{a/A}(\eta_\La, \mur^2)}\,
\frac{\as(\mur^2)}{2\pi} K_{a a'}(z,\mur^2)
\;.
\end{split}
\end{equation}
We are free to choose what $K_{a a'}(z,\mur^2)$ should be. In the second term, the factor in the $\MSbar$ scheme that removes the $1/\epsilon$ singularity from an initial state splitting is\footnote{This is in $4 - 2 \epsilon$ dimensions. Recall that $\as$ is $\epsilon$ dependent and has dimension $(\mathrm{mass})^{-2\epsilon}$ in the dimensionally regulated theory \cite{JCCbook}.}
\begin{equation}
\begin{split}
\label{eq:cZ1form}
\big[\cF_{\!\La}(\mur^2)&\circ\cZ_{F,\La}^{[1]}(\mur^2)\big]
\cF_{\!\La}^{-1}(\mur^2)
\\
={}& \sum_{a'}\int_0^1\!\frac{dz}{z}\,
\frac{f_{a'/A}(\eta_\La/z, \mur^2)}{f_{a/A}(\eta_\La, \mur^2)}
\\&\times
\frac{\as(\mur^2)}{2\pi}
\frac{1}{\epsilon}\frac{(4\pi)^\epsilon}{\Gamma(1-\epsilon)}\,
P_{a \hat a}(z)
\;.
\end{split}
\end{equation}
The operators in Eqs.~(\ref{eq:cKform}) and (\ref{eq:cZ1form}) are proportional to unit operators on the parton color space. In Eq.~(\ref{eq:1D10and01}), $\sbra{1}\cF_{\!\La}\cD^{[1]}_\La \cF_{\!\La}^{-1}$ is expressed as a convolution of the same PDF factor with an operator $\bm P^\La$ that acts nontrivially on the parton color space.  

We write $\cV^{[1]}_\La$ as a convolution with PDFs according to Eq.~(\ref{eq:cV1integral}),
\begin{equation}
\begin{split}
\label{eq:cV1integralbis}
\cV^{[1]}_\mathrm{a}(&\mur^2, \vec\mu_\scS)
\sket{\{p,f,c,c'\}_{m}}
\\={}&
\sket{\{p,f\}_{m}}
\frac{\as(\mur^2)}{2\pi}\,
\sum_{a'}
\int_0^1\!\frac{dz}{z}\,
\frac{
f_{a'/A}(\eta_{\La}/z,\mur^{2})}
{f_{a/A}(\eta_{\La},\mur^{2})}
\\&\times
{\bm V}^\La_{a a'}(z; \{p, f\}_m)
\sket{\{c,c'\}_m}
\;.
\end{split}
\end{equation}
This gives
\begin{equation}
\begin{split}
\label{eq:bmVstart}
{\bm V}^\La_{a a'}(z; \{p, f\}_m) ={}& 
\lim_{\epsilon \to 0}\big\{
K_{a a'}(z,\mur^2)
\\&
+ 
\frac{1}{\epsilon}\frac{(4\pi)^\epsilon}{\Gamma(1-\epsilon)}\,
P_{a \hat a}(z)
\\ & +
{\bm P}^\La_{a a'}(z;\{p\}_{m};\epsilon)
\big\}
\;.
\end{split}
\end{equation}

We now need to examine the structure of ${\bm P}^\La_{a a'}(z;\{p,f\}_{m};\epsilon)$ for $\epsilon  \to 0$. Since we define ${\bm P}^\La_{a a'}$ from its real emission part $\hat{\bm P}^\La_{a a'}$ using the momentum sum rule, Eq.~(\ref{eq:usingMSR}), we can begin with $\hat{\bm P}^\La_{a a'}$. For this, we can use Eq.~(\ref{eq:1PC}):
\begin{equation}
\begin{split}
\label{eq:1PCencore}
\hat {\bm P}^\La_{a a'}(z;&\{p\}_{m};\epsilon)
\\={}&
\frac{z^\epsilon}{(1-z)^{2\epsilon}}
\left[\frac{\mur^2}{Q^2}\right]^{\!\epsilon}
\frac{(4\pi)^\epsilon}{\Gamma(1-\epsilon)}
\int_0^1\!\frac{d\vartheta}{\vartheta}\
[\vartheta(1-\vartheta)]^{-\epsilon}
\\&\times
\Theta\big((z,\vartheta) \in U(\vec \mu_\scS)\big)
\\&\times
\bigg\{\hat P_{a a'}(z, \vartheta,\epsilon)
\\& \
- \int\!\frac{d^{1-2\epsilon}\phi}{S(2-2\epsilon)}
\sum_{k \ne \La}
\frac{2 \delta_{a a'}}{1-z}\, 
W_0\!\left(\xi_{\La k}, z, \vartheta, \phi - \phi_k\right)
\\&\quad \times
\frac{1}{2}\left\{
[\bm T_k \cdot \bm T_\La \otimes 1]
+[1 \otimes \bm T_k \cdot \bm T_\La]
\right\}
\bigg\}
\;.
\end{split}
\end{equation}

We define the unresolved region by saying that $(z,\vartheta) \in U(\vec \mu_\scS)$ if $0<z<1$, $0 < \vartheta < 1$, and
\begin{equation}
\vartheta < a_\mathrm{max}(z, \vec \mu_\scS)
\;,
\end{equation}
where $a_\mathrm{max}(z, \vec \mu_\scS)$ is the combination that appears in Eq.~(\ref{eq:acutamperp}):
\begin{equation}
\begin{split}
\label{eq:amaxdef}
a_\mathrm{max}(z, \vec \mu_\scS)
={}&
\max\left\{a_\mathrm{cut}(z,\vec\mu_\scS), 
a_\perp(z,m_\perp^2(a,\hat a)\right\}
\;.
\end{split}
\end{equation}
We also define
\begin{equation}
\begin{split}
\label{eq:alimitdef}
a_\mathrm{limit}&(z, \vec \mu_\scS)
\\={}&
\max\left\{a_\mathrm{lim}(z,\mu_\mathrm{lim}^2(\vec\mu_\scS)), 
a_\perp(z,m_\perp^2(a,\hat a)\right\}
\;.
\end{split}
\end{equation}
Here, in general, $\mu_\mathrm{lim}$ is a function $\mu_\mathrm{lim}(\vec\mu_\scS)$ of the scales $\vec\mu_\scS$. Then $a_\mathrm{lim}(z,\mu_\mathrm{lim}^2(\vec\mu_\scS(t))$ is the limiting form of $a_\mathrm{cut}(z,\vec\mu_\scS(t))$ at large shower times $t$, as defined in Eq.~(\ref{eq:ylimdef0}). 

With three scales, $\vec \mu_\scS = (\mu_\scE, \mu_\scC, \mu_{\mi\pi})$, the unresolved region for $\hat{\bm D}^{\mathrm{a, LC+}}_{a a'}$ and $\hat{\bm P}_{a a'}^\mathrm{LC+}$ is defined by Eq.~(\ref{eq:ylim1}) with 
\begin{equation}
a_\mathrm{cut}(z,\mu_\scE^2,\mu_\scC^2) = \max\left\{ a_\scE(z,\mu_\scE^2), a_\scC(z, \mu_\scC^2)\right\}
\;.
\end{equation}
On the chosen path shown in Fig.~\ref{fig:evolution-path}, $\mu_\scE$ decreases to zero first, then $\mu_\scE$ decreases to zero. Thus we define $\mu_\scC(\vec \mu_\scS) = \mu_\scC$, and  $a_\mathrm{lim}(z,\mu_\mathrm{lim}^2(\vec\mu_\scS)) = a_\scC(z,\mu_\scC^2)$ in Eq.~(\ref{eq:alimitdef}). This gives us
\begin{equation}
\begin{split}
\label{eq:cutsummary}
a_\mathrm{max}(z, \vec \mu_\scS)
={}&
\max\left\{a_\scE(z,\mu_\scE^2), a_\scC(z, \mu_\scC^2), 
a_\perp(z,m_\perp^2(a,\hat a)\right\}
,
\\
a_\mathrm{limit}(z, \vec \mu_\scS)
={}&
\max\left\{a_\scC(z, \mu_\scC^2), 
a_\perp(z,m_\perp^2(a,\hat a)\right\}
.
\end{split}
\end{equation}

In the term in Eq.~(\ref{eq:1PCencore}) proportional to $W_0$, we can set $\epsilon \to 0$. Then we can use the azimuthal average, $W(\xi, z, \vartheta)$, of $W_0$, defined in Eq.~(\ref{eq:Waveragedef}).

In the term in Eq.~(\ref{eq:1PCencore}) proportional to $\hat P_{a a'}$, we use the notation of Eq.~(\ref{eq:Paanotation}) to write $\hat P_{a a'}(z,\vartheta,\epsilon)$ in the form
\begin{equation}
\begin{split}
\hat P_{a a'}(z,\vartheta,\epsilon) ={}& 
\hat P_{a a'}(z) - \epsilon \hat P_{a a'}^{(\epsilon)}(z)
\\&
+ [\hat P_{a a'}(z,\vartheta,\epsilon) - \hat P_{a a'}(z,0,\epsilon)]
\\&
+ \cO(\epsilon^2)
\;.
\end{split}
\end{equation}
For the term $\hat P_{a a'}(z,\vartheta,\epsilon) - \hat P_{a a'}(z,0,\epsilon)$, it suffices to take the $\epsilon \to 0$ limit inside the integrations. For the first two terms, we need the integral
\begin{equation}
\begin{split}
I ={}& 
\left[\frac{z\mur^2}{(1-z)^2 Q^2}\right]^{\!\epsilon}
\frac{(4\pi)^\epsilon}{\Gamma(1-\epsilon)}
\int_0^1\!\frac{d\vartheta}{\vartheta} 
[\vartheta(1-\vartheta)]^{-\epsilon}
\\&\times
\theta(\vartheta < a_\mathrm{max}(z,\vec \mu_\scS))
\;.
\end{split}
\end{equation}
Performing the integrations gives
\begin{align}
\notag
I ={}& 
- \frac{1}{\epsilon}\,\frac{(4\pi)^\epsilon}{\Gamma(1-\epsilon)}
+ \log\left(\frac{(1-z)^2 Q^2}{z \mur^2}\,
a_\mathrm{max}(z,\vec \mu_\scS)\right)
\\ & 
- \int_1^{a_\mathrm{max}(z,\vec \mu_\scS)}\!\frac{d\vartheta}{\vartheta}\,
\theta(1 < \vartheta)
 + \cO(\epsilon)
 \;.
\end{align}

After some rearrangement, this gives us an infrared sensitive contribution and a non-sensitive (NS) contribution:
\begin{equation}
\begin{split}
\label{eq:hatbfPresult}
\hat {\bm P}^\La_{a a'}&(z;\{p\}_{m};\epsilon) 
\\={}& -\frac{1}{\epsilon}\frac{(4\pi)^\epsilon}{\Gamma(1-\epsilon)}\,
\hat P_{a a'}(z)
+ \hat P_{a a'}^{(\epsilon)}(z)
\\&
+ \hat P_{a a'}(z)\,
\log\!\left(\frac{(1-z)^2 Q^2}{z\mur^2}\,a_\mathrm{limit}(z, \vec \mu_\scS)\right)
\\&
+ \hat {\bm P}^\mathrm{a,NS}_{a a'}(z;\{p\}_{m})
+ \cO(\epsilon)
\;.
\end{split}
\end{equation}
The infrared nonsensitive contribution is
\begin{align}
\label{eq:hatbfPNS}
\notag
\hat {\bm P}^\mathrm{a,NS}_{a a'}&(z;\{p\}_{m})
\\\notag ={}& 
\int_0^{a_\mathrm{max}(z,\vec \mu_\scS)}
\!\frac{d\vartheta}{\vartheta}
\Big[\theta(a_\mathrm{limit}(z,\vec \mu_\scS)) < \vartheta )
-\theta(1 < \vartheta)\Big]
\\ \notag & \quad \times
\hat P_{a a'}(z)
\\\notag &+
\int_0^{a_\mathrm{max}(z,\vec \mu_\scS)}\!
\frac{d\vartheta}{\vartheta}\,
\theta(\vartheta < 1)\,
[\hat P_{a a'}(z,\vartheta) - \hat P_{a a'}(z)]
\\\notag &
- 
\int_0^{a_\mathrm{max}(z,\vec \mu_\scS)}\!\frac{d\vartheta}{\vartheta}\,
\theta(\vartheta < 1)
\sum_{k \ne \La}
\frac{2 \delta_{a a'}}{1-z}\,
W\!\left(\xi_{\La k}, z, \vartheta \right)
\\ &\quad \times
\frac{1}{2}\left\{
[\bm T_k \cdot \bm T_\La \otimes 1]
+[1 \otimes \bm T_k \cdot \bm T_\La]
\right\}
\;.
\end{align}
The first term here is $\hat P_{a a'}(z)$ times $[-\log(a_\mathrm{limit}) + \log(a_\mathrm{max})\theta(a_\mathrm{max}<1)]$. For our later purposes, we have written this as an integral.\footnote{The form used in Eq.~(\ref{eq:hatbfPNS}) assumes that $a_\mathrm{limit} \le a_\mathrm{max}$, as is the case when we use Eq.~(\ref{eq:cutsummary}). A slightly different form would apply if $a_\mathrm{limit} > a_\mathrm{max}$.}

Eq.~(\ref{eq:hatbfPresult}) gives $\hat {\bm P}^\La_{a a'}$, the part of ${\bm P}^\La_{a a'}$ associated with real parton splittings. To obtain the full ${\bm P}^\La_{a a'}$, we use the momentum sum rule, Eq.~(\ref{eq:usingMSR}):
\begin{align}
\label{eq:bfPresult}
\notag
{\bm P}^\La_{a a'}&(z;\{p\}_{m};\epsilon) 
\\ \notag ={}& -
\frac{1}{\epsilon}\frac{(4\pi)^\epsilon}{\Gamma(1-\epsilon)}\,
P_{a \hat a}(z)
+ \left[\hat P_{a a'}^{(\epsilon)}(z)\right]_\textsc{msr}
\\ \notag &
+ \left[\hat P_{a a'}(z)\,
\log\!\left(\frac{(1-z)^2 Q^2}{z \mur^2}\,a_\mathrm{limit}(z,\vec \mu_\scS)\right)
\right]_\textsc{msr}
\\  &
+ {\bm P}^\mathrm{a,NS}_{a a'}(z;\{p\}_{m})
+ \cO(\epsilon)
\;,
\end{align}
where
\begin{equation}
\label{eq:bmPNSmsr}
{\bm P}^\mathrm{a,NS}_{a a'}(z;\{p\}_{m})
=
\left[
\hat {\bm P}^\mathrm{a,NS}_{a a'}(z;\{p\}_{m})
\right]_\textsc{msr}
\;.
\end{equation}
Here we have noted that the DGLAP kernel obeys the momentum sum rule: $[\hat P_{a \hat a}(z)]_\textsc{msr} = P_{a \hat a}(z)$. We note from Eq.~(\ref{eq:Wlk-ini-result2}) that $W(\xi_{\La k}, z, \vartheta)$ is proportional to $\vartheta^2$ for $\vartheta \to 0$, so that the integration of $W$ over $\vartheta$ is not singular. The infrared nonsensitive contribution has no $1/\epsilon$ poles and vanishes in the limit of small scales $\vec \mu_\scS$ when $\big[ \hat {\bm P}^\mathrm{a,NS}_{a a'}(z; \{p\}_{m}) \big]_\textsc{msr}$ is integrated against a fixed test function $f(z)$.

We can now insert these results into Eq.~(\ref{eq:bmVstart}), giving
\begin{align}
\label{eq:bfVresultbis}
{\bm V}^\La_{a a'}&(z;\{p\}_{m}) 
\\ \notag ={}&
K_{a a'}(z,\mur^2)
+ \left[\hat P_{a a'}^{(\epsilon)}(z)\right]_\textsc{msr}
\\ \notag &
+ \left[\hat P_{a a'}(z)\,
\log\!\left(
\frac{(1-z)^2 Q^2}{z\mur^2}\,
a_\mathrm{limit}(z,\vec \mu_\scS)) 
\right)\right]_\textsc{msr}
\\ \notag&
+ {\bm P}^\mathrm{a,NS}_{a a'}(z;\{p\}_{m})
\;.
\end{align}
The contribution proportional to $P_{a a'}(z)/\epsilon$ in Eq.~(\ref{eq:bmVstart}) has cancelled an identical singular term in ${\bm P}^\La_{a a'}(z;\{p\}_{m};\epsilon)$ in Eq.~(\ref{eq:bfPresult}). This is the result reported in Eq.~(\ref{eq:bfVresult}) in the main text. We now define $K_{a a'}(z,\mur^2)$ so that it cancels the second and third terms in Eq.~(\ref{eq:bfVresultbis}), leaving
\begin{equation}
\label{eq:vmVisvmPNS}
{\bm V}^\La_{a a'}(z;\{p\}_{m}) =
{\bm P}^\mathrm{a,NS}_{a a'}(z;\{p\}_{m})
\;.
\end{equation}
This is the result reported in Eq.~(\ref{eq:bfVresult2}) in the main text.

\section{Structure of $\cS_\cV$ at first order}
\label{sec:cScV}

We use the operator $\cV(\mur(t),\vec\mu_\scS(t))$ to define the threshold operator $\cU_\cV(t_2, t_1)$ according to Eqs.~(\ref{eq:cUcV}) and (\ref{eq:cScV}):
\begin{equation}
\begin{split}
\label{eq:cUcVencore}
\cU_\cV(t_2,& t_1) 
\\
={}& 
\cV^{-1}(\mur(t_2),\vec\mu_\scS(t_2))
\cV(\mur(t_1),\vec\mu_\scS(t_1))
\\
={}& \mathbb{T}\exp\!\left(\int_{t_1}^{t_2}\!
dt\, \cS_\cV(t)\right)
\;,
\end{split}
\end{equation}
where
\begin{equation}
\label{eq:cScVencore}
\cS_\cV(t)
= 
-\cV^{-1}(\mur(t),\vec\mu_\scS(t)) 
\frac{d}{d t}
\cV(\mur(t),\vec\mu_\scS(t))
\;.
\end{equation}

We let $t = 0$ be the shower time at the start of the shower, corresponding to a renormalization scale parameter $\mu_\scR = \mu_\scH$, where $\mu_\scH$ is comparable to the scale of the hard interaction. Then $t_\Lf$ is the shower time at the end of the shower. Thus we consider
\begin{equation}
\begin{split}
\label{eq:cUcV0tf}
\cU_\cV(t_\Lf, 0) 
={}& \mathbb{T}\exp\!\left(\int_{0}^{t_\Lf}\!
dt\, \cS_\cV(t)\right)
\;.
\end{split}
\end{equation}

At first order,
\begin{equation}
\label{eq:cScV1storder}
\cS_\cV^{[1]}(t)
= 
- 
\frac{d}{d t}
\cV^{[1]}(\mur(t),\vec\mu_\scS(t))
+ \cO(\as^2)
\;.
\end{equation}
There are two contributions to $\cS_\cV^{[1]}(t)$, corresponding to emissions from partons a and b, respectively:
\begin{equation}
\label{eq:cScVab}
\cS_\cV^{[1]}(t) = \cS_{\cV,\La}^{[1]}(t) + \cS_{\cV,\Lb}^{[1]}(t)
\;.
\end{equation}

If we use the three segment path from $t = 0$ to $t = t_\Lf = 3$, then the $\mathbb{T}$ instruction tells us to break $\cU_\cV(3, 0)$ into $\cU_\cV(3, 2)\,\cU_\cV(2, 1)\,\cU_\cV(1, 0)$. We saw in Sec.~\ref{sec:crosssection} that $\cU_\cV(3,2) = 1$. Next, $\cU_\cV(2,1)$ is calculated using the LC+ approximation. No $\mathbb{T}$ instruction is needed within $\cU_\cV(2,1)$ because all of the operators commute. Finally $\cU_\cV(1, 0)$ is determined by the difference between the full color operators and their LC+ approximate version. This contribution is subleading in threshold logarithms and subleading in color. It is not simple to treat $\cU_\cV(1, 0)$ in computer code. We leave that to future work. For the leading order Drell-Yan process investigated in this paper, the parton state just after the hard interaction is very simple and this color difference operator applied to this state gives zero, so that $\cU_\cV(1, 0) = 1$. 

In this appendix, we examine $\cV^{[1]}_\mathrm{a}$ with just one scale $\mu_\scC^2$, which is what we need for the second segment of the path. To evaluate $\cV^{[1]}_\mathrm{a}$, we can start with Eq.~(\ref{eq:cV1integralbis}). We understand $\cV^{[1]}_\mathrm{a}$ to be applied to a state $\sket{\{p,f,c,c'\}_{m}}$ but omit writing out the state. Eq.~(\ref{eq:cV1integralbis}) expresses $\cV^{[1]}_\mathrm{a}$ in terms of the operator ${\bm V}^\La_{a a'}(z;\{p\}_{m})$, which equals ${\bm P}^\mathrm{a,NS}_{a a'}(z;\{p\}_{m})$ according to Eq.~(\ref{eq:vmVisvmPNS}). For ${\bm P}^\mathrm{a,NS}_{a a'}(z;\{p\}_{m})$, we use Eq.~(\ref{eq:hatbfPNS}). Since we consider the second phase of evolution of $\cS_\cV$, we take $\mu_\scE^2 = 0$. Additionally, although an infrared cutoff is needed, $\cV^{[1,0]}_\mathrm{a}$ is not sensitive to the infrared cutoff used. For that reason, we simplify the notation by setting $m_\perp^2(a,\hat a) = 0$ for now. Then $a_\mathrm{max} = a_\mathrm{limit} = a_\scC(z, \mu_\scC^2)$. We set $\mur^{2} =  \mu_\scC^2$. This gives for $\cV^{[1]}_\mathrm{a}$,
\begin{align}
\label{eq:cV10}
\cV^{[1]}_\mathrm{a}(&\mu_\scC^2)
\\ \notag ={}&
\frac{\as(\mu_\scC^2)}{2\pi}\,
\sum_{a'}
\int_0^1\!\frac{dz}{z}\,
\frac{
f_{a'/A}(\eta_{\La}/z,\mu_\scC^{2})}
{f_{a/A}(\eta_{\La},\mu_\scC^{2})}
\\\notag &\times
\Bigg[
\int_0^{a_\scC(z, \mu_\scC^2)}
\!\frac{d\vartheta}{\vartheta}
\Big\{
-\theta(\vartheta > 1)\hat P_{a a'}(z)
\\\notag &+
\theta(\vartheta < 1)\,
[\hat P_{a a'}(z,\vartheta) - \hat P_{a a'}(z)]
\\\notag &
- 
\theta(\vartheta < 1)
\sum_{k \ne \La}
\frac{2 \delta_{a a'}}{1-z}\,
W\!\left(\xi_{\La k}, z, \vartheta \right)
\\ \notag &\quad \times
\frac{1}{2}\left\{
[\bm T_k \cdot \bm T_\La \otimes 1]
+[1 \otimes \bm T_k \cdot \bm T_\La]
\right\}
\Big\}
\Bigg]_\mathrm{MSR}
.
\end{align}
On the second segment of our chosen path, we use the LC+ approximation for the color operators in Eq.~(\ref{eq:cV10}), but we do not specify that choice here.

We need the derivative of this with respect to $t$:
\begin{equation}
\label{eq:cScV1storderbis}
\cS_\cV^{[1]}(t)
= 
- 
\frac{d\mu_\scC^2(t)}{d t}\
\frac{d}{d \mu_\scC^2}\,
\cV^{[1]}(\mu_\scC^2)
+ \cO(\as^2)
\;.
\end{equation}
When we differentiate with respect to $\mu_\scC^2$, we differentiate with respect to the upper endpoint of the $\vartheta$ integral, but we do not differentiate $\as(\mu_\scC^2)$, $f_{a'/A}(\eta_{\La}/z,\mu_\scC^{2})$, or $f_{a/A}(\eta_{\La},\mu_\scC^{2})$ because the derivatives of these objects are of order $\as$ and we want only the first order contribution to $\cS_\cV^{[1]}(t)$. Then, using the fact that $a_\scC(z, \mu_\scC^2) \propto \mu_\scC^2$, we obtain
\begin{align}
\label{eq:cScV10A}
\notag
\cS^{[1]}_{\cV,\mathrm{a}}(&t)
\\ \notag ={}&
\frac{1}{\mu_\scC^2(t)}
\frac{d\mu_\scC^2(t)}{dt}
\\ \notag &\times
\sum_{a'}
\int_0^1\!\frac{dz}{z}\,
\frac{
f_{a'/A}(\eta_{\La}/z,\mu_\scC^{2}(t))}
{f_{a/A}(\eta_{\La},\mu_\scC^{2}(t))}\,
\frac{\as(\mu_\scC(t)^2)}{2\pi}
\\\notag &\times
\Bigg[
\theta(\vartheta(z,t) > 1)\hat P_{a a'}(z)
\\\notag &
\quad -
\theta(\vartheta(z,t) < 1)\,
[\hat P_{a a'}(z,\vartheta(z,t)) - \hat P_{a a'}(z)]
\\\notag &
\quad + 
\theta(\vartheta(z,t) < 1)
\sum_{k \ne \La}
\frac{2 \delta_{a a'}}{1-z}\,
W\!\left(\xi_{\La k}, z, \vartheta(z,t) \right)
\\ &\quad \times
\frac{1}{2}\left\{
[\bm T_k \cdot \bm T_\La \otimes 1]
+[1 \otimes \bm T_k \cdot \bm T_\La]
\right\}
\Bigg]_\mathrm{MSR}
\;.
\end{align}
In Eq.~(\ref{eq:cScV10A}), we evaluate $\vartheta$ as
\begin{equation}
\vartheta(z,t) = a_\scC(z, \mu_\scC^2(t))
\;.
\end{equation}

The operator $\cU_\cV(t_\Lf, 0)$ has the effect of summing threshold logarithms. In order to improve this summation, we can add some contributions to $\cS^{[1]}_{\cV,\mathrm{a}}(t)$ that are higher order in $\as$ and thus beyond the order of approximation that is controlled in our derivation. We replace
\begin{equation}
\label{eq:newas}
\as(\mu_\scC^2) \to \as^{(K_\Lg)}(k_\LT^2)\,
\theta(k_\LT^2 > m_\perp^2(a,\hat a))
\;,
\end{equation}
where $k_\LT^2$ is the squared transverse momentum associated with an initial state parton splitting,
\begin{equation}
k_\LT^2(z,t) = \frac{(1-z)^2}{z}\,Q^2\,\vartheta(z,t)
\;,
\end{equation}
as in Eq.~(\ref{eq:thetatoyandktsq}), and
\begin{equation}
\begin{split}
\label{eq:asKgbis}
\as^{(K_\Lg)}(\mu^2) = {}&
\as(\mu^2)\left[
1 + K_\Lg \frac{\as(\mu^2)}{2\pi}
\right]
\;,
\end{split}
\end{equation}
as in Eq.~(\ref{eq:asKg}). Here $K_\Lg$ is the standard factor \cite{CMW},
\begin{equation}
\label{eq:Kgbis}
K_\Lg = C_\LA \frac{67 - 3 \pi^2}{18} - T_\LR \frac{10 n_\Lf}{9}
\;,
\end{equation}
with $n_\Lf$ being the number of active quark flavors at scale $\mu^2$. The theta function in Eq.~(\ref{eq:newas}) is included to avoid reaching a singularity in $\as$. It also prevents reaching a point where the parton distribution functions are not smooth. Using Eq.~(\ref{eq:cScV10A}) with these enhancements, we have
\begin{align}
\label{eq:cScV10B}
\notag
\cS^{[1]}_{\cV,\mathrm{a}}(&t)
\\ \notag ={}&
\frac{1}{\mu_\scC^2(t)}
\frac{d\mu_\scC^2(t)}{dt}
\sum_{a'}
\int_0^1\!\frac{dz}{z}\,
\frac{
f_{a'/A}(\eta_{\La}/z,\mu_\scC^{2}(t))}
{f_{a/A}(\eta_{\La},\mu_\scC^{2}(t))}
\\ \notag &\times
\frac{\as^{(K_\Lg)}(k_\LT^2(z,t))}{2\pi}\,
\theta(k_\LT^2(z,t) > m_\perp^2(a,\hat a))
\\\notag &\times
\Bigg[
\theta(\vartheta(z,t) > 1)\hat P_{a a'}(z)
\\\notag &
\quad -
\theta(\vartheta(z,t) < 1)\,
[\hat P_{a a'}(z,\vartheta(z,t)) - \hat P_{a a'}(z)]
\\\notag &
\quad + 
\theta(\vartheta(z,t) < 1)
\sum_{k \ne \La}
\frac{2 \delta_{a a'}}{1-z}\,
W\!\left(\xi_{\La k}, z, \vartheta(z,t) \right)
\\ &\quad \times
\frac{1}{2}\left\{
[\bm T_k \cdot \bm T_\La \otimes 1]
+[1 \otimes \bm T_k \cdot \bm T_\La]
\right\}
\Bigg]_\mathrm{MSR}
.
\end{align}
%

\section{Summation of threshold logs}
\label{sec:thresholdsum}

In the shower cross section repeated in Eq.~(\ref{eq:sigmaJshower}) from Ref.~\cite{NSAllOrder}, the first two operator factors applied to the hard scattering statistical state $\sket{\rho_\scH}$ are $\cF(\muh^2)$ and $\cU_\cV(t_\Lf, t_{\scH})$. (For application to the second segment of our three segment path, the $\cU_\cV$ factor would be $\cU_\cV(2, 1)$.) The operator $\cU_\cV(t_\Lf, t_{\scH})$ leaves the number of partons and their momenta and flavors unchanged but multiplies by certain factors that relate to how the shower splitting functions match to the evolution of parton distribution functions \cite{ISevolve}. The operator $\cF(\muh^2)$ multiplies by some standard normalization factors and by parton distribution functions $f_{a/A}(\eta_\La,\muh^2)$ and $f_{b/B}(\eta_\Lb,\muh^2)$. These are shower oriented parton distribution functions, which can be different from the normal $\MSbar$ PDFs. 

When the cross section to be computed involves a very hard scattering, so that $\eta_\La$ and $\eta_\Lb$ are close to 1, the operator $\cU_\cV(t_\Lf, t_{\scH})$ applied to $\sket{\rho_\scH}$ can be large. Additionally, the ratios $f_{a/A}(\eta_\La,\muh^2)/f_{a/A}^{\msbar}(\eta_\La,\muh^2)$ and $f_{b/B}(\eta_\Lb,\muh^2)/f_{b/B}^{\msbar}(\eta_\Lb,\muh^2)$ can be large. In this case, there are big effects because Eq.~(\ref{eq:sigmaJshower}) is summing what are called threshold logarithms. The summation of threshold logarithms was first analyzed by Sterman in 1987 \cite{Sterman1987}. There is a large literature on this summation using analytical calculations.\footnote{There has been substantial recent interest in the summation of large logarithms by parton shower algorithms \cite{Hoeche2018, Dasgupta2018, angular2020, DasguptaShowerSum, HamiltonShowerSum, NSLogSum}. However, to our knowledge the summation of threshold logarithms is not included in parton shower algorithms except for \textsc{Deductor} \cite{NSThreshold, NSAllOrder, NSThresholdII}.} Ref.~\cite{NSThreshold} cites some of this literature. In an analytical approach, one normally analyzes a suitable Mellin transform of the cross section. Then the transformed cross section is a function of a Mellin moment variable $N$ and we are interested in the limit $N\to \infty$. In this appendix, we apply the Mellin transform approach to the operators that appear in one of the simple single scale versions of the shower cross section introduced in Sec.~\ref{sec:simpleordering}. Our aim is to compare the parton shower version of threshold summation to the analytical results, extending the less general analysis in Ref.~\cite{NSThreshold}.

We find that $\cU_\cV(t_\Lf,  t_\scH)$ becomes an exponential
\begin{equation}
\cU_\cV(t_\Lf, t_\scH) \to 
\exp(\cE_{\cV,\La}(N) + \cE_{\cV,\Lb}(N))
\;,
\end{equation}
where $\mathcal{E}_{\cV,\La}(N)$ and $\mathcal{E}_{\cV,\Lb}(N)$ can be expanded in a series of powers of $\as(\muh^2)$ and powers of $\log(N)$. Similarly, the PDF factors become exponentials
\begin{equation}
\begin{split}
\label{eq:pdffactor}
\frac{f_{a/A}(\eta_\La,\muh^2)}{f_{a/A}^{\msbar}(\eta_\La,\muh^2)}
\to {}& \exp(\cE_\mathrm{pdf,a}(N))
\;,
\\
\frac{f_{b/B}(\eta_\La,\muh^2)}{f_{b/B}^{\msbar}(\eta_\La,\muh^2)}
\to {}& \exp(\cE_\mathrm{pdf,b}(N))
\;.
\end{split}
\end{equation}
The leading approximation obtained with analytical approaches takes the form
\begin{equation}
\mathrm{threshold\ factor} = 
\exp(\cE_{\La}(N) + \cE_{\Lb}(N))
\;.
\end{equation}
With the shower approach in this appendix, we find
\begin{equation}
\begin{split}
\cE_{\La}(N) ={}& 
\cE_{\cV,\La}(N) + \cE_\mathrm{pdf,a}(N)
\;,
\\
\cE_{\Lb}(N) ={}& 
\cE_{\cV,\Lb}(N) + \cE_\mathrm{pdf,b}(N)
\;.
\end{split}
\end{equation}

We begin with the operator $\cU_\cV(t_\Lf, 0)$. Using Eqs.~(\ref{eq:cUcV0tf}) and (\ref{eq:cScVab}), we write
\begin{equation}
\cU_\cV(t_\Lf, 0) =  \mathbb{T}\exp\!\left(
\cE_{\cV,\La} + \cE_{\cV,\Lb} \right)
\end{equation}
with contributions from the two initial state parton legs, ``a'' and ``b.'' For initial parton leg ``a,'' we have
\begin{equation}
\cE_{\cV,\La} = \int_{t_\scH}^{t_\Lf}\!dt\ \cS_{\cV,\La}(t)
\;,
\end{equation}
where $\cS_{\cV,\La}(t)$ is given by Eq.~(\ref{eq:cScV10B}). In Eq.~(\ref{eq:cScV10B}), the unresolved region is defined using scale parameters $\mu_\scC^2$, where $\LC = \perp$ for $k_T$ ordering (Eq.~(\ref{eq:aPerpdef})), $\LC = \Lambda$ for $\Lambda$ ordering (Eq.~(\ref{eq:aLambdadef})), and $\LC = \angle$ for $\vartheta$ ordering (Eq.~(\ref{eq:aAngledef})).

For the purpose of summing the leading threshold logarithms, the only term that matters in $\cS_{\cV,\La}$, in Eq.~(\ref{eq:cScV10B}), is the first, proportional to $\theta(\vartheta(z,t) > 1) \hat P_{a a'}(z)$. Furthermore, it suffices to approximate $\hat P_{a a'}(z)$ by its most singular term,
\begin{equation}
\hat P_{a a'}(z)
\to \delta_{a a'} \frac{2 C_a}{1-z}
\;.
\end{equation}
We choose the argument of $\as$ to be $k_\LT^2 = (1-z)^2 \vartheta Q^2/z$ as in Eq.~(\ref{eq:newas}), but we can omit the $\as^2$ term in Eq.~(\ref{eq:asKgbis}). Additionally, we can set the infrared cutoffs $m_\perp^2(a,\hat a)$ and $m_\perp^2$ to zero with the understanding that we work term by term in the expansion of $\as(k_\LT^2)$ in powers of $\as(\mu_\scH^2)$, where $\mu_\scH^2$ is the scale of the hard scattering. We use $k_\LT^2$ as the integration variable instead of $t$. The theta function in the first term in Eq.~(\ref{eq:cScV10B}) limits the integration to $1 < \vartheta < a_\scC(z,\muh^2)$ or
\begin{equation}
\label{eq:kTcuts}
\frac{(1-z)^2 Q^2}{z}
< k_\LT^2 < 
\frac{(1-z)^2 Q^2}{z}\,a_\scC(z,\muh^2)
\;.
\end{equation}
We identify the square of the momentum of the partons just after the hard scattering, $Q^2$, with $\mu_\scH^2$. For the leading behavior of the integral, we are interested only in the $(1-z)\to 0$ behavior of the integrand. Thus we can replace the factor $1/z$ in Eq.~(\ref{eq:kTcuts}) and the factor $z$ in Eq.~(\ref{eq:aLambdadef}) for $a_\Lambda(z,\mu^2)$ by just 1. We can also replace the factor $r_\La = \eta_\La/\eta_\La^{(0)}$ in Eq.~(\ref{eq:aLambdadef}) for $a_\Lambda(z,\mu^2)$ by just 1 because in Eq.~(\ref{eq:sigmaJshower}), $\cU_\cV(t_\Lf, 0)$ is applied to the hard scattering state $\sket{\rho_\scH}$, for which  $\eta_\La = \eta_\La^{(0)}$. 

With these replacements, the cuts in Eq.~(\ref{eq:kTcuts}) become
\begin{equation}
\label{eq:kTcutsII}
(1-z)^2 \mu_\scH^2
< k_\LT^2 < 
(1-z)^n \mu_\scH^2
\end{equation}
with
\begin{equation}
n = 
\begin{cases}
0 & k_\LT\ \mathrm{ordering} \\
1 & \Lambda\ \mathrm{ordering} \\
2 & \vartheta\ \mathrm{ordering}
\end{cases}
\;.
\end{equation}
This gives us
\begin{equation}
\begin{split}
\cE_{\cV,\La} ={}& 
-\int_0^1\!dz\
\frac{
f_{a/A}(\eta_{\La}/z,\mu_\scH^{2})}
{z f_{a/A}(\eta_{\La},\mu_\scH^{2})}
\\ &\times
\Bigg[\int^{(1-z)^n \mu_\scH^2}_{(1-z)^2 \muh^2}\!\frac{dk_\LT^2}{k_\LT}\ 
\frac{2 C_a}{1-z}
\frac{\as(k_\LT^2)}{2\pi}
\Bigg]_\mathrm{MSR}
\;.
\end{split}
\end{equation}
We can write the MSR instruction explicitly as a subtraction at $z = 1$ using the definition (\ref{eq:MSR2}). This gives
\begin{equation}
\begin{split}
\cE_{\cV,\La} ={}& 
-\int_0^1\!dz\
\left[\frac{
f_{a/A}(\eta_{\La}/z,\mu_\scH^{2})}
{z^2 f_{a/A}(\eta_{\La},\mu_\scH^{2})} - 1\right]
\frac{2 z C_a}{1-z}
\\ &\times
\int^{(1-z)^n\muh^2}_{(1-z)^2 \muh^2}\!\frac{dk_\LT^2}{k_\LT^2}\ 
\frac{\as(k_\LT^2)}{2\pi}
\;.
\end{split}
\end{equation}

In order to compare to standard analytical results that work with a Mellin transform of the cross section, we need to relate this to the Mellin transformed parton distributions. We can use the ``single power approximation'' \cite{} in which we replace  
\begin{equation}
\frac{f_{a/A}(\eta_{\La}/z,\mu_\scH^{2})}
{f_{a/A}(\eta_{\La},\mu_\scH^{2})}
\to z^N
\;.
\end{equation}
This gives us an exponent
\begin{equation}
\begin{split}
\mathcal{E}_{\cV,\La}(N) ={}& 
-\int_0^1\!dz\
\left[z^{N-2} - 1\right]
\frac{2 z C_a}{1-z}
\\  &\times
\int^{(1-z)^n\muh^2}_{(1-z)^2 \muh^2}\!\frac{dk_\LT^2}{k_\LT^2}\ 
\frac{\as(k_\LT^2)}{2\pi}
\;.
\end{split}
\end{equation}

Now we can look at the PDF factor, Eq.~(\ref{eq:pdffactor}). We use the Mellin transform of the parton distribution, at the scale $\muh^2$ of the hard scattering that starts the shower, as a function of the parameter $\lambda$ in Sec.~\ref{sec:PDFs},
\begin{equation}
\begin{split}
g_{a/A}(N,\muh^2;\lambda) ={}& \int_0^1\!\frac{d\eta}{\eta}\ \eta^N\,
f_{a/A}(\eta,\muh^2;\lambda)
\;.
\end{split}
\end{equation}
We need the Mellin transform of the shower oriented PDF, $f_{a/A}(\eta,\mu^2;1)$ with $\lambda = 1$. The shower oriented PDF with $\lambda = 0$ corresponds to $k_\LT$ ordering and differs from the $\MSbar$ PDF only by a small adjustment specified in Eq.~(\ref{eq:Kdef0}). We obtain $f_{a/A}(\eta,\muh^2;1)$ from $f_{a/A}(\eta,\muh^2;0)$ by solving the differential equation (\ref{eq:dtildefdlambda}) with the first order kernel $d\widetilde K/d\lambda$ given by Eq.~(\ref{eq:lambdageneratorA}) for $\Lambda$ ordering or Eq.~(\ref{eq:A1angular}) for $\vartheta$ ordering.

We are interested only in the leading threshold logarithms in $g_{a/A}(N,\mu^2;1)$, so we make some small adjustments in the differential equation, as we did in the previous analysis of $\cU_\cV$. We set the fixed infrared cutoff to $m_\perp^2(a,a') = 0$. We replace the DGLAP kernel in the differential equation by its leading singularity as $z \to 1$,
\begin{equation}
\hat P_{a a'}(z)
\to \delta_{a a'} \frac{2 C_a}{1-z}
\;.
\end{equation}
Since only the $(1-z) \to 0$ limit affects the threshold logarithms, we replace $z^A(1-z)^B$ by just $(1-z)^B$ in the differential equation. For $\Lambda$ ordering, there is a factor $r_\La$ in the scale choice that relates $f_{a/A}(\eta,\muh^2;\lambda)$ to an adjusted function $\tilde f$. Since $r_\La$ is just 1 in the hard scattering state for which the PDF is measured, we set $r_\La = 1$. The argument of $\as$ in the first order kernel is not determined by strictly lowest order considerations. In order to incorporate leading logarithms beyond leading order in $\as(\muh^2)$, we follow the choice generally made in the \textsc{Deductor} code and take the argument of $\as$ to be
\begin{equation}
\label{eq:PDFkT2}
k_\LT^2 = [(1-z)^n]^\lambda \muh^2
\;.
\end{equation}
Here we understand that $\as(k_\LT^2)$ is to be expanded in powers of $\as(\mu_\scH^2)$. With these adjustments, the differential equation for $g_{a/A}(N,\muh^2;\lambda)$ is
\begin{align}
\frac{dg_{a/A}(N,\muh^2;\lambda)}{d\lambda} \hskip - 1.5 cm&
\\ \notag = {}& g_{a/A}(N,\muh^2;\lambda)
\int_0^1\!\frac{dz}{z}\,z^N
\\ \notag &\times
\Bigg[
\frac{\as((1-z)^{n\lambda} \muh^2)}{2\pi}\,
\frac{2 C_a}{1-z}\,
\log((1-z)^n)
\Bigg]_\mathrm{MSR}
\;.
\end{align}

The solution of this is
\begin{align}
g_{a/A}(N,&\muh^2;1)
\\ \notag = {}& g_{a/A}(N,\muh^2;0)
\exp\Bigg\{
\int_0^1\!\frac{dz}{z}\,z^N
\Bigg[\int_0^1\! d\lambda
\\ \notag &\times
\frac{\as((1-z)^{n\lambda} \muh^2)}{2\pi}\,
\frac{2 C_a}{1-z}\,
\log((1-z)^n)
\Bigg]_\mathrm{MSR}
\Bigg\}
\;.
\end{align}
We can change integration variables from $\lambda$ to $k_\LT^2$ specified in Eq.~(\ref{eq:PDFkT2}):
\begin{align}
g_{a/A}(N,&\muh^2;1)
\\ \notag = {}& g_{a/A}(N,\muh^2;0)
\exp\Bigg\{
-\int_0^1\!\frac{dz}{z}\,z^N
\\ \notag &\times
\Bigg[\int_{(1-z)^n \muh^2}^{\muh^2}\! \frac{dk_\LT^2}{k_\LT^2}
\frac{\as(k_\LT^2)}{2\pi}\,
\frac{2 C_a}{1-z}\,
\Bigg]_\mathrm{MSR}
\Bigg\}
\;.
\end{align}
We can write the MSR instruction explicitly as a subtraction at $z = 1$ using the definition (\ref{eq:MSR2}). This gives
\begin{equation}
g_{a/A}(N,\muh^2;1) = g_{a/A}(N,\muh^2;0)
\exp[\mathcal{E}_\mathrm{pdf,a}(N)]
\;,
\end{equation}
where
\begin{equation}
\begin{split}
\mathcal{E}_\mathrm{pdf,a}(N)
= {}&
-\int_0^1\!dz\,\left[z^{N-2} - 1\right] 
\frac{2 z C_a}{1-z}
\\  &\times
\int_{(1-z)^n \muh^2}^{\muh^2}\! \frac{dk_\LT^2}{k_\LT^2}
\frac{\as(k_\LT^2)}{2\pi}\,
\;.
\end{split}
\end{equation}

The sum of the $\cU_\cV$ and PDF contributions is
\begin{align}
\mathcal{E}_{\cV,\La}(&N) + \mathcal{E}_\mathrm{pdf,a}(N)
\\ \notag
={}& 
-\int_0^1\!dz \left[z^{N-2} - 1\right] 
\frac{2 z C_a}{1-z}
\int_{(1-z)^2 \muh^2}^{\muh^2}\! \frac{dk_\LT^2}{k_\LT^2}
\frac{\as(k_\LT^2)}{2\pi}
.
\end{align}
Notice that the net result is the same for $k_\LT$ ordering ($n = 0$), $\Lambda$ ordering ($n = 1$), and $\vartheta$ ordering ($n = 2$). For $k_\LT$ ordering, the entire result comes from $\mathcal{E}_{\cV,\La}(N)$. For $\vartheta$ ordering, the entire result comes from $\mathcal{E}_\mathrm{pdf,a}(N)$. For $\Lambda$ ordering, both $\mathcal{E}_{\cV,\La}(N)$ and $\mathcal{E}_\mathrm{pdf,a}(N)$ contribute.

The comparable standard formula given in Ref.~\cite{StermanZeng} is
\begin{align}
\mathcal{E}_{\La}(N)&
\\ \notag
={}& 
-\int_0^1\!dz  \left[z^{N-1} - 1\right] 
\frac{2 C_a}{1-z}
\int_{(1-z)^2 \muh^2}^{\muh^2}\! \frac{dk_\LT^2}{k_\LT^2}
\frac{\as(k_\LT^2)}{2\pi}
.
\end{align}
Here we have $z^{N-1}$ instead of $z^{N-2}$ and $2 C_a$ instead of $2 z C_a$, but these differences are not important in the $N \to \infty$ limit.

\section{Structure of operators with more scales}
\label{sec:twoscales}

In this appendix, we provide details about the operators $\hat{\bm D}^\La_{a a'}$ and $\hat{\bm P}^\La_{a a'}$ needed in Sec.~\ref{sec:softsplittings} when we use three scales, $(\mu_\scE^2, \mu_\scC^2, \mu_{\mi\pi}^2)$. Here $\mu_\scE^2$ specifies a cut on $(1-z)$ according to Eq.~(\ref{eq:aEdef}) and $\mu_\scC^2$ specifies the unresolved region for $k_\LT$ ordering for $\mathrm{C} = \perp$ (Eq.~(\ref{eq:aPerpdef})) or $\Lambda$ ordering for $\mathrm{C} = \Lambda$ (Eq.~(\ref{eq:aLambdadef})) or $\vartheta Q^2$ ordering for $\mathrm{C} = \angle$ (Eq.~(\ref{eq:aAngledef})). We do not need to consider the scale $\mu_{\mi\pi}^2$ since $\hat{\bm D}^\La_{a a'}$ and $\hat{\bm P}^\La_{a a'}$ do not depend on $\mu_{\mi\pi}^2$, which controls $\mathrm{Im}\,\cD^{[0,1]}$.

We first need to specify in detail the two versions of the unresolved region defined in Sec.~\ref{sec:softsplittings}: the region $U(\mu_\scE, \mu_\scC)$ that we use for  $\hat{\bm D}^{\mathrm{a, LC+}}_{a a'}$ and $\hat{\bm P}_{a a'}^\mathrm{LC+}$ and the  region $U(\mathrm{soft};\mu_\scE)$ that we use for  $\hat{\bm D}^{\mathrm{a, soft}}_{a a'}$ and $\hat{\bm P}_{a a'}^\mathrm{soft}$.

\subsection{$U(\mu_\mathrm{E}, \mu_\mathrm{C})$}
\label{sec:cUmuEmuC}

With scales $\mu_\scE$ and $\mu_\scC$, the unresolved region is defined in Eq.~(\ref{eq:ylim1}) and Eq.~(\ref{eqYcutdef}): $(z,\vartheta) \in U(\mu_\scE, \mu_\scC)$ if $0<z<1$, $0 < \vartheta < 1$ and
\begin{equation}
\label{eq:varthetabound}
\vartheta < a_\mathrm{max}(z, \mu_\scE^2,\mu_\scC^2)
\;,
\end{equation}
where we define
\begin{align}
\label{eq:amaxdefagain}
\notag
a_\mathrm{max}(z, &\mu_\scE^2,\mu_\scC^2)
\\ ={}&
\max\left\{a_\mathrm{cut}(z,\mu_\scE^2,\mu_\scC^2), 
a_\perp(z,m_\perp^2(a,\hat a)\right\}
\end{align}
with
\begin{equation}
\label{eqYcutdefebcore}
a_\mathrm{cut}(z,\mu_\scE^2,\mu_\scC^2) =
\max\left\{
a_\scE(z,\mu_\scE^2),
a_\scC(z, \mu_\scC^2)\right\}
\;.
\end{equation}
When both scales $\mu_\scE^2$ and $\mu_\scC^2$ tend to zero, the limiting form of $a_\mathrm{cut}(z,\mu_\scE^2,\mu_\scC^2)$ at fixed $z$ is $a_\scC(z, \mu_\scC^2)$. That is,  $a_\scC(z, \mu_\scC^2)$ is the limiting function $a_\mathrm{lim}$  defining the unresolved region with small scales, as described in Eq.~(\ref{eq:ylimdef0}). Thus the function $a_\mathrm{limit}(z, \vec \mu_\scS)$ defined in Eq.~(\ref{eq:alimdef}) is
\begin{equation}
\begin{split}
\label{eq:alimexample}
a_\mathrm{limit}(z,\mu_\scC^2)
={}&
\max\left\{a_\scC(z,\mu_\scC^2), 
a_\perp(z,m_\perp^2(a,a')\right\}
\;,
\end{split}
\end{equation}
as in Eq.~(\ref{eq:cutsummary}). Consistently with our choices in Eqs.~(\ref{eq:aPerpdef}), (\ref{eq:aLambdadef}), and (\ref{eq:aAngledef}), we take the renormalization scale to be
\begin{equation}
\label{eq:murchoice}
\mur^2(\mu_\scE^2, \mu_\scC^2) = \mu_\scC^2
\;.
\end{equation}

We use the path $(\mu_\scE^2(t), \mu_\scC^2(t), \mu_{\mi\pi}^2(t))$ specified in Sec.~\ref{sec:pathchoice}. On the second segment of the path, we have $\mu_\scE^2(t) = 0$ and $a_\scE(z,0) = 0$, so Eq.~(\ref{eq:varthetabound}) becomes
\begin{equation}
\label{eq:alimsegment2}
\vartheta <
\max\!\left\{
a_\scC(z, \muh^2),
a_\perp(z,m_\perp^2(a,\hat a))\right\}
\;.
\end{equation}
This gives us just the evolution for a $k_\LT$-, $\Lambda$-, or $\vartheta$-ordered shower according to our choice of $\LC \in \{\perp, \Lambda, \angle\}$.

For the first segment of the path, $0 < t < 1$, some analysis is needed. On this path segment, $\mu_\scE^2$ is evaluated at a variable scale with $0 < \mu_\scE^2(t) < \mu_\scE^2(0)$, while $\mu_\scC^2$ is evaluated at the fixed scale $\mu_\scC^2(0) = \muh^2$. Then Eq.~(\ref{eq:varthetabound}) becomes
\begin{equation}
\vartheta < a_\mathrm{total}(z,\mu_\scE^2(t))
\;,
\end{equation}
where
\begin{align}
a_\mathrm{total}&(z,\mu_\scE^2(t)) 
\\ \notag ={}&
\min\!\Big[ 1,
\\ \notag &\
\max\!\left\{
a_\scE(z,\mu_\scE^2(t)),
a_\scC(z, \muh^2),
a_\perp(z,m_\perp^2(a,\hat a))\right\}\!
\Big]
\;,
\end{align}
as in Eq.~(\ref{eq:atotaldef}). We have taken the minimum of $a_\mathrm{max}$ and 1 because $\vartheta$ is at most 1, so that values of these functions that are greater than 1 do not affect $U(\mu_\textsc{e}, \mu_\scC)$.

Choose a limiting value $z_\scH$ of $z$ such that
\begin{equation}
a_\scC(z, \muh^2)  > 1
\end{equation}
for $1 - z < 1 - z_\scH$. For $\LC = \angle$, we can simply take $1 - z_\scH = 1$. For $\LC = \Lambda$ or $\LC = \perp$, we can let $z_\scH$ be the solution of $a_\scC(z_\scH, \muh^2)  = 1$ as in Eq.~(\ref{eq:zHdef}). Then 
\begin{equation}
a_\mathrm{total}(z,\mu_\scE^2(t))
= 1
\end{equation}
for $1-z < 1 - z_\scH$ and for all $(a,\hat a)$.

We now define the starting $\mu_\scE$ scale by Eq.~(\ref{eq:muE0}),
\begin{equation}
\mu_\scE^2(0) = (1-z_\scH)\,Q_0^2
\;.
\end{equation}
Using the definition (\ref{eq:aEdef}) we have $a_\scE(z,\mu_\scE^2(t)) = 0$ for $(1-z) > \mu_\scE^2(t)/Q^2$. But $\mu_\scE^2(t) < \mu_\scE^2(0)$ and $Q^2 > Q_0^2$, so also $a_\scE(z,\mu_\scE^2(t)) = 0$ for  $(1-z) > \mu_\scE^2(0)/Q_0^2$. That is, 
\begin{equation}
a_\scE(z,\mu_\scE^2(t)) = 0\, \mathrm{\ for\ }
(1-z) > (1-z_\scH)
\;.
\end{equation}
This gives us
\begin{equation}
\begin{split}
a_\mathrm{total}&(z,\mu_\scE^2(t)) 
\\
={}&
\min\!\big[ 1,
\max\!\left\{
a_\scC(z, \muh^2),
a_\perp(z,m_\perp^2(a,\hat a))\right\}\!
\big]
\end{split}
\end{equation}
for $1 - z > 1 - z_\scH$. That is
\begin{equation}
a_\mathrm{total}(z,\mu_\scE^2(t)) 
=
a_\mathrm{total}(z,\mu_\scE^2(0))
\end{equation}
for $1 - z > 1 - z_\scH$. Thus

\begin{equation}
\label{eq:atotalresult}
a_\mathrm{total}(z,\mu_\scE^2(t))
= 
\begin{cases}
1 & 1-z < 1 - z_\scH \\
a_\mathrm{total}(z,\mu_\scE^2(0)) & 
1 - z_\scH < 1-z
\end{cases}
\;,
\end{equation}
as was reported in Eq.~(\ref{eq:atotal}). The important point about Eq.~(\ref{eq:atotalresult}) is that $a_\mathrm{total}(z,\mu_\scE^2(t))$ is independent of $t$ along the first segment of the path.

\subsection{$U(\mathrm{soft}, \mu_\mathrm{E})$}
\label{sec:cUsoft}

We have $(z,\vartheta) \in U(\mathrm{soft}, \mu_\scE)$ if 
\begin{equation}
\vartheta < a_\mathrm{total}^\mathrm{soft}(z,\mu_\scE^2(t))
\;,
\end{equation}
where $a_\mathrm{total}^\mathrm{soft}$ is obtained from $a_\mathrm{total}$ by omitting $a_\scC(z, \muh^2)$:
\begin{align}
a_\mathrm{total}^\mathrm{soft}&(z,\mu_\scE^2(t)) 
\\ \notag ={}&
\min\!\Big[ 1,
\max\!\left\{
a_\scE(z,\mu_\scE^2(t)),
a_\perp(z,m_\perp^2(a,\hat a))\right\}\!
\Big]
.
\end{align}
Using Eq.~(\ref{eq:aEdef}) for $a_\scE(z,\mu_\scE^2)$ and Eq.~(\ref{eq:aPerpdef}) for $a_\perp(z,k_\LT^2)$, this is
\begin{align}
\label{eq:atotalsoft}
\notag
a_\mathrm{total}^\mathrm{soft}(z,\mu_\scE) ={}& 
\theta((1-z) Q^2 < \mu_\scE^2)
\\&\notag 
+
\theta(\mu_\scE^2 < (1-z) Q^2)
\\  &\quad\times
\min\left\{1, \frac{z\, m_\perp^2(a,\hat a)}{(1-z)^2Q^2}\right\}
 \;.
\end{align}

We need $a_\mathrm{total}^\mathrm{soft}(z,\mu_\scE(t))$ on two segments of the path specified in Sec.~\ref{sec:pathchoice}. For $1 < t < 2$, we have $\mu_\scE = 0$, so we have a fixed unresolved region given by $a_\mathrm{total}^\mathrm{soft}(z,0)$. For $0 < t < 1$, $\mu_\scE(t)$ varies and we can use Eq.~(\ref{eq:atotalsoft}) for $a_\mathrm{total}^\mathrm{soft}(z,\mu_\scE(t))$.

\subsection{Decomposition of $\hat {\bm P}^\mathrm{a,NS}_{a a'}$ and $\hat {\bm D}^\La_{\hat a a}$}
\label{sec:bmP}

We now record the details of the operators $\hat {\bm P}^\mathrm{a,NS}_{a a'}$ and $\hat {\bm D}^\La_{\hat a a}$ when we use two scales, $\mu_\scE$ and $\mu_\scC$, that affect the definition of the unresolved region. We define $a_\mathrm{max}(z,\mu_\scE^2,\mu_\scC^2)$ and $a_\mathrm{limit}(z,\mu_\scE^2,\mu_\scC^2)$ using Eq.~(\ref{eq:cutsummary}).

We use Eq.~(\ref{eq:hatbfPresult}) to obtain $\hat {\bm P}^\La_{a a'}$. In the third term on the right-hand-side of Eq.~(\ref{eq:hatbfPresult}), we use Eq.~(\ref{eq:cutsummary}). In the last term, we write\footnote{Note that we have dropped ``NS'' in the superscripts on the right hand side of this equation in order to avoid clutter.}
\begin{equation}
\begin{split}
\hat {\bm P}^\mathrm{a,NS}_{a a'}(z;\{p\}_{m})
={}&
\hat {\bm P}^\mathrm{a,LC+}_{a a'}(z;\{p\}_{m})
\\&+ 
\hat {\bm P}^\mathrm{a,soft}_{a a'}( z;\{p\}_{m})
\;.
\end{split}
\end{equation}
Then $\hat {\bm P}^\mathrm{a,LC+}_{a a'}$ is given by Eq.~(\ref{eq:hatbfPNS}) with the LC+ approximation applied to the color operator in the third term:

\begin{widetext}
\begin{align}
\label{eq:hatbfPNSLCplus}
\hat {\bm P}^\mathrm{a,LC+}_{a a'}(z;\{p\}_{m})
={}& 
\int_0^{a_\mathrm{max}(z,\mu_\scE^2,\mu_\scC^2)}\!
\frac{d\vartheta}{\vartheta }\,
\Big[\theta(a_\mathrm{limit}(z,\mu_\scE^2,\mu_\scC^2)) < \vartheta )
-\theta(1 < \vartheta)\Big]\,
\hat P_{a a'}(z)
\\\notag &+
\int_0^{a_\mathrm{max}(z,\mu_\scE^2,\mu_\scC^2)}\!
\frac{d\vartheta}{\vartheta}\,
\theta(\vartheta < 1)
[\hat P_{a a'}(z,\vartheta) - \hat P_{a a'}(z)]
\\\notag &
- \int_0^{a_\mathrm{max}(z,\mu_\scE^2,\mu_\scC^2) }\!\frac{d\vartheta}{\vartheta}\,
\theta(\vartheta < 1)
\sum_{k \ne \La}
\frac{\delta_{a a'}}{1-z}\,
W\!\left(\xi_{\La k}, z, \vartheta \right)
\left\{
[\bm T_k \cdot \bm T_\La \otimes 1]
+[1 \otimes \bm T_k \cdot \bm T_\La]
\right\}_\mathrm{LC+}
.
\end{align}
The operator $\hat {\bm P}^\mathrm{a,LC+}_{a a'}(z;\{p\}_{m})$ multiplies $\as(\mur^2)$. In order to provide an improved summation of large logarithms, when this formula is applied in \textsc{Deductor}, $\as$ is placed inside the $\vartheta$ integration and replaced by $\as^{(K_\Lg)}(k_\LT^2)\,\theta(k_\LT^2 > m_\perp^2)$, with $k_\LT^2 = (1-z)^2 \vartheta Q^2/z$, as in Eq.~(\ref{eq:asKg}). The same replacement applies for other formulas in this section.

For $\hat {\bm P}^\mathrm{a,soft}_{a a'}$, we take the difference between the $W$ term in $\hat {\bm P}^\mathrm{a,NS}_{a a'}$ with the exact color operator and the same term with the LC+ approximation for the color operator. Then we use the unresolved region specified by Eq.~(\ref{eq:ylim2}):
\begin{equation}
\vartheta < a_\mathrm{max}^\mathrm{soft}(z,\mu_\scE^2)
\end{equation}
with
\begin{equation}
\label{eq:asoftmax}
a_\mathrm{max}^\mathrm{soft}(z,\mu_\scE^2)
=
\max\left\{
a_\scE(z,\mu_\scE^2),
a_\perp(z,m_\perp^2(a,a'))\right\}
.
\end{equation}
This gives us 
\begin{align}
\label{eq:hatbfPNSsoft}
\hat {\bm P}^\mathrm{a,soft}_{a a'}(z;\{p\}_{m})
={}&
- \sum_{k \ne \La}
\frac{\delta_{a a'}}{1-z}\,
\int_0^{a_\mathrm{max}^\mathrm{soft}(z,\mu_\scE^2)}\!\frac{d\vartheta}{\vartheta}\,
\theta(\vartheta < 1)\,
W(\xi_{\La k}, z, \vartheta)
\\\notag & \quad\times
\big[\left\{
[\bm T_k \cdot \bm T_\La \otimes 1]
+[1 \otimes \bm T_k \cdot \bm T_\La]
\right\}
-
\left\{
[\bm T_k \cdot \bm T_\La \otimes 1]
+[1 \otimes \bm T_k \cdot \bm T_\La]
\right\}_\mathrm{LC+}\big]
\;.
\end{align}

From Eq.~(\ref{eq:D10a4}), the derivative of the operator $\hat {\bm D}^\La_{\hat a a}$ in four dimensions but using two scales, $\mu_\scE$ and $\mu_\scC$, is, with the LC+ approximation,
\begin{equation}
\begin{split}
\label{eq:D10aLCplus}
\frac{d}{dt}
\hat {\bm D}^{\mathrm{a, LC+}}_{a \hat a}(z;&\{\hat p,\hat f\}_\mpone,\{p,f\}_{m};\vec \mu_\scS(t))
\\={}&
\int_0^1\!\frac{d\vartheta}{\vartheta}
\int\!\frac{d\phi}{2\pi}\,
\delta\big(\{\hat p, \hat f\}_\mpone - R_\La(z,\vartheta,\phi, 
 \hat a;\{p,f\}_{m})\big)\,
\frac{d}{dt}\Theta\big((z,\vartheta) \in U(\mu_\scE^2(t),\mu_\scC^2(t))\big)\,
\\&\times
\sum_{k}\bigg[
\theta(k = \La)\,
\frac{1}{N(a ,\hat a)}\, \hat P_{a \hat a}(z, \vartheta)
-\theta(k\ne \La)\,
\delta_{a \hat a}\,
\frac{2}{1-z}\, 
W_0\!\left(\xi_{\La k}, z, \vartheta, \phi - \phi_k\right)
\bigg]
\\&\times
\frac{1}{2}\left\{
t^\dagger_\La(f_\La \to \hat f_\La\!+\!\hat f_\mpone)\otimes
t_k(f_k \to \hat f_k \!+\! \hat f_\mpone)
+
t^\dagger_k(f_k \to \hat f_k \!+\! \hat f_\mpone)\otimes
t_\La(f_\La \to \hat f_\La \!+\! \hat f_\mpone)
\right\}_\mathrm{LC+}
.
\end{split}
\end{equation}
We define 
\begin{equation}
\hat {\bm D}^\mathrm{a}_{a a'}
=
\hat {\bm D}^\mathrm{a,LC+}_{a a'}
+ 
\hat {\bm D}^\mathrm{a,soft}_{a a'}
\;,
\end{equation}
where $\hat {\bm D}^\mathrm{a,soft}_{a a'}$ uses the difference of the color operators with full color and with the LC+ approximation. Then $\hat {\bm D}^\mathrm{a,soft}_{a a'}$ has soft singularities but no collinear singularities, so we define the unresolved region for this operator using $U(\mathrm{soft},\mu_\scE^2(t))$. Thus
\begin{equation}
\begin{split}
\label{eq:D10soft}
\frac{d}{dt}\hat {\bm D}^{\mathrm{a, soft}}_{a \hat a}(& z;\{\hat p,\hat f\}_\mpone,\{p,f\}_{m};\vec \mu_\scS(t))
\\={}&
- \int_0^1\!\frac{d\vartheta}{\vartheta}
\int_0^{2\pi}\!\frac{d\phi}{2\pi}\
\delta\big(\{\hat p, \hat f\}_\mpone 
- R_\La(z,\vartheta,\phi, 
 \hat a;\{p,f\}_{m})\big)\,
\frac{d}{dt}\Theta\big((z,\vartheta) \in U(\mathrm{soft},\mu_\scE^2(t)\big)
\\&\times
\sum_{k \ne \La}
\frac{\delta_{a \hat a}}{1-z}\, 
W_0\!\left(\xi_{\La k}, z, \vartheta, \phi - \phi_k\right)
\\&\times\Big[
\big\{
t^\dagger_\La(f_\La \!\to\! \hat f_\La\!+\!\hat f_\mpone)\otimes
t_k(f_k \!\to\! \hat f_k \!+\! \hat f_\mpone)
+ t^\dagger_k(f_k \!\to\! \hat f_k \!+\! \hat f_\mpone)\otimes
t_\La(f_\La \!\to\! \hat f_\La \!+\! \hat f_\mpone)
\big\}
\\&\quad
-\big\{
t^\dagger_\La(f_\La \!\to\! \hat f_\La\!+\!\hat f_\mpone)\otimes
t_k(f_k \!\to\! \hat f_k \!+\! \hat f_\mpone)
+ t^\dagger_k(f_k \!\to\! \hat f_k \!+\! \hat f_\mpone)\otimes
t_\La(f_\La \!\to\! \hat f_\La \!+\! \hat f_\mpone)
\big\}_\mathrm{LC+}
\Big]
.
\end{split}
\end{equation}
\end{widetext}





\begin{thebibliography}{99}

\bibitem{angleFS} 
  Z.~Nagy and D.~E.~Soper,
  {\em Multivariable evolution in final state parton shower algorithms},
  \href{http://dx.doi.org/10.1103/PhysRevD.105.054012}
  {Phys. Rev. D \textbf{105}, 054012 (2022)}
  [\href{http://inspirehep.net/search?p=find+doi+10.1103/PhysRevD.105.054012}
  {\textsc{inSPIRE}}].

\bibitem{NSAllOrder} 
  Z.~Nagy and D.~E.~Soper,
  {\em What is a parton shower?},
  \href{http://dx.doi.org/10.1103/PhysRevD.98.014034}
  {Phys.\ Rev.\ D {\bf 98}, 014034 (2018)}
  [\href{http://inspirehep.net/search?p=find+doi+10.1103/PhysRevD.98.014034}
  {\textsc{inSPIRE}}].

\bibitem{NSI}
  Z.~Nagy and D.~E.~Soper,
  {\em Parton showers with quantum interference},
  \href{http://dx.doi.org/10.1088/1126-6708/2007/09/114}
  {JHEP {\bf 0709} (2007) 114}
  [\href{http://inspirehep.net/search?p=find+doi+10.1088/1126-6708/2007/09/114}
  {\textsc{inSPIRE}}].
  
\bibitem{NScolor}
  Z.~Nagy and D.~E.~Soper,
  {\em Parton shower evolution with subleading color},
  \href{http://dx.doi.org/10.1007/JHEP06(2012)044}
  {JHEP {\bf 1206} (2012) 044}
  [\href{http://inspirehep.net/search?p=find+doi+10.1007/JHEP06(2012)044}
  {\textsc{inSPIRE}}].

\bibitem{JCCspin}
  J.~C.~Collins,
  {\em Spin correlations in Monte Carlo event generators},
  \href{http://dx.doi.org/10.1016/0550-3213(88)90654-2}
  {Nucl.\ Phys.\  B {\bf 304}, 794 (1988)}
  [\href{http://inspirehep.net/search?p=find+doi+10.1016/0550-3213(88)90654-2}
  {\textsc{inSPIRE}}].

\bibitem{NSspin}
  Z.~Nagy and D.~E.~Soper,
  {\em Parton showers with quantum interference: Leading color, with spin},
  \href{http://dx.doi.org/10.1088/1126-6708/2008/07/025}
  {JHEP \textbf{07}, 025 (2008)}
  [\href{http://inspirehep.net/search?p=find+doi+10.1088/1126-6708/2008/07/025}
  {\textsc{inSPIRE}}].

\bibitem{NSLogSum}
  Z.~Nagy and D.~E.~Soper,
  {\em Summations of large logarithms by parton showers},
  \href{http://dx.doi.org/10.1103/PhysRevD.104.054049}
  {Phys. Rev. D \textbf{104}, 054049 (2021)}
  [\href{http://inspirehep.net/search?p=find+doi+10.1103/PhysRevD.104.054049}
  {\textsc{inSPIRE}}].
  
\bibitem{ISevolve}
  Z.~Nagy and D.~E.~Soper,
  {\em Evolution of parton showers and parton distribution functions},
  \href{http://dx.doi.org/10.1103/PhysRevD.102.014025}
  {Phys. Rev. D \textbf{102}, 014025 (2020)}
  [\href{http://inspirehep.net/search?p=find+doi+10.1103/PhysRevD.102.014025}
  {\textsc{inSPIRE}}].
  
\bibitem{NSThreshold}
  Z.~Nagy and D.~E.~Soper,
  {\em Summing threshold logs in a parton shower},
  \href{http://dx.doi.org/10.1007/JHEP10(2016)019}
  {JHEP {\bf 1610} (2016) 019}
  [\href{http://inspirehep.net/search?p=find+doi+10.1007/JHEP10(2016)019}
  {\textsc{inSPIRE}}].
   
\bibitem{NSThresholdII}
  Z.~Nagy and D.~E.~Soper,
  {\em Jets and threshold summation in Deductor},
  \href{http://dx.doi.org/10.1103/PhysRevD.98.014035}
  {Phys.\ Rev.\ D {\bf 98}, 014035 (2018)}
  [\href{http://inspirehep.net/search?p=find+doi+10.1103/PhysRevD.98.014035}
  {\textsc{inSPIRE}}].

\bibitem{NSColoriPi}
Z.~Nagy and D.~E.~Soper,
{\em Exponentiating virtual imaginary contributions in a parton shower},
\href{http://dx.doi.org/10.1103/PhysRevD.100.074005}
{Phys. Rev. D \textbf{100}, 074005 (2019)}
[\href{http://inspirehep.net/search?p=find+doi+10.1103/PhysRevD.100.074005}
{\textsc{inSPIRE}}].

\bibitem{backwardevolution} 
  T.~Sjostrand,
  {\em A Model for Initial State Parton Showers}
  \href{http://dx.doi.org/10.1016/0370-2693(85)90674-4}
  {Phys. Lett. B \textbf{157}, 321 (1985)}
  [\href{http://inspirehep.net/search?p=find+doi+10.1016/0370-2693(85)90674-4}
  {\textsc{inSPIRE}}].
  
\bibitem{NSpartons} 
  Z.~Nagy and D.~E.~Soper,
  {\em Parton distribution functions in the context of parton showers},
  \href{http://dx.doi.org/10.1007/JHEP06(2014)179}
  {JHEP \textbf{06}, 179 (2014)}
  [\href{http://inspirehep.net/search?p=find+doi+10.1007/JHEP06(2014)179}
  {\textsc{inSPIRE}}].

\bibitem{ShowerTime}
  Z.~Nagy and D.~E.~Soper,
  {\em Ordering variable for parton showers},
  \href{http://dx.doi.org/10.1007/JHEP06(2014)178}
  {JHEP {\bf 1406} (2014) 178}
  [\href{http://inspirehep.net/search?p=find+doi+10.1007/JHEP06(2014)178}
  {\textsc{inSPIRE}}].
  
\bibitem{Jadach:2015mza} 
  S.~Jadach, W.~P\l{}aczek, S.~Sapeta, A.~Si\'odmok and M.~Skrzypek,
  {\em Matching NLO QCD with parton shower in Monte Carlo scheme \textemdash{} 
  the KrkNLO method},
  \href{http://dx.doi.org/10.1007/JHEP10(2015)052}
  {JHEP \textbf{10}, 052 (2015)}
  [\href{http://inspirehep.net/search?p=find+doi+10.1007/JHEP10(2015)052}
  {\textsc{inSPIRE}}].
  
\bibitem{CMW} 
  S.~Catani, B.~R.~Webber and G.~Marchesini,
  {\em QCD coherent branching and semiinclusive processes at large x},
  \href{http://dx.doi.org/10.1016/0550-3213(91)90390-J}
  {Nucl.\ Phys.\ B {\bf 349} (1991) 635}
  [\href{http://inspirehep.net/search?p=find+doi+10.1016/0550-3213(91)90390-J}
  {\textsc{inSPIRE}}].

\bibitem{NSNewColor}
  Z.~Nagy and D.~E.~Soper,
  {\em Parton showers with more exact color evolution},
  \href{http://dx.doi.org/10.1103/PhysRevD.99.054009}
  {Phys. Rev. D \textbf{99}, 054009 (2019)}
  [\href{http://inspirehep.net/search?p=find+doi+10.1103/PhysRevD.99.054009}
  {\textsc{inSPIRE}}].
  
\bibitem{GapColor}
  Z.~Nagy and D.~E.~Soper,
  {\em Effect of color on rapidity gap survival},
  \href{http://dx.doi.org/10.1103/PhysRevD.100.074012}
  {Phys. Rev. D \textbf{100}, 074012 (2019)}
  [\href{http://inspirehep.net/search?p=find+doi+10.1103/PhysRevD.100.074012}
  {\textsc{inSPIRE}}].
 
\bibitem{CT14} 
  S.~Dulat {\it et al.},
  {\em New parton distribution functions from a global analysis 
  of quantum chromodynamics}
  \href{http://dx.doi.org/10.1103/PhysRevD.93.033006}
  {Phys.\ Rev.\ D {\bf 93},  033006 (2016)}
  [\href{http://inspirehep.net/search?p=find+doi+10.1103/PhysRevD.93.033006}
  {\textsc{inSPIRE}}].

\bibitem{BNX} 
  T.~Becher, M.~Neubert and G.~Xu,
  {\em Dynamical Threshold Enhancement and Resummation in 
  Drell-Yan Production},
  \href{http://dx.doi.org/10.1088/1126-6708/2008/07/030}
  {JHEP {\bf 0807} (2008) 030}
  [\href{http://inspirehep.net/search?p=find+doi+10.1088/1126-6708/2008/07/030}
  {\textsc{inSPIRE}}].

\bibitem{MCFM} 
  J.~M.~Campbell and R.~K.~Ellis,
  {\em Radiative corrections to Z b anti-b production},
  \href{http://dx.doi.org/10.1103/PhysRevD.62.114012}
  {Phys.\ Rev.\ D {\bf 62} (2000) 114012}
  [\href{http://inspirehep.net/search?p=find+doi+10.1103/PhysRevD.62.114012}
  {\textsc{inSPIRE}}].

\bibitem{CSSpT}
  J.~C.~Collins, D.~E.~Soper and G.~F.~Sterman,
  {\em Transverse Momentum Distribution in Drell-Yan Pair and 
  W and Z Boson Production},
  \href{http://dx.doi.org/10.1016/0550-3213(85)90479-1}
  {Nucl.\ Phys.\ B {\bf 250} (1985) 199}
  [\href{http://inspirehep.net/search?p=find+doi+10.1016/0550-3213(85)90479-1}
  {\textsc{inSPIRE}}].
  
\bibitem{ResBos1}
  G.~A.~Ladinsky and C.~P.~Yuan,
  {\em The Nonperturbative regime in QCD resummation for gauge 
  boson production at hadron colliders},
  \href{http://dx.doi.org/10.1103/PhysRevD.50.R4239}
  {Phys.\ Rev.\ D {\bf 50} (1994) 4239}
  [\href{http://inspirehep.net/search?p=find+doi+10.1103/PhysRevD.50.R4239}
  {\textsc{inSPIRE}}].
  
\bibitem{ResBos}
  F.~Landry, R.~Brock, P.~M.~Nadolsky and C.~P.~Yuan,
  {\em Tevatron Run-1 $Z$ boson data and Collins-Soper-Sterman 
  resummation formalism},
  \href{http://dx.doi.org/10.1103/PhysRevD.67.073016}
  {Phys.\ Rev.\ D {\bf 67} (2003) 073016}
  [\href{http://inspirehep.net/search?p=find+doi+10.1103/PhysRevD.67.073016}
  {\textsc{inSPIRE}}]

\bibitem{NSZpT}
  Z.~Nagy and D.~E.~Soper,
  {On the transverse momentum in Z-boson production in a virtuality 
  ordered parton shower},
  \href{http://dx.doi.org/10.1007/JHEP03(2010)097}
  {JHEP \textbf{03}, 097 (2010)}
  [\href{http://inspirehep.net/search?p=find+doi+10.1007/JHEP03(2010)097}
  {\textsc{inSPIRE}}]
  
\bibitem{JCCbook}
  John Collins,
  {\em Foundations of Perturbative QCD},
  Cambridge University Press, Cambridge U.K.\ (2011).

\bibitem{Sterman1987} 
  G.~F.~Sterman,
  {\em Summation of Large Corrections to Short Distance Hadronic 
  Cross-Sections},
  \href{http://dx.doi.org/10.1016/0550-3213(87)90258-6}
  {Nucl.\ Phys.\ B {\bf 281} (1987) 310}\
  [\href{http://inspirehep.net/search?p=find+doi+10.1016/0550-3213(87)90258-6}
  {\textsc{inSPIRE}}].

\bibitem{Hoeche2018}
  S.~H\"oche, D.~Reichelt and F.~Siegert,
  {\em Momentum conservation and unitarity in parton showers and NLL resummation},
  \href{http://dx.doi.org/10.1007/JHEP01(2018)118}
  {JHEP \textbf{01}, 118 (2018)}
  [\href{http://inspirehep.net/search?p=find+doi+10.1007/JHEP01(2018)118}
  {\textsc{inSPIRE}}].
  
  \bibitem{Dasgupta2018}
  M.~Dasgupta, F.~A.~Dreyer, K.~Hamilton, P.~F.~Monni and G.~P.~Salam,
  {\em Logarithmic accuracy of parton showers: a fixed-order study},
  \href{http://dx.doi.org/10.1007/JHEP09(2018)033}
  {JHEP \textbf{09}, 033 (2018)
  [erratum: JHEP \textbf{03}, 083 (2020)]}
  [\href{http://inspirehep.net/search?p=find+doi+10.1007/JHEP09(2018)033}
  {\textsc{inSPIRE}}].

\bibitem{angular2020}
  G.~Bewick, S.~Ferrario Ravasio, P.~Richardson and M.~H.~Seymour,
  {\em Logarithmic accuracy of angular-ordered parton showers},
  \href{http://dx.doi.org/10.1007/JHEP04(2020)019}
  {JHEP \textbf{04}, 019 (2020)}
  [\href{http://inspirehep.net/search?p=find+doi+10.1007/JHEP04(2020)019}
  {\textsc{inSPIRE}}].

\bibitem{DasguptaShowerSum} 
  M.~Dasgupta, F.~A.~Dreyer, K.~Hamilton, P.~F.~Monni, G.~P.~Salam and G.~Soyez,
  {\em Parton showers beyond leading logarithmic accuracy},
  \href{http://dx.doi.org/10.1103/PhysRevLett.125.052002}
  {Phys. Rev. Lett. \textbf{125}, 052002 (2020)}
  [\href{http://inspirehep.net/search?p=find+doi:10.1103/PhysRevLett.125.052002}
  {\textsc{inSPIRE}}].
  
\bibitem{HamiltonShowerSum}
  K.~Hamilton, R.~Medves, G.~P.~Salam, L.~Scyboz and G.~Soyez,
  {\em Colour and logarithmic accuracy in final-state parton showers},
  \href{https://arxiv.org/abs/2011.10054}
  {[arXiv:2011.10054 [hep-ph]]}.

\bibitem{StermanZeng} 
  G.~Sterman and M.~Zeng,
  {\em Quantifying Comparisons of Threshold Resummations},
  \href{http://dx.doi.org/10.1007/JHEP05(2014)132}
  {JHEP {\bf 1405}, 132 (2014)}
  [\href{http://inspirehep.net/search?p=find+doi+10.1007/JHEP05(2014)132}
  {\textsc{inSPIRE}}]



\end{thebibliography}
\end{document}